\documentclass[12pt]{article}
\usepackage{amsmath,amssymb,epsfig,times}
\usepackage[nolists]{endfloat}
\usepackage{subfigure}
\usepackage{natbib}
\usepackage{times}
\usepackage[usenames]{color}
\bibpunct{(}{)}{;}{a}{,}{,}

\newcommand{\bC}{\mathbf{C}}

\def\N{\mathbb{N}}
\def\R{\mathbb{R}}

\def\topint{\operatorname{int}}
\def\topcl{\operatorname{cl}}
\def\cvxhull{\operatorname{cvxhull}}

\def\Gn{G_n}
\def\Ginf{G_{\infty}}

\def\Gbn{G^{(n)}}

\def\ainf{a_{\infty}}
\def\binf{b_{\infty}}

\def\bst{b^{\star}}

\def\So{S_0}
\def\bGo{\bar{G}_0}

\def\pist{\pi^{\star}}
\newcommand{\set}[1]{\{#1\}}
\newcommand{\Lset}[1]{\left\{#1\right\}}

\def\eps{\varepsilon}
\def\Th{\Theta}
\def\tTh{\bar{\Th}}

\def\tThinf{\tTh_{\infty}}
\def\tThn{\tTh_{n}}

\def\th{\theta}

\def\mBTh{\mathcal{B}(\Th)}

\def\Go{G_0}
\def\Co{C^0}
\def\Cob{\Co_b}
\def\CobTh{\Cob(\Th)}
\def\CoTh{\Co(\Th)}

\def\Wp{W_p}

\def\mP{\mathcal{P}}

\def\mPTh{\mP(\Theta)}
\def\mPpTh{\mP_p(\Theta)}
\def\mPOTh{\mP_1(\Theta)}
\def\mPbTh{\mP_*(\Theta)}

\def\supp{\operatorname{supp}}
\def\Prob{\operatorname{\mathbb{P}}}
\def\ind{\mathbb{I}}

\def\Gbn{G^{(n)}}
\def\Gbon{\Go^{(n)}}
\def\Gn{G_n}

\def\bG{\bar{G}}

\def\bGwn{\bGw_n}
\def\bGbon{\bGo^{(n)}}

\def\Bor{\mathcal{B}}

 \def\bGwn{\bar{G}(\omega)}

\def\mX{\mathcal{X}}

\def\bphi{\bar{\phi}}
\def\bpi{\bar{\pi}}
\newcommand{\nrmInf}[1]{\|#1\|_{\infty}}
\newcommand{\LnrmInf}[1]{\left\|#1\right\|_{\infty}}
\def\bmu{\bar{\mu}}

\def\gam{\gamma}
\def\Gam{\Gamma}
\def\bgam{\bar{\gamma}}

\def\ITh{I_{\Th}}
\def\IPh{I_{\Phi}}

\def\mPp{\mathcal{P}_p}

\def\mB{\mathcal{B}}
\def\mY{\mathcal{Y}}
\def\mD{\mathcal{D}}
\def\mDY{\mathcal{D}(\mY)}
\def\mDoY{\mathcal{D}^0(\mY)}
\def\mDocY{\mathcal{D}_c^0(\mY)}
\def\mDs{\mathcal{D}_*}
\def\mDsY{\mDs(\mY)}
\def\im{\operatorname{im}}

\def\adj{^*}

\def\mB{\mathcal{B}}
\def\mF{\mathcal{F}}

\def\mY{\mathcal{Y}}
\def\mYs{\mY^*}
\def\mD{\mathcal{D}}
\def\mDY{\mathcal{D}(\mY)}
\def\mDoY{\mathcal{D}^0(\mY)}
\def\mDocY{\mathcal{D}_c^0(\mY)}
\def\mDs{\mathcal{D}_*}
\def\mDsY{\mDs(\mY)}
\def\im{\operatorname{im}}

\def\adj{^*}
\def\Ths{\Th^*}
\def\Phis{\Phi^*}
\def\Hd{\mathcal{H}_d}

\def\uod{u_0^{\delta}}

\newtheorem{definition}{Definition}
\newtheorem{proposition}{Proposition}
\newtheorem{theorem}{Theorem}
\newtheorem{property}{Property}

\newtheorem{lemma}{Lemma}

\allowdisplaybreaks

\title{\bf Discrete Sequential Barycenter Arrays: Representation, Approximation, and Modeling of Probability Measures}

\author{{\sc Alejandro Jara and Carlos Sing-Long}}

\begin{document}
\date{\today}
\maketitle \footnotetext[1]{Alejandro Jara is Full Professor,
Department of Statistics, Pontificia Universidad Cat\'{o}lica de
Chile, Santiago, Chile (E-mail~: atjara@uc.cl). Supported by Fondecyt grant 1220907. Carlos Sing-Long is Associate Professor,
Institute for Mathematical and Computational Engineering Pontificia Universidad Cat\'{o}lica de Chile, Santiago, Chile. Supported by grants Puente 2025, Open Seed Fund 2024, and by the National Center for Artificial Intelligence CENIA FB210017, Basal ANID.}

\begin{abstract}
Constructing flexible probability models that respect constraints on key functionals --such as the mean-- is a fundamental problem in nonparametric statistics. Existing approaches lack systematic tools for enforcing such constraints while retaining full modeling flexibility. This paper introduces a new representation for univariate probability measures based on discrete sequential barycenter arrays (SBA). We study structural properties of SBA representations and establish new approximation results. In particular, we show that for any target distribution, its SBA-based discrete approximations converge in both the weak topology and in Wasserstein distances, and that the representation is exact for all distributions with finite discrete support. We further characterize a broad class of measures whose SBA partitions exhibit regularity and induce increasingly fine meshes, and we prove that this class is dense in standard probabilistic topologies. These theoretical results enable the construction of probability models that preserve prescribed values --or full distributions-- of the mean while maintaining large support. As an application, we derive a mixture model for density estimation whose induced mixing distribution has a fixed or user-specified mean. The resulting framework provides a principled mechanism for incorporating mean constraints in nonparametric modeling while preserving strong approximation properties. The approach is illustrated using both simulated and real data.\\

\noindent\emph{Keywords:} Prior elicitation; Mixture models; Mean constraints; Random sequential barycenter arrays.
\end{abstract}

\section{Introduction}

Discrete mixture models of continuous parametric distributions underpin many approaches to density estimation and modelling. Notable examples include kernel-based techniques \citep[see, e.g.,][]{silverman;81}, nonparametric maximum likelihood methods \citep[see, e.g.,][]{lindsay;83}, and Bayesian approaches based on finite or infinite mixtures \citep[see, e.g.,][and references therein]{mueller;quintana;jara;hanson;2015}. A mixture model for a density on a sample space $\mathcal{Y}$ takes the form
\begin{equation*}
f(y \mid G) = \int_{\Theta} k(y \mid \theta) G(d \theta),
\end{equation*}
where $k(\cdot \mid \cdot)$ is a fixed, non-negative, continuous kernel defined on $(\mathcal{Y} \otimes \Theta,  \mathcal{B}(\mathcal{Y}) \otimes \mathcal{B}(\Theta))$, such that for each $\theta \in \Theta$, $\int_{\mathcal{Y}} k(y \mid \theta)\, dy = 1$, and for each $y \in \mathcal{Y}$, $\int_{\Theta} k(y \mid \theta)\, dG(\theta) < \infty$. Here, $\mathcal{Y}$ and $\Theta$ are Borel subsets of Euclidean spaces, and $\mathcal{B}(\mathcal{Y})$ and $\mathcal{B}(\Theta)$ denote the corresponding Borel $\sigma$--fields. 
 In the Bayesian framework the mixing distribution $G$ is typically modeled as a random discrete probability measure,
\[
G(\cdot) = \sum_{l=1}^m w_l \delta_{\theta_l}(\cdot),
\]
where $1 \leq m \leq \infty$, the weights $(w_1,\ldots,w_n)$ are non-negative and sum to one, $(\theta_1,\ldots,\theta_m)$ are component-specific parameters, and $\delta_{\theta}(\cdot)$ denotes the Dirac measure at $\theta$. Popular models for $G$ include the Dirichlet–Multinomial allocation model \citep[see, e.g.,][]{green;richardson;97}, the Dirichlet process \citep{ferguson;73}, the Poisson–Dirichlet process \citep{pitman;yor;97}, normalized random measures \citep[see, e.g.,][]{nieto;pruenster;walker;2004, LijoiPrunster2010}, and more general stick-breaking priors \citep[see, e.g.,][]{GilLeyva2023}, among others.

A rich theoretical literature supports the use of discrete mixtures for density estimation and more general hierarchical models, establishing strong results on approximation, support, posterior consistency, and convergence rates \citep[see, e.g.,][and references therein]{mueller;quintana;jara;hanson;2015,GhosalVanDerVaart2017}. However, in many applications, domain knowledge is naturally expressed in terms of constraints or partial information about statistical functionals, most notably the mean, and incorporating such information in a principled manner is challenging for most existing mixture model frameworks. For example, practitioners may have prior information about certain functionals, often derived from historical data or expert opinion, which should be directly incorporated into the model. Additionally, hard constraints on functionals may be necessary to ensure model identifiability. For instance, in regression or hierarchical models, such constraints allow for meaningful interpretations of location parameters. 

The main challenge in the construction of flexible probability models that can incorporate structural information about specific functionals arises from the difficulty in deriving the induced distribution of them \citep[see, e.g.,][]{cifarelli;regazzini;90,james;2005,james;lijoi;pruesnter;2008,LijoiRegazzini2004,lijoi;pruenster;2009, GaffiLijoiPrunster2025}. \cite{kessler;hoff;dunson;2015} proposed a general method for constructing marginally specified models, based on standard Bayesian nonparametric priors and an importance sampling scheme. However, their approach requires knowledge of the induced distribution on the functional of interest. While this distribution can be numerically approximated, the method also relies on the induced and target distributions sharing a common dominating measure--a condition not satisfied when hard constraints on the functional are needed. This challenge arises irrespective of the inferential paradigm, affecting both likelihood-based and Bayesian approaches.

In this work, we develop a representation-based solution to this problem using \emph{sequential barycenter arrays} (SBA). The barycenter construction is an intuitive geometric device for representing probability measures through a sequence of nested partitions and their conditional means. Building on ideas originating in \cite{hill;monticino;98} and on more recent developments in Wasserstein geometry \citep{Villani2003, villani2008optimal}, the SBA representation encodes a univariate probability measure through a triangular array of cumulative barycenters. We focus on the \emph{discrete} SBA representation obtained by truncating the infinite array at depth $n$, and study several structural and approximation properties of this representation. In particular, we show that discrete SBA representations determine exactly all distributions with finite support, and that the discrete SBA approximation converges to the target distribution in both the weak topology and in Wasserstein distances as the array depth increases. We demonstrate that the proposed class of distributions is dense in standard probabilistic topologies.

These approximation results serve as the basis for a constrained modeling framework in which the mean of the resulting measure can be fixed or assigned a user-specified distribution. The SBA structure allows the construction of discrete probability measures whose mean is preserved exactly, leading to a family of distributions that retain full nonparametric flexibility while satisfying functional constraints. We refer to such distributions as \emph{discrete SBA} (DSBA) measures. The DSBA family is rich, enjoys large support properties, and provides a principled bridge between structural constraints and flexible nonparametric modeling.

We then consider the density estimation via location and location-scale mixture models. Replacing the mixing distribution with its discrete SBA approximation yields a class of mixture models whose induced densities inherit the approximation properties of the SBA representation and preserve the mean of the mixing distribution. While we develop a Bayesian implementation, due to its computational convenience and the interpretability of prior specification, the representation is not restricted to the Bayesian paradigm. In fact, the finite mixture structure induced by the discrete SBA array admits a latent variable formulation that makes nonparametric maximum likelihood estimation (NPMLE) feasible and constitutes an appealing direction for future research and highlights the generality of the SBA framework.

The main contributions of this paper are as follows. First, we introduce and study discrete SBA representations of probability measures, establishing new structural and approximation results. Second, we develop a constrained modeling framework that preserves the mean while retaining the flexibility of nonparametric mixtures. Third, we present an application to density estimation via discrete SBA mixtures, illustrating the practical utility of the approach and its theoretical advantages. Collectively, these results demonstrate that SBA-based representations provide a principled, theoretically grounded, and computationally tractable mechanism for integrating functional constraints into nonparametric modeling, with implications that extend beyond Bayesian inference.

The paper is organized as follows. Section~\ref{section2} reviews SBA probability measures. Section~\ref{section3} introduces the proposed model and its properties. 
Section~\ref{section4} provides details for posterior computation. Section~\ref{section5} presents applications to simulated and real data. Finally, Section~\ref{section6} offers concluding remarks and directions for future work. The proofs of all mathematical results are provided in the Appendix section, which also includes a preliminaries section containing the additional definitions and technical ingredients required for the development of the proofs (Appendix \ref{Appendix0}). We have tried to make the appendix self-contained, providing detailed proofs of our claims.

\section{The SBA Construction}\label{section2}

\subsection{SBA definition}

The following definitions are slight modifications of those provided by \cite{hill;monticino;98}. From now on, we assume that \(\Th\) is either the real line, a closed half-line, or a compact interval. Let $G$ be a probability measure defined on the measurable space $(\Theta, \mathcal{B}(\Theta))$. From now on, we do not differentiate between \(G\) and its cumulative distribution function.

\begin{definition}\label{def1} 
For every $a_1, a_2 \in \Theta$ such that $a_2 \ge a_1$, the $G$--barycenter of the interval $(a_1, a_2]$, denoted $b_G(a_1, a_2]$, is defined as
\[
b_G(a_1, a_2] = 
\begin{cases}
\displaystyle \frac{\int_{(a_1,a_2]} \theta \, dG(\theta)}{G(a_2) - G(a_1)}, & \text{if } G(a_2) > G(a_1), \\
a_1, & \text{if } G(a_2) = G(a_1).
\end{cases}
\]
\end{definition}

Thus, the $G$--barycenter of $(a_1, a_2]$ corresponds to the conditional expectation of $G$ over that interval. The SBA of a probability measure $G$ is a triangular array where each entry represents the barycenter of a specific subinterval constructed recursively to encode the structure of $G$.
\begin{definition}\label{def2}
The SBA of $G$, denoted $\{\mu_{j,l}\}_{j=1,\, l=1}^{\infty,\, 2^j - 1}$, is defined inductively as follows:
\begin{itemize}
\item[(i)] $\mu_{1,1} = b_G(\Theta) = \int_{\Theta} \theta \, dG(\theta)$,
\item[(ii)] for  $j =2, 3, \ldots$  and $l = 1, \ldots, 2^{j-1} - 1$, $\mu_{j, 2\cdot l} = \mu_{j-1, l}$, 
\item[(iii)]  for $j =2, 3, \ldots$ and $l = 1, \ldots, 2^{j-1}$, $\mu_{j, 2\cdot l-1} = b_G(\mu_{j-1, l-1}, \mu_{j-1, l}]$, 
\end{itemize}
with the convention that $\mu_{j,0} = \inf \Theta$ and $\mu_{j,2^j} = \sup \Theta$.
\end{definition}

Note that item (iii) defines $\mu_{j, 2\cdot l-1}$ as the $G$--conditional barycenter of the interval $$(\mu_{j-1, l-1}, \mu_{j-1, l}],$$ for $j =2, 3,  \ldots$. The structure can be visualized as an infinite binary tree. \cite{hill;monticino;98} showed how $G$ is determined by its SBA and provided an inversion formula. Given the SBA, the values of the cumulative distribution function of $G$ at the barycenters can be computed via a recursive formula. This result is key to reconstructing $G$ from its SBA.

\begin{property}[\cite{hill;monticino;98}]\label{property1}
Let $\{\mu_{j,l}\}_{j=1,\, l=1}^{\infty,\, 2^j - 1}$ be the SBA of $G$. Then, the values of the cumulative distribution function at the points in the SBA can be computed recursively as follows:
\begin{itemize}
\item[(i)] $G(\mu_{1,1}) = \frac{\mu_{2,3} - \mu_{2,2}}{\mu_{2,3} - \mu_{2,1}}$, 
\item[(ii)] and, for  $j =2, 3, \ldots$ and  $l = 1, \ldots, 2^{j-1}$,
\[
G(\mu_{j,0}) = 0, \quad G(\mu_{j,2^j}) = 1,
\]
and
\[
G(\mu_{j,2\cdot l - 1}) = G(\mu_{j-1,l-1}) + \left[ G(\mu_{j-1,l}) - G(\mu_{j-1,l-1}) \right] 
\cdot 
\left[ \frac{ \mu_{j+1,4 \cdot l -1} - \mu_{j+1,4 \cdot l-2} }{ \mu_{j+1,4 \cdot l - 1} - \mu_{j+1,4 l-3} } \right],\]
with the convention that $0/0 = 1$, which corresponds to assigning full conditional mass to a degenerate subinterval when the numerator and denominator simultaneously vanish.
\end{itemize}
\end{property}

\cite{hill;monticino;98} also provided necessary and sufficient conditions for a triangular array to be the SBA of some probability measure.

\begin{property}[\cite{hill;monticino;98}]\label{property2}
A triangular array $\{\mu_{j,l}\}_{j=1,\, l=1}^{\infty,\, 2^j - 1}$ is the SBA of a probability measure $G$ if and only if:
\begin{itemize}
\item[(i)] $\mu_{j,2\cdot l} = \mu_{j-1,l}$, for all $j =2,3, \ldots$ and $l = 1, \ldots, 2^{j-1} - 1$,
\item[(ii)] $\mu_{j,l-1} \le \mu_{j,l}$, for all $j =2,3,\ldots$ and $l = 1, \ldots, 2^j$,
\item[(iii)] $\mu_{j,4\cdot l-3} = \mu_{j, 4 \cdot l-2}$ if and only if $\mu_{j,4\cdot l-1} = \mu_{j, 4\cdot l-2}$ for all $j =2, 3, \ldots$ and $l = 1, \ldots, 2^{j-2}$.
\end{itemize}
\end{property}

Based on these results, \cite{hill;monticino;98}  proposed generating random probability measures on the unit interval $[0,1]$ by randomly generating triangular arrays that are an SBA almost surely, and then reconstructing the probability measure via the inversion formula. Since the distribution of the initial barycenter $\mu_{1,1} = \int_{\Theta} \theta \, dG(\theta)$ can be arbitrarily specified, the construction allows for the generation of random probability measures with a prescribed mean or even a prescribed distribution for the mean. Although this approach can be extended to generate random probability measures supported on more general subsets of $\mathbb{R}$, using it for nonparametric modeling of smooth densities would require sampling infinite-dimensional SBAs, which poses additional challenges.

\subsection{Regular SBA}
\label{sec:regSBA}

A probability measure \(G\) has an SBA if its expected value is well defined. In this case, the construction of the SBA suggests that for any level \(n\) there are \(2^n - 1\) barycenters 
 \(\set{\mu_{n,l}: l=1, \ldots, 2^n - 1}\). However, several of them may coincide and thus, in practice, a measure has at most \(2^n - 1\) {\em distinct} barycenters at level \(n\). Repeated barycenters require a careful handling in the reconstruction formula and thus it is of interest to identify a class of measures for which its barycenters at all levels are 
distinct. Here we will focus on the case where the SBA  has  \(2^n - 1\) distinct barycenters up to level \(n\). 

\begin{definition}\label{def:regSBA}
Let \(\mPTh\) be set of all probability measures defined on \((\Th,\mBTh)\). Let $\mPOTh$  be set of all probability measures defined on \((\Th,\mBTh)\) and with finite first moment. 
    Let \(G\in \mPOTh\) and let \(n \in \N\). We say that \(G\) it has a {\em regular level \(n\) SBA} or, alternatively, that {\em the level \(n\) SBA is regular} if its barycenters at level \(n\), \(\set{\mu_{n,l}: l=1, \ldots, 2^n - 1}\), are distinct. We say that \(G\) has a {\em regular SBA} if it has a regular level \(n\) SBA for all \(n\in \N\). 
\end{definition}

When a measure \(G\) has a regular SBA, its level \(n\) SBA induces the \(2^n\) non-degenerate disjoint intervals
\begin{equation*}
    \Th_{n,l} = \begin{cases}
        [\mu_{n,0}, \mu_{n,1}],                  & l = 1,\,\, \mu_{n, 0} > -\infty,\\
        (\mu_{n,0}, \mu_{n,1}],                  & l = 1,\,\, \mu_{n, 0} = -\infty,\\
        (\mu_{n, l-1}, \mu_{n, l}],              & l\in\set{1,\ldots, 2^n - 1},\\
        (\mu_{n, 2^n - 1}, \mu_{n, 2^n}],        & l = 2^n,\,\, \mu_{n,2^n} < \infty,\\
        (\mu_{n, 2^n - 1}, \mu_{n, 2^n}),        & l = 2^n,\,\, \mu_{n,2^n} = \infty,
    \end{cases}
\end{equation*}
which form a partition of \(\Th\). However, that they do {\em not} form a partition of the support of \(G\), but rather a covering. This is because barycenters do not have to belong to the support of \(G\). Consider the following example
\begin{equation*}
    G(\cdot) = \frac{1}{4}\mathcal{U}_{[-2, -1]}(\cdot) + \frac{1}{4}\delta_{-1}(\cdot) + \frac{1}{4} \delta_{+1}(\cdot) + \frac{1}{4}\mathcal{U}_{[+1, +2]}(\cdot),
\end{equation*}
where $\mathcal{U}_{A}(\cdot)$ refers to the continuous uniform distribution on the set $A$. It is apparent that \(0 = \mu_{1, 1} = \mu_{n, 2^{n-1}}\) does not belong to the support, and that for \(n > 1\) the interval \(\Th_{n, 2^{n-1}} = (\mu_{n,2^{n-1} - 1}, \mu_{n, 2^{n-1}}]\) is not contained in the support of \(G\). In this regard, a pathological example is the uniform measure on the Cantor set when \(\Th = [0, 1]\). Due to the symmetry and self-similarity of the Cantor set, it is straightforward to see that {\em no} barycenter is on the support of the measure. The measure in the previous example is also useful to illustrate that it is {\em not} true that the measures of the sets \(\set{\Th_{n,l}}_{j=0}^{2^n}\) tend to zero as \(n\) tends to infinity, as we have that   
$$
G(\Th_{n, 2^{n-1}}) > \frac{1}{4},
$$
for every $n > 1$. Finally, it is {\em not} true that their {\em Lebesgue measure} tends to zero, as we also have that
\[
 |\Th_{n, 2^{n-1}}| \geq 1,
\]
for every $n > 1$. For reasons that shall become clear, we now identify a class of measures with a regular SBA, and such that its SBA induces a partition of their support for which the Lebesgue measure of each part tends to zero as \(n\) tends to infinity. From now on, if \(G\) has a regular SBA, we let
\begin{equation}\label{eq:regSBA:partitionOfSupport}
    \tTh_{n, l} = \begin{cases}
        \Th_{n, 1} \cap \supp(G),       & l = 1,\\
        \Th_{n, l},                     & l\in\set{2,\ldots, 2^n - 1},\\
        \Th_{n, 2^{n}} \cap \supp(G),   & l = 2^n.
    \end{cases}
\end{equation}
The following lemma, proved in Appendix \ref{AppendixA1}, shows that it suffices to assume that \(G\) is supported on a non-degenerate interval to conclude that these sets are non-degenerate intervals that partition \(\Th\).

\begin{lemma}\label{lem:regSBAPartitionSupport}
    Let \(G\in \mPOTh\) be such that \(\supp(G)\) is a non-degenerate interval. Then, \(G\) has a regular SBA. Furthermore, for every \(n \in\N\) the collection of intervals \(\set{\tTh_{n,l}}_{l= 1}^{2^{n}}\) defined in~\eqref{eq:regSBA:partitionOfSupport} are non-degenerate and form a partition of \(\supp(G)\).
\end{lemma}

This lemma motivates us to define the class
\[
    \mPbTh := \set{G\in \mPTh:\,\, \mbox{\(\supp(G)\) is a non-degenerate compact interval}}.
\]
Interestingly, these measures have barycentric decompositions for which both the measure and the Lebesgue measure of the intervals \(\set{\tTh_{n,l}}_{l=1}^{2^n}\) tend to zero as \(n\) tends to infinity. The proof of the following lemma is provided in Appendix \ref{AppendixA2}.

\begin{lemma}\label{lem:regSBAIntervalReduction}
    Let \(G\in\mPbTh\) and let \(\set{\tTh_{n,l}}_{n=1,l=1}^{\infty, 2^n}\) be the sequence of intervals defined in~\eqref{eq:regSBA:partitionOfSupport}. For any \(l:\N\to \N\), such that
    \[
        l(n) \in \set{1,\ldots, 2^{n}},
    \]
    and such that the sequence \(\set{\tTh_{n, l(n)}}_{n\in\N}\) is decreasing, we have that
    \[
        \lim_{n\to\infty}\,\, |\tTh_{n,l(n)}| = 0.
    \]
    Furthermore, it follows from this that
    \[
        \lim_{n\to\infty} \sup_{l\in \set{0,\ldots, 2^n}}\,\, |\tTh_{n,l}| = 0.
    \]
\end{lemma}

Intuitively, the measures in \(\mPbTh\) have a SBA that is well-behaved. Surprisingly, this behavior is {\em generic} in the weak topology, i.e., \(\mPbTh\) is dense in \(\mPTh\) in the weak topology. The proof of the following lemma is provided in Appendix \ref{AppendixA3}.

\begin{lemma}\label{lem:regSBADenseWeakStar}
    The space \(\mPbTh\) is dense in \(\mPTh\). 
\end{lemma}

It is also interesting to consider the density of this set on the space \(\mPpTh\) of probability measures with finite \(p\)-th moment, for \(p\in [1,\infty)\). In fact, remark that \(\mPbTh\in \mPpTh\) for any \(p\in [1, \infty)\). Our next result shows that the behavior exhibited by the measures in \(\mPbTh\) is also generic in \(\mPpTh\) in the Wasserstein distance of order \(p\). The proof of the following lemma is provided in Appendix \ref{AppendixA4}.

\begin{lemma}\label{lem:regSBADenseWasserstein}
    Let \(p\in [1,\infty)\). The space \(\mPbTh\) is dense in \(\mPpTh\).
\end{lemma}

These results show that it suffices to focus on constructing approximations of measures that have a regular SBA, and that induce partitions of its support, to approximate any measure in \(\mPTh\) endowed with the weak topology, or in \(\mPpTh\) endowed with the Wasserstein distance of order \(p\) for \(p\in [1,\infty)\).

\section{Discrete SBA mixture models}\label{section3}

\subsection{Mean mixture models}

We begin the construction by considering a mean mixture model of the form
\begin{eqnarray}\nonumber
f(\cdot \mid \phi, G) &=&\int_{\Theta} k\left (\cdot \mid \theta, \phi \right) dG( \theta),
\end{eqnarray}
where $k(\cdot\mid \theta,\phi)$ is the density of an absolutely continuous distribution with mean
$\theta \in \Theta$ and dispersion parameter $\phi \in \mathbb{R}_+$, defined on the appropriate sample space $(\mathcal{Y},\mathcal{B}(\mathcal{Y}))$, with $\mathcal{Y} \subseteq \mathbb{R}$, and $G$ is a discrete mixing distribution defined on
$(\Theta, \mathcal{B}(\Theta))$, such that
\begin{eqnarray}\nonumber
G(\cdot) = \sum_{l=1}^m w_l \delta_{\theta_l}(\cdot),
\end{eqnarray}
with $1 \leq m \leq \infty$, $(w_1,\ldots,w_m)$ being a point in the $(m-1)$-dimensional simplex, and $\delta_{\theta}(\cdot)$ being the Dirac measure at $\theta$. The choice of the appropriate distribution $k(\cdot\mid \theta,\phi)$ depends on the underlying sample space. If the density function $f(\cdot \mid \phi, G)$ is defined on the entire real line, a Gaussian distribution, $\mathcal{N}(\theta,\phi)$, is the standard starting point for $k(\cdot\mid \theta,\phi)$. On the unit interval, a Beta distribution parameterized in terms of its mean $\theta$ and dispersion $\phi$, 
\begin{eqnarray}\nonumber
k(y \mid \theta,\phi)= \frac{\Gamma\left(\phi \right)}{\Gamma\left(\theta\phi \right)\Gamma\left((1-\theta)\phi\right)} y^{\theta \phi -1} \left(1-y\right)^{(1-\theta)\phi-1},
\end{eqnarray}
where $\theta \in (0,1)$ and $\phi \in \mathbb{R}_+$, form a flexible two parameter family. On the positive half line, the use
of a Gamma distribution,
\begin{eqnarray}\nonumber
k(y\mid \theta,\phi)=
\frac{(\phi/\theta)^{\phi}}{\Gamma(\phi)}\,y^{\phi-1}\exp\!\left\{-\tfrac{\phi}{\theta}y\right\},
\end{eqnarray}
where $\theta \in \mathbb{R}_+$, is a sensible option.

With the chosen parameterizations, the mean of the resulting mixture distribution coincides with the mean of the mixing distribution,
\[
\int_{\mathcal{Y}} y f(y \mid \phi, G)\, dy = \int_{\Theta} \theta\, dG(\theta).
\]
Thus, to specify a prior over densities with a fixed or prescribed distribution for the mean, it is enough to define a random discrete mixing distribution $G$ with that desired property. We propose a Bayesian model that achieves this through a discrete approximation to a random probability measure via random SBA. Specifically, we assume
\[
G \mid n,  \mathcal{H}_n \sim \mbox{DSBA}(n, \mathcal{H}_n),
\]
where DSBA denotes a discrete SBA random probability measure with parameters $(n, \mathcal{H}_n)$,
with $n \in \mathbb{N}$ and $\mathcal{H}_n$ being a collection of probability measures defined on $(\Theta, \mathcal{B}(\Theta))$.

\subsection{Discrete SBA probability measures}

We propose a discrete approximation of a measure $G\in\mPbTh$ based on the partition of the support $\Theta$, induced by its SBA. For \(n > 1\), given the partition $\left\{\Theta_{n,l} \right\}_{l=1}^{2^{n}}$ of $\Theta$, consider the decomposition of the probability distribution $G$ given by
\begin{equation}\nonumber
G(\cdot) = \sum_{l=1}^{2^{n}} G\left(\Theta_{n,l}\right) G\left(\cdot \mid \Theta_{n,l} \right),
\end{equation}
where $G\left(\Theta_{n,l}\right)$ is the $G$-measure of the interval $\Theta_{n,l}$, and $G\left(\cdot \mid \Theta_{n,l} \right)$ is the restriction of $G$ to the set $\Theta_{n,l}$, defined by
$G(B \mid \Theta_{n,l}) = G(B \cap \Theta_{n,l}) / G(\Theta_{n,l})$, for every $B \in \mathcal{B}(\Theta)$.

A discrete approximation to $G$, denoted by $G^{(n)}$, is proposed by replacing the restriction
of $G$ to the partition set $\Theta_{n,l}$ by the Dirac measure at the $G$-barycenter of that set, in the previous level \(n\) SBA approximation of $G$. That is, we consider
\begin{eqnarray}\nonumber
G^{(n)}(\cdot) &=& \sum_{l=1}^{2^{n}} G(\Theta_{n,l}) \delta_{\mu_{n+1,2\cdot l-1}}(\cdot),\\\nonumber
 &=& \sum_{l=1}^{2^{n}} w_{n,l} \delta_{\mu_{n+1,2\cdot l-1}}(\cdot),
\end{eqnarray}
where $w_{n,l} = G(\Theta_{n,l})$ and $\mu_{n+1,2\cdot  l-1}$ is the $G$-barycenter of the interval $\Theta_{n,l}$, i.e., 
$$\mu_{n+1,2\cdot l-1} = b_G((\mu_{n, l - 1}, \mu_{n, l}]),$$
for $l = 1, \ldots, 2^{n}-1$ and
$$\mu_{n+1,2^{n+1}-1} = b_G((\mu_{n,2^{n-1}}, \mu_{n,2^n})).$$
 
The probability measure $G^{(n)}$ is referred to as the level \(n\) SBA approximation of $G$. Notice that, by using the inversion formula of Property~\ref{property1}, the quantities $\{G(\Theta_{n,l})\}_{l=1}^{2^{n}}$ can be computed from the first $n+1$ rows of the SBA of $G$, that is, without explicitly knowing $G$. In particular, the level \(n\) SBA approximation satisfies
\[
    G^{(n)}(\Th_{n,l}) = G(\Th_{n,l}).
\]
for every \(l\in\set{1,\ldots, 2^n}\).

The SBA approximation of $G$ retains important properties of the original probability measure. Specifically, it can be verified that both probability measures have the same mean:
\begin{equation*}
\mu_{1,1} = \int_{\Theta} \theta \, dG(\theta) = \int_{\Theta} \theta \, dG^{(n)}(\theta).
\end{equation*}

The SBA approximation is exact for an important class of probability distributions. The following theorem, proved in Appendix~\ref{AppendixB1}, shows that for a probability measure supported on a finite set of elements, its level \(n\) SBA approximation is exact for $n$ large enough.
\begin{theorem}\label{theorem1}
Let $G$ be a discrete probability measure with support on a finite set of points $\{\theta^*_1, \ldots, \theta^*_k\} \in \Theta^k$, where $k \ge 1$ . For every $n \geq 1$, let $G^{(n)}$
be the level \(n\) SBA approximation to $G$. Then, for every
$n \geq k$, the discrete probability measure and its level \(n\) SBA approximation are the same, that is, 
$G = G^{(n)}$.
\end{theorem}

In many topologies the measures of finite support are dense on \(\mPTh\). Since each measure of finite support coincides with its level \(n\) SBA approximation, one may conclude from this that any measure with an expected value can be approximated by its level \(n\) SBA for sufficiently large \(n\). However, our method relies on the level \(n\) SBA approximation of measures in \(\mPbTh\). Our next result provides shows that measure in \(\mPOTh \subset \mPTh\) can be approximated, for sufficiently large \(n\), by the level \(n\) SBA of some measure close to it in the weak topology. The proof of the following theorem is provided in Appendix \ref{AppendixB2}.



\begin{theorem}\label{thm:SBAApproximatingMeasureDenseWeakStar}
    Let \(\Go\in \mPOTh\). Then, for any weak neighborhood \(V\) of \(\Go\) there exists a measure \(G\in \mPTh\) for which there exists \(n\in \N\) such that \(\Gbn \in V\). If \(\Go\in\mPbTh\), then we may choose \(G = \Go\) and in this case \(\set{\Gbn}_{n\in\N}\) converges to \(G\) in the weak topology. 
\end{theorem}

This approximation property of the level \(n\) SBA also holds true in \(\mPpTh\) endowed with the Wasserstein distance of order \(p\). The proof of the following theorem is provided in Appendix \ref{AppendixB3}.
\begin{theorem}\label{thm:SBAApproximatingMeasureDenseWasserstein}
    Let \(p\in [1, \infty)\) and let \(\Go\in \mPpTh\). Then, for any neighborhood \(V\) of \(\Go\) in the Wasserstein distance of order \(p\), there exists \(G\in V\) and \(n\in \N\) such that \(\Gbn \in V\). If \(\Go\in\mPbTh\), then we may choose \(G = \Go\) and, in this case, \(\set{\Gbn}_{n\in\N}\) converges to \(G\) in the Wasserstein distance of order \(p\). 
\end{theorem}

\subsection{Random discrete SBA probability measures}

We formally define the proposed model next.  
\begin{definition}\label{DBSAdef1}
For a given integer $n \geq 1$, set $m = 2^{n}$ and let $\mathcal{H}_n = \{ H_{j,2\cdot l-1}\}_{j=1, l=1}^{n+1 , 2^j-1}$ be a collection of probability measures defined on $(\Theta, \mathcal{B}(\Theta))$. As before, set $\Theta_{n,l} = (\mu_{n,l - 1}, \mu_{n, l}]$ for $l = 1, \ldots, 2^{n}-1$, and
set $\Theta_{n,2^{n}} = (\mu_{n,2^{n-1}}, \mu_{n,2^n})$, with the convention $\mu_{n,0} = \inf \Theta$ and $\mu_{n,2^n} = \sup \Theta$. The random function $G$  defined on the measurable space $(\Theta, \mathcal{B}(\Theta))$ is said to be  
a discrete SBA random probability measure with parameters $(n, \mathcal{H}_n)$,  denoted as
$G \mid n, \mathcal{H}_n \sim \mbox{DSBA}(n, \mathcal{H}_n)$, if:
\begin{itemize}
\item[(i)] $\mu_{1,1} \sim H_{1,1}$,
    
\item[(ii)] $\mu_{j, 2\cdot l} = \mu_{j-1, l}$,  for all $j =2, 3, \ldots,n$  and $l = 1, \ldots, 2^{j-1} - 1$, 

\item[(iii)] $\mu_{j, 2\cdot l-1} \mid \mu_{j-1, l-1}, \mu_{j-1, l} 
\overset{ind.}{\sim} H_{j, 2\cdot l-1} |_{(\mu_{j-1,l-1}, \mu_{j-1,l}]},$
$j =2, 3, \ldots,n+1 $ and $l = 1, \ldots, 2^{j-1}$, where $H|_A$ denotes the restriction of $H$ to the set $A$, and

\item[(iv)] $G(\cdot) \overset{a.s.}{=} \sum_{l=1}^{m} w_{n,l} \delta_{\theta_l}(\cdot)$, where
$\theta_l \overset{a.s.}{=}  \mu_{n+1,2\cdot l-1}$, $w_{n,l} \overset{a.s.}{=}  G(\Theta_{n,l})$, $l=1,\ldots,m$, and $G(\Theta_{n,l})$ is computed using (i) -- (iii) in Property~\ref{property1}.
\end{itemize}
\end{definition}

For a given $n \geq 1$ set $\mathcal{C}_n = \Theta \times \Theta^3 \times \cdots \times \Theta^{2^{n+1}-1}$. Let $\mathcal{BC}_n \subset \mathcal{C}_n$ be the Borel set of all valid sequential barycenter arrays (SBAs) at level $n$. Let $\mathcal{T}_n$ be the mapping that sends an array $\{\mu_{j,l}\}_{j=1, l=1}^{n+1, 2^j-1} \in \mathcal{BC}_n$  to its associated function $G$, using (i) -- (iii) in Property~\ref{property1}. A discrete SBA random probability measure is a stochastic process which trajectories are probability measures defined on $\Theta$ and with law given by $\mathbb{Q}_{\mathcal{H}_n} \circ \mathcal{T}_n^{-1}$, where $\mathbb{Q}_{\mathcal{H}_n}$ is the probability law of $M^{(n)} = \{\mu_{j,l}\}_{j=1, l=1}^{n+1, 2^j-1}$, the random array defined in Definition~\ref{DBSAdef1}. Notice, that 
\[
\mathbb{Q}_{\mathcal{H}_n}\left( M^{(n)} \in \mathcal{BC}_n  \right) = 1.
\]
By construction, the set $\mathcal{BC}_n  \subset \mathcal{C}_n$ contains all arrays  $\{\mu_{j,l}\}_{j=1, l=1}^{n+1, 2^j-1}$ that satisfy the recursive SBA structure. In particular, for all $j = 2, \ldots, n$ and $l = 1, \ldots, 2^{j-1}$, the elements of a valid SBA array satisfy the following properties:
\begin{itemize}
    \item[(a)] $\mu_{j,2\cdot l} = \mu_{j-1,l}$ (even-indexed nodes are inherited from the previous level), and
    \item[(b)] $\mu_{j,2\cdot l-1} \in (\mu_{j-1,l-1}, \mu_{j-1,l}]$ (odd-indexed nodes lie in open-right intervals between parent nodes).
\end{itemize}
These structural constraints are enforced directly in the definition of $M^{(n)}$. Condition (ii) of Definition~\ref{DBSAdef1} imposes the even-indexed identities deterministically. Condition (iii) specifies that $\mu_{j,2\cdot l-1}$ is sampled from $H_{2\cdot l-1}$ restricted to the interval $(\mu_{j-1,l-1}, \mu_{j-1,l}]$, which ensures that $\mu_{j,2\cdot l-1}$ lies in that interval almost surely. 


Definition~\ref{DBSAdef1} always induces a well defined random barycentric array \(M^{(n)}\) on \(\mathcal{BC}_n\). However, to prove it generates a valid stochastic process on \(\mPTh\) and \(\mPpTh\), we need to show that $\mathcal{T}_n$, the mapping that sends an array $\{\mu_{j,l}\}_{j=1, l=1}^{n+1, 2^j-1} \in \mathcal{BC}_n$ to its associated probability function $G$, is measurable as a map \(\mathcal{T}_n:\mathcal{BC}_n\to \mPTh\) under the Borel $\sigma$-field generated by the weak topology, and as a map \(\mathcal{T}_n:\mathcal{BC}_n\to \mPpTh\) under the Borel $\sigma$-field generated by the metric topology induced by the Wasserstein distance of order \(p\). The proof of the following lemma is given in Appendix~\ref{AppendixB5}.


\begin{lemma}\label{lem:DSBAmeasurability} 
Let \(n\in \N\). Suppose that \(\mathcal{H}_n\) is such that the random barycenters \(M^{(n)}\) correspond to a regular level \(n\) SBA almost surely. Then, $\mbox{DSBA}(n, \mathcal{H}_n)$ is a valid stochastic process both on \(\mPTh\) and on \(\mPpTh\) for every \(p\in [1,\infty)\).
\end{lemma}

Large support is an important and basic property that any flexible Bayesian model should possess. In fact, assigning positive prior probability mass to neighborhoods of any probability distribution is a minimum requirement for a Bayesian nonparametric model to be considered ``nonparametric''. This property is also important because it is typically a required condition for consistency of the posterior distribution. To obtain a large support, we need to be able to choose \(n\) arbitrarily large, and thus we select \(n\) at random. The proof of the following theorem is given in Appendix~\ref{AppendixB6}.

\begin{theorem} \label{thm:dsbaFullSupport}
    Let $n$ be a random variable with full support on $\mathbb{N}$ and, given $n$, let $$\mathcal{H}_n = \{ H_{j,2\cdot l-1}\}_{j=1, l=1}^{n+1 , 2^j-1},$$ be a collection of absolutely continuous probability measures on $(\Theta, \mathcal{B}(\Theta))$, each with full support on $\Theta$, and such that \(H_{j, 1}, H_{j, 2^{j + 1} -3}\in \mPOTh\). Assume that, given $n$, $G$ is a DSBA with parameters $(n, \mathcal{H}_n)$, that is $G \mid n, \mathcal{H}_n \sim \mathrm{DSBA}(n, \mathcal{H}_n)$. Then, the following assertions hold:
    \begin{itemize}
        \item[(i)]{$\mathcal{P}(\Theta)$ is the support of $G$ under the weak topology.
        }
        \item[(ii)]{For every \(p\in [1, \infty)\), $\mathcal{P}_p(\Theta)$ is the support of $G$ under the metric topology induced by the Wasserstein distance of order \(p\).
        }
    \end{itemize}
\end{theorem}

\subsection{Discrete location-scale SBA mixture models}

Mixture models that incorporate both location and scale parameters, of the form
\begin{eqnarray}\nonumber
f(\cdot \mid G) &=&\int_{\Theta \times \mathbb{R}_+ } k\left (\cdot \mid \theta, \phi \right) dG( \theta,  \phi),
\end{eqnarray}
offer greater flexibility and accuracy in capturing the heterogeneity of real-world data compared to models based solely on location \citep[see, e.g.,][and references therein]{mueller;quintana;jara;hanson;2015}. While location mixtures can model multimodality and shifts in central tendency, they often fall short in accounting for varying degrees of dispersion across subpopulations. Including a scale parameter in the mixing distribution allows the model to adapt to local variability, accommodate skewness or heavy tails, and better represent complex data structures. 

A possible  construction would consider two independent discrete random mixing distributions, $G_1$ and $G_2$, such that
\begin{eqnarray}\nonumber
f(\cdot \mid G) &=&\int_{\Theta} \int_{\mathbb{R}_+ } k\left (\cdot \mid \theta, \phi \right) dG(\theta, \phi),
\end{eqnarray}
where $G(\cdot) = G_1(\cdot )\times G_2(\cdot )$, $G_1 \mid n,  \mathcal{H}_{n} \sim \mbox{DSBA}(n, \mathcal{H}_{n})$, and $G_2$ is a random discrete probability distribution defined on the positive real line. However, it is possible to show that the product of independent random measures having full weak support marginally, does not have full weak on the space of probability measures defined in the corresponding product space. We propose a general construction having an appealing support property next.

\begin{definition}\label{DBSAdef2}
For a given $n \geq 1$, set $m_1 = 2^{n}$ and let $\mathcal{H}_n = \{ H_{j,2\cdot l-1}\}_{j=1, l=1}^{n+1 , 2^j-1}$ be a collection of probability measures defined on $(\Theta, \mathcal{B}(\Theta))$. Let $\boldsymbol{\eta}$ be a finite-dimensional parameter and $P_{\boldsymbol{\eta}}$ be a distribution defined on $(\mathbb{R}_+, \mathcal{B}(\mathbb{R}_+))$.  For a given $m_2 \geq 1$, let $\boldsymbol{\alpha}_{m_2} \in \mathbb{R}_+^{m_2}$. Set 
 $\Theta_{n,l} = (\mu_{n,l - 1}, \mu_{n, l}]$ for $l = 1, \ldots, 2^{n}-1$, and
set $\Theta_{n,2^{n}} = (\mu_{n,2^{n-1}}, \mu_{n,2^n})$, with the convention $\mu_{n,0} = \inf \Theta$ and $\mu_{n,2^n} = \sup \Theta$. The random function $G$  defined on the measurable space $(\Theta \otimes \mathbb{R}_+, \mathcal{B}(\Theta) \otimes  \mathcal{B}(\mathbb{R}_+))$ is said to be  
a discrete SBA scale  general  random probability measure with parameters $(n, \mathcal{H}_n, \boldsymbol{\eta}, m_2, \boldsymbol{\alpha}_{m_2})$,  denoted as
$G \mid n, \mathcal{H}_n, \boldsymbol{\eta}, m_2, \boldsymbol{\alpha}_{m_2}  \sim \mbox{DSBASg}(n, \mathcal{H}_n, \boldsymbol{\eta}, m_2, \boldsymbol{\alpha}_{m_2})$, if the following holds:
\begin{itemize}
\item[(i)] $\mu_{1,1} \sim H_{1,1}$,
    
\item[(ii)] $\mu_{j, 2\cdot l} = \mu_{j-1, l}$,  $j =2, 3, \ldots,n$  and $l = 1, \ldots, 2^{j-1} - 1$, 

\item[(iii)] $\mu_{j, 2\cdot l-1} \mid \mu_{j-1, l-1}, \mu_{j-1, l} 
\overset{ind.}{\sim} H_{j, 2\cdot l-1} |_{(\mu_{j-1,l-1}, \mu_{j-1,l}]},$ $j =2, 3, \ldots,n+1 $ and $l = 1, \ldots, 2^{j-1}$, where $H|_A$ denotes the restriction of $H$ to the set $A$,

\item[(iv)] $\phi_j \mid \boldsymbol{\eta} \overset{ind.}{\sim}  P_{\boldsymbol{\eta}}$, $j =1, \ldots m_2$,

\item[(v)] $\left( w^{\phi}_{j_1,1}, \ldots, w^{\phi}_{j_1,m_2} \right) \mid \boldsymbol{\alpha}_{m_2} \overset{ind.}{\sim} \mbox{Dirichlet}\left(\boldsymbol{\alpha}_{m_2} \right)$, $j_1 = 1,\ldots, m_1$, and

\item[(vi)] $G(\cdot) \overset{a.s.}{=} \sum_{l_1=1}^{m_1} \sum_{l_2=1}^{m_2} w^{\theta}_{n,l_1} \times w^{\phi}_{j_1, l_2}  \times \delta_{(\theta_l, \phi_l)}(\cdot)$, where
$\theta_l \overset{a.s.}{=}  \mu_{n+1,2\cdot l-1}$, $w^{\theta}_{n,l} \overset{a.s.}{=}  G_1(\Theta_{n,l})$, $l=1,\ldots,m$, and $G_1(\Theta_{n,l})$ is computed using (i) -- (iii) in Property~\ref{property1}.
\end{itemize}
\end{definition}

The same arguments that we used to prove that the DSBA defines a valid stochastic process on \(\mPTh\) can be used with minor modifications to show that the DSBAg and DSBAp define a valid stochastic process on \(\mP(\Th\times \R_+)\). The location-scale generalization retains the full weak support of the original construction. The proof of the following theorem is given in Appendix~\ref{AppendixC1}.

\begin{theorem} \label{thm:dsbagFullSupport}
    Let $n$ and $m_2$ be random variables with full support on $\mathbb{N}$ and, given $n$ , let $$\mathcal{H}_n = \{ H_{j,2\cdot l-1}\}_{j=1, l=1}^{n+1 , 2^j-1},$$ be a collection of absolutely continuous probability measures on $(\Theta, \mathcal{B}(\Theta))$, each with full support on $\Theta$, and such that \(H_{j, 1}, H_{j, 2^{j + 1} -3}\in \mPOTh\). 
 Assume that $P_{\boldsymbol{\eta}}$ has full support on $\mathbb{R}_+$. Finally, assume that, given $n$ and $m_2$ , $G$  is a discrete SBA scale  general  random probability measure with parameters $(n, \mathcal{H}_n, \boldsymbol{\eta}, m_2, \boldsymbol{\alpha}_{m_2})$.  Then, $\mathcal{P}(\Theta)$ is the support of $G$ under the weak topology.
\end{theorem}

We also consider a parsimonious generalization of the DSBA construction to account for the dispersion parameter, that retains the desirable induced distribution on the marginal mean for the location coordinate and the appealing support property.

\begin{definition}\label{DBSAdef3}
For a given $n \geq 1$, set $m = 2^{n}$ and let $\mathcal{H}_n = \{ H_{j,2\cdot l-1}\}_{j=1, l=1}^{n+1 , 2^j-1}$ be a collection of probability measures defined on $(\Theta, \mathcal{B}(\Theta))$. Let $\boldsymbol{\eta}$ be a finite-dimensional parameter and $P_{\boldsymbol{\eta}}$ be a distribution defined on $(\mathbb{R}_+, \mathcal{B}(\mathbb{R}_+))$. Set 
 $\Theta_{n,l} = (\mu_{n,l - 1}, \mu_{n, l}]$ for $l = 1, \ldots, 2^{n}-1$, and
set $\Theta_{n,2^{n}} = (\mu_{n,2^{n-1}}, \mu_{n,2^n})$, with the convention $\mu_{n,0} = \inf \Theta$ and $\mu_{n,2^n} = \sup \Theta$. The random probability measure $G$  defined on the measurable space $(\Theta \otimes \mathbb{R}_+, \mathcal{B}(\Theta) \otimes  \mathcal{B}(\mathbb{R}_+))$ is said to be  
a discrete SBA scale parsimonious random probability measure with parameters $(n, \mathcal{H}_n, \boldsymbol{\eta})$,  denoted as
$G \mid n, \mathcal{H}_n, \boldsymbol{\eta}  \sim \mbox{DSBASp}(n, \mathcal{H}_n, \boldsymbol{\eta})$, if:
\begin{itemize}
\item[(i)] $\mu_{1,1} \sim H_{1,1}$,
    
\item[(ii)] $\mu_{j, 2\cdot l} = \mu_{j-1, l}$,  $j =2, 3, \ldots,n$  and $l = 1, \ldots, 2^{j-1} - 1$, 

\item[(iii)] $\mu_{j, 2\cdot l-1} \mid \mu_{j-1, l-1}, \mu_{j-1, l} 
\overset{ind.}{\sim} H_{j, 2\cdot l-1} |_{(\mu_{j-1,l-1}, \mu_{j-1,l}]},$
$j =2, 3, \ldots,n+1$ and $l = 1, \ldots, 2^{j-1}$, where $H|_A$ denotes the restriction of $H$ to the set $A$,

\item[(iv)] $\phi_j \mid \boldsymbol{\eta} \overset{ind.}{\sim}  P_{\boldsymbol{\eta}}$, $j =1, \ldots m$, and

\item[(v)] $G(\cdot) \overset{a.s.}{=} \sum_{l=1}^{m} w_{n,l} \delta_{(\theta_l, \phi_l)}(\cdot)$, where
$\theta_l \overset{a.s.}{=}  \mu_{n+1,2\cdot l-1}$, $w_{n,l} \overset{a.s.}{=}  G(\Theta_{n,l})$, $l=1,\ldots,m$, and $G(\Theta_{n,l})$ is computed using (i) -- (iii) in Property~\ref{property1}.
\end{itemize}
\end{definition}

The  parsimonious generalization also retains the full weak support of the original construction. The proof of the following theorem is given in Appendix~\ref{AppendixC2}.

\begin{theorem} \label{thm:dsbapFullSupport}
    Let $n$ be random variable with full support on $\mathbb{N}$ and, given $n$ , let $$\mathcal{H}_n = \{ H_{j,2\cdot l-1}\}_{j=1, l=1}^{n+1 , 2^j-1},$$ be a collection of absolutely continuous probability measures on $(\Theta, \mathcal{B}(\Theta))$, each with full support on $\Theta$, and such that \(H_{j, 1}, H_{j, 2^{j + 1} -3}\in \mPOTh\). 
 Assume that $P_{\boldsymbol{\eta}}$ has full support on $\mathbb{R}_+$. Finally, assume that, given $n$ , $G$  is a discrete SBA scale parsimonious random probability measure with parameters $(n, \mathcal{H}_n, \boldsymbol{\eta})$.  Then, $\mathcal{P}(\Theta)$ is the support of $G$ under the weak topology.
\end{theorem}

Finally, both the general and parsimonious generalizations induce mixture models with appealing support properties. We focus here in the cases where the sample space is either $\mathcal{Y} = \mathbb{R}$, $\mathcal{Y} = \mathbb{R}_+$, or $\mathcal{Y} = [0,1]$, and consider the Gaussian, Gamma, and Beta kernels, respectively. Let $\mathcal{D}(\mathcal{Y})$ be the space of all probability distributions defined on $\mathcal{Y}$ that admit a density with respect to Lebesgue measure. The proof of the following theorem is given in Appendix~\ref{AppendixC3}.

\begin{theorem}\label{thm:hellinger:support} Consider the mixture model 
\begin{eqnarray}\nonumber
f(\cdot \mid G) &=&\int_{\Theta \times \mathbb{R}_+ } k\left (\cdot \mid \theta, \phi \right) dG( \theta,  \phi),
\end{eqnarray}
where $k$ is the Gaussian, Gamma, or Beta kernel, $n$ is a random variable with full support on $\mathbb{N}$, and 
\[
    G \mid n, \mathcal{H}_n, \boldsymbol{\eta}, m_2, \boldsymbol{\alpha}_{m_2}  \sim \mbox{DSBASg}(n, \mathcal{H}_n, \boldsymbol{\eta}, m_2, \boldsymbol{\alpha}_{m_2}),
\]
and the conditions of Theorem \ref{thm:dsbagFullSupport} are satisfied, or 
\[
    G \mid n, \mathcal{H}_n, \boldsymbol{\eta}  \sim \mbox{DSBASp}(n, \mathcal{H}_n, \boldsymbol{\eta}),
\]
and the conditions of Theorem \ref{thm:dsbapFullSupport} are satisfied. Then, \(\mathcal{D}(\mathcal{Y})\) is support of the process, that is, the induced probability law of the measure-valued stochastic process with trajectories in $\mathcal{D}(\mathcal{Y})$, denoted as $\Pi$, is such that
\[
    P\in\mDY,\, \eps > 0:\,\, \Pi(B_{\Hd}(P, \epsilon)) > 0,
\]
where \(B_{\Hd}(P, \epsilon)\) denotes the open ball in the Hellinger distance centered at \(P\) of radius \(\epsilon\).
\end{theorem}

\section{Posterior computation for the DSBAS mixture models}\label{section4}

We assume observed data \(\mathbf{y} = (y_1, \ldots, y_d)\) are independent draws from the mean-scale mixture model
\[
f(y_i \mid G) = \int_\Theta k(y_i \mid \theta, \phi) \, dG(\theta, \phi),
\]
where
$$G \mid n, \mathcal{H}_n, \boldsymbol{\eta}  \sim \mbox{DSBASp}(n, \mathcal{H}_n, \boldsymbol{\eta}),$$
or 
 $$G \mid n, \mathcal{H}_n, \boldsymbol{\eta}, m_2, \boldsymbol{\alpha}_{m_2}  \sim \mbox{DSBASg}(n, \mathcal{H}_n, \boldsymbol{\eta}, m_2, \boldsymbol{\alpha}_{m_2}).$$

In the rest of this section, we provide details on some of the resulting conditional distributions and the implementation of a Gibbs sampler algorithm for fixed $n$ and $m_2$. The development of a trans-dimensional algorithm for dealing with random $n$ and $m_2$ is the subject of ongoing research. Functions implementing the described algorithms were implemented in the R library ``DPpackage'' \citep{jara;2007,jara;etal;2011}. Since DPpackage is not longer available for newer versions of R, R functions based on JAGS \citep{Plummer_JAGS_2003} are also provided.

\subsection{Parsimonious DSBAS}\label{section4.1}
Here we assume that
\[
G \mid n, \mathcal{H}_n, \boldsymbol{\eta}  \sim \mbox{DSBASp}(n, \mathcal{H}_n, \boldsymbol{\eta}),
\]
where $k(\cdot \mid \theta, \phi)$ is an appropriate kernel. For a given $n$ and $m = 2^{n-1}$, a hierarchical representation of the model, introducing latent allocation variables, is given by
\begin{eqnarray}\nonumber
y_i \mid z_i,  \{\mu_{j,l}\}_{j=1, l=1}^{n+1, 2^j-1} , \phi_1,\ldots, \phi_m \overset{ind.}{\sim}  k(\cdot \mid \theta_i, \phi_i),\quad i=1, \ldots, d,
\end{eqnarray}
\begin{eqnarray}\nonumber
z_i \mid  \{\mu_{j,l}\}_{j=1, l=1}^{n+1, 2^j-1} \overset{i.i.d.}{\sim}  \mathrm{Discrete}(w_{n,1}, \ldots, w_{n,m}),\quad i=1,\ldots,d,
\end{eqnarray}
\begin{eqnarray}\nonumber
\mu_{1,1} \sim H_{1,1},
\end{eqnarray}
\begin{eqnarray}\nonumber
\mu_{j, 2\cdot l-1} \mid \mu_{j-1, l-1}, \mu_{j-1, l} 
\overset{ind.}{\sim} H_{j, 2\cdot l-1} |_{(\mu_{j-1,l-1}, \mu_{j-1,l}]}, \quad j =2, 3, \ldots,n+1,  \mbox{ and } l = 1, \ldots, 2^{j-1},
\end{eqnarray}
\begin{eqnarray}\nonumber
\mu_{j, 2\cdot l} = \mu_{j-1, l},\ \ \ j =2, 3, \ldots,n+1, \mbox{ and } l = 1, \ldots, 2^{j-1} - 1,
\end{eqnarray}
and
\begin{eqnarray}\nonumber
\phi_j \mid \boldsymbol{\eta} \overset{ind.}{\sim}  P_{\boldsymbol{\eta}}, \quad  j =1, \ldots,m.
\end{eqnarray}
 
\subsubsection{Updating the latent allocations \(\{z_i\}\)}\label{section4.1.1}

For each \(i=1,\ldots,d\) and \( j=1,\ldots,m\), compute the posterior probabilities
\[
\tilde{w}_{i,j} = \mathbb{P}(z_i = j \mid \mathbf{y}, \{\mu_{j,l}\}_{j=1, l=1}^{n+1, 2^j-1} , \phi_1,\ldots, \phi_m ) \propto w_{n,l} \, k(y_i \mid \theta_j, \phi_j),
\]
and sample \(z_i\) from the corresponding discrete distribution,
\[
z_i \mid   \mathbf{y}, \{\mu_{j,l}\}_{j=1, l=1}^{n, 2^j-1} , \phi_1,\ldots, \phi_m \overset{ind.}{\sim}  \mathrm{Discrete}(\tilde{w}_{i,1}, \ldots, \tilde{w}_{i,m}).
\]

\subsubsection{Updating dispersion parameters \(\{\phi_j\}\)}\label{section4.1.2}

We update each $\phi_j$ conditionally on the remaining parameters and the observed data by sampling from the corresponding full conditional distribution. Given the model specification, the full conditional distribution for $\phi_j$ is proportional to
\[
\pi(\phi_j \mid \mathbf{y}, \{\mu_{j,l}\}_{j=1, l=1}^{n+1, 2^j-1}) \ \propto\ p_{\boldsymbol{\eta} }(\phi_j)\ 
\prod_{i: z_i = j} k(y_i \mid \theta_j, \phi_j),
\]
where $p_{\boldsymbol{\eta} }(\cdot)$ is the prior density for $\phi_j$, which specific form depends on the kernel and prior $p_{\boldsymbol{\eta}}$.

\noindent {\bf Gaussian kernel}:  Consider the Gaussian kernel
\[
k(y_i \mid \theta_j, \phi_j) = \frac{1}{\sqrt{2\pi \phi_j}}
\exp\left\{ -\frac{(y_i - \theta_j)^2}{2\phi_j} \right\},
\]
where \(\phi_j > 0\) denotes the dispersion parameter associated with the \(j\)-th component. Let \(S_j = \{i : z_i = j\}\) be the set of indices assigned to component \(j\), with \(n_j = |S_j|\) the corresponding cluster size, and let
\[
\bar{y}_j = \frac{1}{n_j} \sum_{i \in S_j} y_i,
\]
denote the cluster mean. Assuming that \(\phi_j\mid a_\phi, b_\phi \overset{i.i.d.}{\sim} \text{Inverse-Gamma}(a_\phi, b_\phi)\), the full conditional distribution is given by
\[
\phi_j \mid \mathbf{y}, \theta_j, z_1, \ldots, z_d  \;\sim\;
\text{Inverse-Gamma}\left(
a_\phi + \frac{n_j}{2},
\;
b_\phi + \frac{1}{2} \sum_{i \in S_j} (y_i - \theta_j)^2
\right).
\]
Hence, \(\phi_j\) can be updated via a direct Gibbs step. If no conjugacy is assumed (e.g., under a log-normal prior), we may update \(\phi_j\) via a Metropolis–Hastings random walk on the log scale, or by employing a slice sampler \citep{neal;2003}. In either case, the full conditional density is proportional to
\[
\pi(\phi_j \mid \cdot) \;\propto\;
\phi_j^{-n_j/2}
\exp\left\{ -\frac{1}{2\phi_j} \sum_{i \in S_j} (y_i - \theta_j)^2 \right\}
\; p_{\boldsymbol{\eta}}(\phi_j).
\]

\noindent {\bf Beta kernel}: Consider now the beta kernel parameterized by the mean \(\mu_j\) and dispersion \(\phi_j\):
\[
k(y_i \mid \mu_j, \phi_j) =
\frac{\Gamma(\phi_j)}{\Gamma(\mu_j\phi_j)\,\Gamma((1-\mu_j)\phi_j)}
y_i^{\mu_j\phi_j - 1} (1-y_i)^{(1-\mu_j)\phi_j - 1}, \quad 0<y_i<1.
\]
The log-likelihood contribution of \(\phi_j\) is
\begin{eqnarray}\nonumber
\ell(\phi_j) &=&
\sum_{i \in S_j} \Big[ \log\Gamma(\phi_j)
- \log\Gamma(\mu_j \phi_j)
- \log\Gamma((1-\mu_j)\phi_j)+
\\\nonumber
&&
 (\mu_j \phi_j - 1) \log y_i
+ ((1-\mu_j)\phi_j - 1) \log (1-y_i) \Big],
\end{eqnarray}
where \(S_j = \{i : z_i = j\}\) denote the indices assigned to component \(j\), with \(n_j = |S_j|\).  Assuming a prior \(\phi_j \sim \mathrm{Gamma}(a_\phi,b_\phi)\) (shape–rate),
the log-full conditional is given by
\[
\log \pi(\phi_j \mid \mathbf{y}, \theta_j, z_1, \ldots, z_d ) \propto
\ell(\phi_j) + (a_\phi - 1) \log \phi_j - b_\phi \phi_j.
\]
Since this distribution is non-conjugate in \(\phi_j\), we can update it using a Metropolis–Hastings step on the log scale or via slice sampling.

\noindent {\bf Gamma kernel}:  Finally, consider the gamma kernel given by
\begin{eqnarray}\nonumber
k(y_i \mid \theta_j,\phi_j)= \left( \frac{\phi_j}{\theta_j}\right)^{\phi_j} y^{\phi_j-1} \exp\left\{-\frac{\phi_j}{\theta_j} y_i\right\}.
\end{eqnarray}
where $\theta_j > 0$ is the mean parameter and $\phi_j > 0$ is the shape (or dispersion) parameter. The log-likelihood contribution of $\phi_j$ given $\theta_j$ and the data $\{y_i : i \in S_j\}$ is given by
\begin{equation}\nonumber
\ell(\phi_j) 
= n_j \phi_j \log\left( \frac{\phi_j}{\theta_j} \right)
+ (\phi_j - 1) \sum_{i \in S_j} \log y_i
- \frac{\phi_j}{\theta_j} \sum_{i \in S_j} y_i,
\end{equation}
where $S_j = \{ i : z_i = j \}$ denote the set of observations assigned to component $j$ and $n_j = |S_j|$ its size.   Assuming that $\phi_j \sim p_{\boldsymbol{\eta}}(\phi_j)$, then the log-full conditional distribution is given by
\begin{equation}\nonumber
\log \pi(\phi_j \mid \mathbf{y}, \theta_j, z_1, \ldots, z_d ) 
\propto \ell(\phi_j) + \log p_{\boldsymbol{\eta}}(\phi_j).
\end{equation}
Since $\phi_j$ appears both inside a logarithm and multiplying $\log \phi_j$, the posterior is not of standard form, so we can update $\phi_j$ using a Metropolis–Hastings step or via slice sampling. For instance, a random-walk proposal on the log-scale
\begin{equation}\nonumber
\log \phi_j^{\star} = \log \phi_j + \tau_\phi \,\epsilon, 
\quad \epsilon \sim N(0,1),
\end{equation}
is convenient to ensure positivity. The acceptance probability is given by
\begin{equation}\nonumber
\alpha = \min\left\{
1,\,
\frac{\pi(\phi_j^{\star} \mid  \mathbf{y}, \theta_j, z_1, \ldots, z_d )}
     {\pi(\phi_j \mid  \mathbf{y}, \theta_j, z_1, \ldots, z_d )}
\right\},
\end{equation}
and the tuning parameter $\tau_\phi$ is chosen to yield an adequate acceptance rate.

\subsubsection{Updating  $\{\mu_{j,l}\}_{j=1, l=1}^{n+1, 2^j-1}$}\label{section4.1.3}

The SBA parameters $\{\mu_{j,l}\}_{j=1, l=1}^{n+1, 2^j-1}$ define the weights and support points of the random probability measure \(G\) and are updated conditional on the observed data and other model parameters. Recall that  $m=2^{n}$, $\theta_j=\mu_{n+1,2\cdot j-1}$, for $j=1,\ldots,m$, and that even nodes are deterministic
\[
\mu_{j,2 \cdot l}=\mu_{j-1,l},\qquad j=2,\ldots,n+1,\; l=1,\ldots,2^{j-1}-1,
\]
so we only sample the odd nodes $\{\mu_{j,2\cdot l-1}\}$. Due to the nature of the problem, the feasible set for $\mu_{j,l}$, $I_{j,l}$, depends on its position in the SBA. 
For $j=1$ and $l=1$,  
\[
I_{1,1} = (\,\mu_{2,1},\; \mu_{2,3}\,]
\]
since the two odd nodes at level $2$ straddle the root. For interior nodes, $2 \le j \le n-1$,
\[
I_{j,l} = (\,\mu_{j+1,\,4 \cdot l-3},\; \mu_{j+1,\,4\cdot l-1}\,]
\]
because both odd children lie inside the parent interval $(\mu_{j-1,l-1},\mu_{j-1,l}]$ and bound the parent from below and above.  
This child-induced interval is always contained in the parent interval. Finally, for $j=n$,  
\[
I_{n,l} = (\mu_{n-1,l-1},\; \mu_{n-1,l}]
\]
since they have no children. Changing any $\mu_{j,2l-1}$ modifies both the support points $\theta_h$ and the weights $w_{n,h}$ of \emph{all} mixture components, through the recursion in Property~\ref{property1}. Thus, the full conditional for the odd nodes $\mu_{j,2l-1}$, given all other parameters, is given by
\[
\pi(\mu_{j,2\cdot l-1} \mid \cdots)
\ \propto\
h_{j,2\cdot l-1}(\mu_{j,2\cdot l-1})\ \mathbb{I}(\mu_{j,2\cdot l-1})_{I_{j,l}}
\ \prod_{i=1}^d \left[ \sum_{h=1}^m w_{n,h}\ k\!\left(y_i \mid \theta_h, \phi_h\right) \right],
\]
where $h_{j,2\cdot l-1}(\cdot)$ is the density w.r.t. Lebesgue measure of $H_{j,2\cdot l-1}$, 
 $\mathbb{I}(\cdot)_{A}$ is the indicator function for the set $A$, and $w_{n,h}$ and $\theta_h$ are computed recursively from the full SBA array $\{\mu_{j,l}\}_{j=1, l=1}^{n+1, 2^j-1}$. We propose update $\mu_{j,2l-1}$ by slice sampling restricted to $I_{j,l}$.

\subsection{General DSBAS}\label{section4.2}

Here we assume that
\[
G \mid n, \mathcal{H}_n, \boldsymbol{\eta}, m_2, \boldsymbol{\alpha}_{m_2}  
\sim \mbox{DSBASg}(n, \mathcal{H}_n, \boldsymbol{\eta}, m_2, \boldsymbol{\alpha}_{m_2}).
\]
For fixed $n$ and $m_2$, we obtain the following hierarchical representation of the model:

\begin{eqnarray}\nonumber
y_i \mid z^{\theta}_i, z^{\phi}_i,
\{\mu_{j,l}\}_{j=1,l=1}^{n+1,2^j-1}, \{\phi_j\}_{j=1}^{m_2}
&\overset{ind.}{\sim}&
k\!\left( y_i \mid \theta_{z^{\theta}_i}, \phi_{z^{\phi}_i} \right), 
\quad i=1,\ldots,d,
\end{eqnarray}
\begin{eqnarray}\nonumber
z^{\theta}_i \mid \{\mu_{j,l}\}_{j=1,l=1}^{n+1,2^j-1}
&\overset{i.i.d.}{\sim}&
\mathrm{Discrete}(w^{\theta}_{n,1}, \ldots, w^{\theta}_{n,m_1}), \quad i=1,\ldots,d,
\end{eqnarray}
\begin{eqnarray}\nonumber
z^{\phi}_i \mid \boldsymbol{\alpha}_{m_2}, z^{\theta}_i 
&\overset{i.i.d.}{\sim}&
\mathrm{Discrete}(w^{\phi}_{z^{\theta}_i ,1}, \ldots, w^{\phi}_{z^{\theta}_i , m_2}), \quad i=1,\ldots,d,
\end{eqnarray}
\begin{eqnarray}\nonumber
\mu_{1,1} &\sim& H_{1,1}, 
\end{eqnarray}
\begin{eqnarray}\nonumber
\mu_{j,2\cdot  l-1} \mid \mu_{j-1,l-1}, \mu_{j-1,l}
&\overset{ind.}{\sim}& H_{j,2\cdot  l-1}\big|_{(\mu_{j-1,l-1},\mu_{j-1,l}]}, 
\quad j=2,\ldots,n+1, \ l=1,\ldots,2^{j-1}, 
\end{eqnarray}
\begin{eqnarray}\nonumber
\mu_{j,2\cdot  l} &=& \mu_{j-1,l}, 
\quad j=2,\ldots,n+1, \ l=1,\ldots,2^{j-1}-1, 
\end{eqnarray}
\begin{eqnarray}\nonumber
\phi_j \mid \boldsymbol{\eta}
&\overset{ind.}{\sim}& P_{\boldsymbol{\eta}}, \quad j=1,\ldots,m_2,
\end{eqnarray}
\begin{eqnarray}\nonumber
(w^{\phi}_1,\ldots,w^{\phi}_{m_2}) \mid \boldsymbol{\alpha}_{m_2}
&\sim& \mathrm{Dirichlet}(\boldsymbol{\alpha}_{m_2}).
\end{eqnarray}
Here $m_1=2^{n}$ and $\theta_j=\mu_{n+1,2\cdot j-1}$ for $j=1,\ldots,m_1$.
Thus, each observation selects a location index $z^{\theta}_i$ and
an independent scale index $z^{\phi}_i$, with the corresponding
kernel parameter pair $(\theta_{z^{\theta}_i}, \phi_{z^{\phi}_i})$.

\subsubsection{Updating latent allocations}\label{section4.2.1}

For each observation $i=1,\ldots,d$, the joint posterior distribution of
$(z^{\theta}_i, z^{\phi}_i)$ is given by
\[
\mathbb{P}(z^{\theta}_i=j_1, z^{\phi}_i=j_2 \mid \cdot)
\;\propto\;
w^{\theta}_{n,l_1}\, w^{\phi}_{j_1, j_2}\,
k(y_i \mid \theta_{j_1}, \phi_{j_2}),
\quad j_1=1,\ldots,m_1,\ j_2=1,\ldots,m_2.
\]
Hence $(z^{\theta}_i, z^{\phi}_i)$ can be sampled directly from
the discrete distribution over the $m_1\times m_2$ pairs.

\subsubsection{Updating scale weights}\label{section4.2.2}

Let
\[
S^{\phi}_{j_2}=\{\, i:\ z^{\phi}_i=j_2 \,\},\qquad 
n^{\phi}_{j_2}=|S^{\phi}_{j_2}|,
\qquad 
n_{j_1 j_2}=\big|\{\, i:\ z^{\theta}_i=j_1,\ z^{\phi}_i=j_2 \,\}\big|.
\]
Conditionally on the allocations, the vector of weights attached to location \(j_1\) has the conjugate full conditional distribution
\[
\big(w^{\phi}_{j_1,1},\ldots,w^{\phi}_{j_1,m_2}\big)
\;\big|\;\boldsymbol{\alpha}_{m_2},\ \{z^{\theta}_i,z^{\phi}_i\}_{i=1}^d
\;\sim\;
\mathrm{Dirichlet}\!\left(\alpha_1+n_{j_1 1},\,\ldots,\,\alpha_{m_2}+n_{j_1 m_2}\right),
\]
$j_1=1,\ldots,m_1$.

\subsubsection{Updating global scale atoms}\label{section4.2.3}

Each \(\phi_{j_2}\) pools information across all locations:
\[
\pi(\phi_{j_2}\mid \mathbf{y},\{z^\theta_i,z^\phi_i\},\{\theta_{j_1}\})
\;\propto\;
p_{\boldsymbol{\eta}}(\phi_{j_2})\,
\prod_{i\in S^{\phi}_{j_2}} k\!\left(y_i \mid \theta_{z^{\theta}_i},\ \phi_{j_2}\right),
\quad j_2=1,\ldots,m_2.
\]

\noindent Updates are kernel--specific:

\smallskip
\noindent\emph{Gaussian kernel.} Assume \(\phi_{j_2}\sim \mathrm{Inverse\text{-}Gamma}(a_\phi,b_\phi)\). Then
\[
\phi_{j_2}\mid \cdot \;\sim\;
\mathrm{Inverse\text{-}Gamma}\!\left(
a_\phi+\tfrac{n^{\phi}_{j_2}}{2},\;
b_\phi+\tfrac{1}{2}\sum_{i\in S^{\phi}_{j_2}}(y_i-\theta_{z^{\theta}_i})^2
\right).
\]

\noindent\emph{Beta kernel.} With prior \(\phi_{j_2}\sim \mathrm{Gamma}(a_\phi,b_\phi)\) (shape--rate), the log--full conditional is
\begin{eqnarray}\nonumber
\log \pi(\phi_{j_2}\mid \cdot)\ &\propto\ &
\sum_{i\in S^{\phi}_{j_2}}
\Big\{
\log\Gamma(\phi_{j_2})
-\log\Gamma(\theta_{z^{\theta}_i}\phi_{j_2})
- \\\nonumber
&&
\log\Gamma((1-\theta_{z^{\theta}_i})\phi_{j_2})
+(\theta_{z^{\theta}_i}\phi_{j_2}-1)\log y_i
+((1-\theta_{z^{\theta}_i})\phi_{j_2}-1)\log(1-y_i)
\Big\}
+ \\\nonumber
&&
(a_\phi-1)\log\phi_{j_2}-b_\phi \phi_{j_2}.
\end{eqnarray}
This non--conjugate update can be carried out using MH on the log--scale or slice sampling.

\noindent\emph{Gamma kernel.} With mean--shape parameterization
\[
k(y\mid \theta,\phi)=
\frac{(\phi/\theta)^{\phi}}{\Gamma(\phi)}\,y^{\phi-1}\exp\!\left(-\tfrac{\phi}{\theta}y\right),
\]
the log--likelihood for \(\phi_{j_2}\) is
\[
\ell(\phi_{j_2})=
\sum_{i\in S^{\phi}_{j_2}}
\Big[
\phi_{j_2}\log(\phi_{j_2}/\theta_{z^{\theta}_i})
-\log\Gamma(\phi_{j_2})
+(\phi_{j_2}-1)\log y_i
-\tfrac{\phi_{j_2}}{\theta_{z^{\theta}_i}}\,y_i
\Big].
\]
Then
\[
\log \pi(\phi_{j_2}\mid \cdot)\ \propto\ \ell(\phi_{j_2})+\log p_{\boldsymbol{\eta}}(\phi_{j_2}),
\]
and we update \(\log\phi_{j_2}\) using random--walk MH or slice sampling.

\subsubsection{Updating location nodes}\label{section4.2.4}

As in Section~\ref{section4.1.3}, only the odd nodes are stochastic:
\[
\mu_{j,2\cdot l}=\mu_{j-1,l},\qquad j=2,\ldots,n+1,\;\; l=1,\ldots,2^{j-1}-1.
\]
We update \(\{\mu_{j,2l-1}\}\) subject to their feasibility intervals $$I_{1,1}=(\mu_{2,1},\mu_{2,3}],$$ $$I_{j,l}=(\mu_{j+1,4l-3},\mu_{j+1,4l-1}],\; 2\le j\le n-1,$$ and
$$I_{n,l}=(\mu_{n-1,l-1},\mu_{n-1,l}].$$

When an odd node \(\mu_{j,2l-1}\) is modified, both the support points \(\{\theta_{j_1}\}\) and the location weights \(\{w^{\theta}_{n,l_1}\}\) change through the SBA recursion. A robust strategy is to marginalize over allocations \(\{z^{\theta}_i,z^{\phi}_i\}\), leading to the full conditional
\[
\pi(\mu_{j,2l-1}\mid \cdots)\ \propto\
h_{j,2l-1}(\mu_{j,2l-1})\,
\mathbb{I}\!\left(\mu_{j,2l-1}\in I_{j,l}\right)\,
\prod_{i=1}^d
\left[
\sum_{j_1=1}^{m_1}\sum_{j_2=1}^{m_2}
w^{\theta}_{n,l_1}\, w^{\phi}_{j_1,j_2}\,
k\!\left(y_i \mid \theta_{j_1}, \phi_{j_2}\right)
\right],
\]
where \(h_{j,2l-1}(\cdot)\) is the density of \(H_{j,2l-1}\).
We update each odd node \(\mu_{j,2l-1}\) using slice sampling restricted to \(I_{j,l}\), recomputing 
\(\{\theta_{j_1}\}_{j_1=1}^{m_1}\) and \(\{w^{\theta}_{n,l_1}\}_{j_1=1}^{m_1}\) via Property~\ref{property1} at every proposal.

\section{Illustrations}\label{section5}

We now present simulated and real--life examples to illustrate the application of the proposed model. 
Both simulated and well--known benchmark datasets are used to assess performance.  
Through these analyses, we emphasize different aspects of the inferential problem, showing that the proposed approach can both incorporate prior information and deliver accurate density estimation.  
Model comparison is carried out using the widely applicable Bayesian information criterion (WAIC) proposed by \cite{waic} and the log pseudo--marginal likelihood (LPML) based on the conditional predictive ordinates \citep{geisser;eddy;79}.

\subsection{Density estimation}

As a first illustration, we reanalyze the galaxy dataset from \cite{roeder;1990},  which contains the velocities of 82 galaxies  from six well-separated conic sections of  the Corona Borealis region, reported in km/second. As discussed by \cite{roeder;1990}, there is strong evidence that the modes in these data correspond to clumped galaxies, and this is why this dataset is  often used for demonstrating mixture modeling. We base our prior on the \emph{mean velocity} on external astronomical knowledge, to illustrate how scientific information can guide prior specification.  Galaxy velocities can be decomposed into \emph{recession velocities}  (Hubble flow; the expansion of the universe, with 
$v = H_0 d \approx cz$ at low redshifts, where  $v$ denotes the recession velocity,  $H_0$ the Hubble constant,  $d$ the proper distance to the galaxy,  $c$ the speed of light, and 
$z$ the cosmological redshift) and \emph{peculiar velocities} (local deviations due to gravitational clustering, typically a few hundred km/second) 
\citep[see, e.g.,][]{Peebles1980}.

The galaxy dataset contains recession velocities. Large galaxy redshift surveys provide benchmarks for expected mean velocities. Sloan Digital Sky Survey (SDSS) reported a median redshift $z \approx 0.104$, corresponding to $\sim$31{,}000 km/s \citep{York2000, Strauss2002}. 2dF Galaxy Redshift Survey (2dFGRS) reported a median redshift $z \approx 0.11$, corresponding to $\sim$33{,}000 km/second \citep{Colless2001}. For low redshifts, the approximation $v \approx cz$ holds. Thus, galaxy samples from magnitude-limited surveys often have central velocities in the 30{,}000--33{,}000 km/second range. From SDSS and 2dF, we select a prior mean for the \emph{mean velocity} of 30{,}000 km/second, reflecting typical depths of magnitude-limited surveys of nearby galaxies. We assume that the true mean velocity of a comparable survey could plausibly range from 15{,}000 km/second (shallower samples) up to 45{,}000 km/second (deeper samples). This interval represents our 95\% prior credible interval. Thus, we select our prior standard deviation, $\sigma$, by solving  $1.96 \times \sigma = 15{,}000$, which gives $\sigma \approx 7{,}650$. Therefore, we assume an $\mathcal{N}(30{,}000, \; 7{,}650^2)$ prior for \emph{mean velocity}, where the prior mean of 30{,}000 km/second ties directly to empirical medians from SDSS and 2dF.  The variance reflects realistic survey depth variation, while still concentrating mass around the central value.  

We fit the two versions of the DSBAS mixture of normal model to the  velocities (in $1{,}000$ km/second), considering $n = 4$, $5$, and $6$, 
$m_2 = n$, and $\boldsymbol{\alpha}_{m_2} = \mathbf{1}_{m_2}$. In all cases, we set $H_{j,2\cdot l-1} \sim \mathcal{N}(30.0, 7.65^2)$ and \(\phi_j\mid a_\phi, b_\phi \overset{i.i.d.}{\sim} \text{Inverse-Gamma}(1.0/2, 3.0/2)\). For comparison, we also fit a finite--dimensional approximation to the celebrated Dirichlet process mixture (DPM) model \citep{ferguson;73}:
\[
f(y_i \mid G) = \sum_{l=1}^L W_l \,\phi(y_i \mid \mu_l, \sigma_l^2),
\]
where
\[
 \mu_l \overset{i.i.d.}{\sim} \mathcal{N}(30, 7.65^2), \quad l=1,\ldots,L,
\]
\[
\sigma_l^2 \overset{i.i.d.}{\sim} \text{Inverse-Gamma}(1/2, 3/2), \quad l=1,\ldots,L,
\]
\[
V_l \overset{i.i.d.}{\sim} \text{Beta}(1, 1), \quad l=1,\ldots,L-1,
\]
and
\[
W_l = V_l \prod_{o=1}^{l-1} (1 - V_o), \quad l=1,\ldots,L,
\]
with $V_L = 1$.  We considered $L = 8$, $16$ and $32$, so that each DPM version had the same number of support points as the corresponding DSBAS version.

\begin{figure}
\centering
\subfigure[]{\includegraphics[width=5cm]{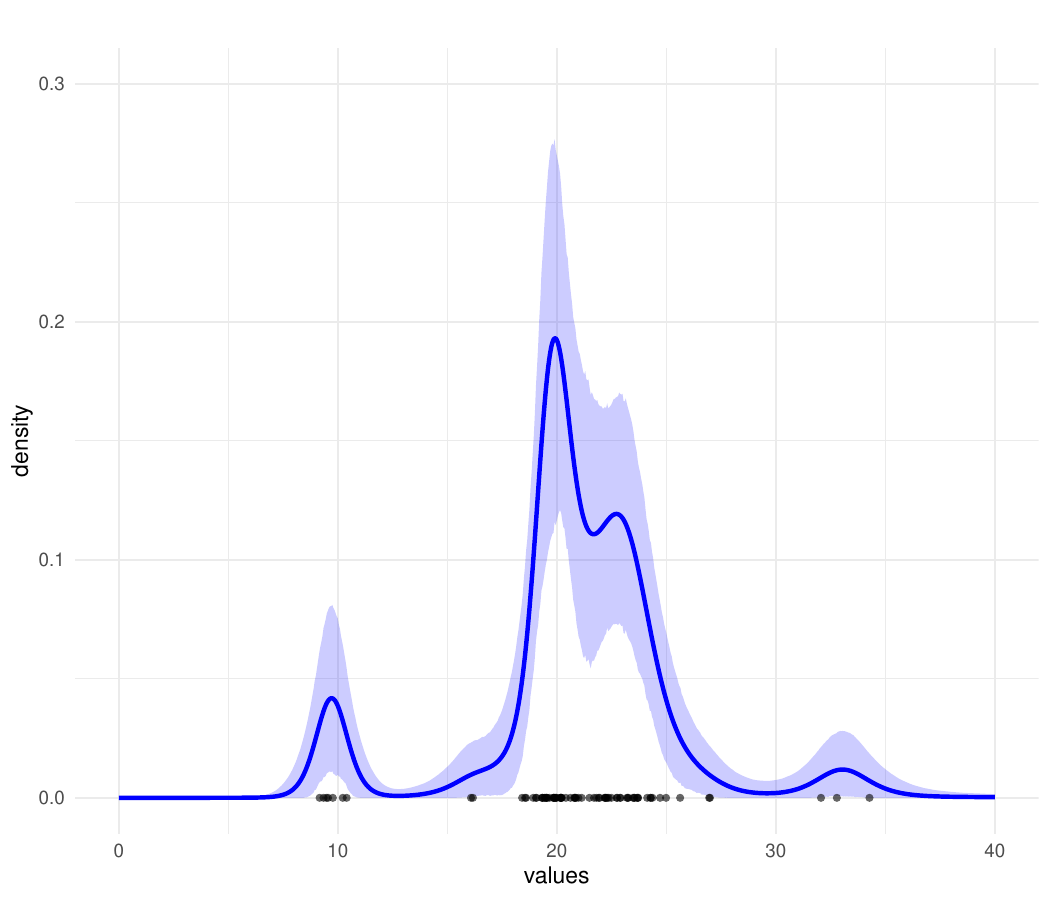}}
\subfigure[]{\includegraphics[width=5cm]{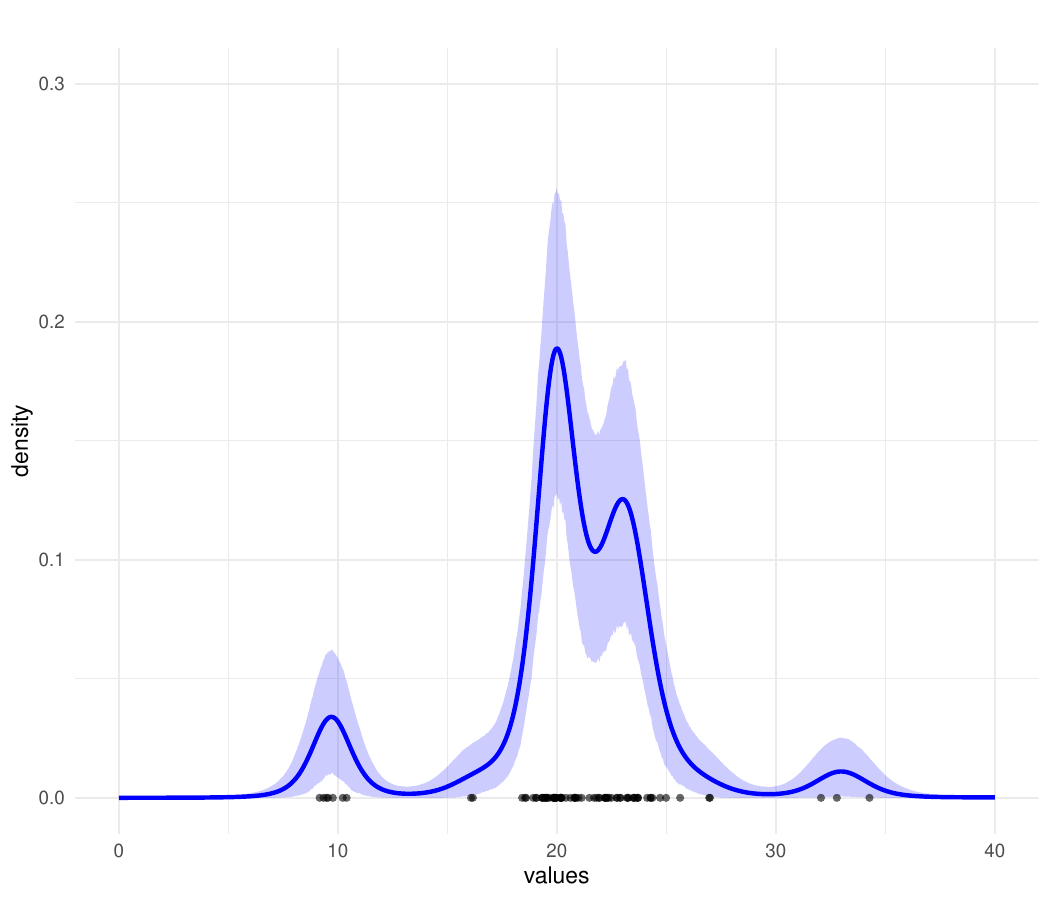}}
\subfigure[]{\includegraphics[width=5cm]{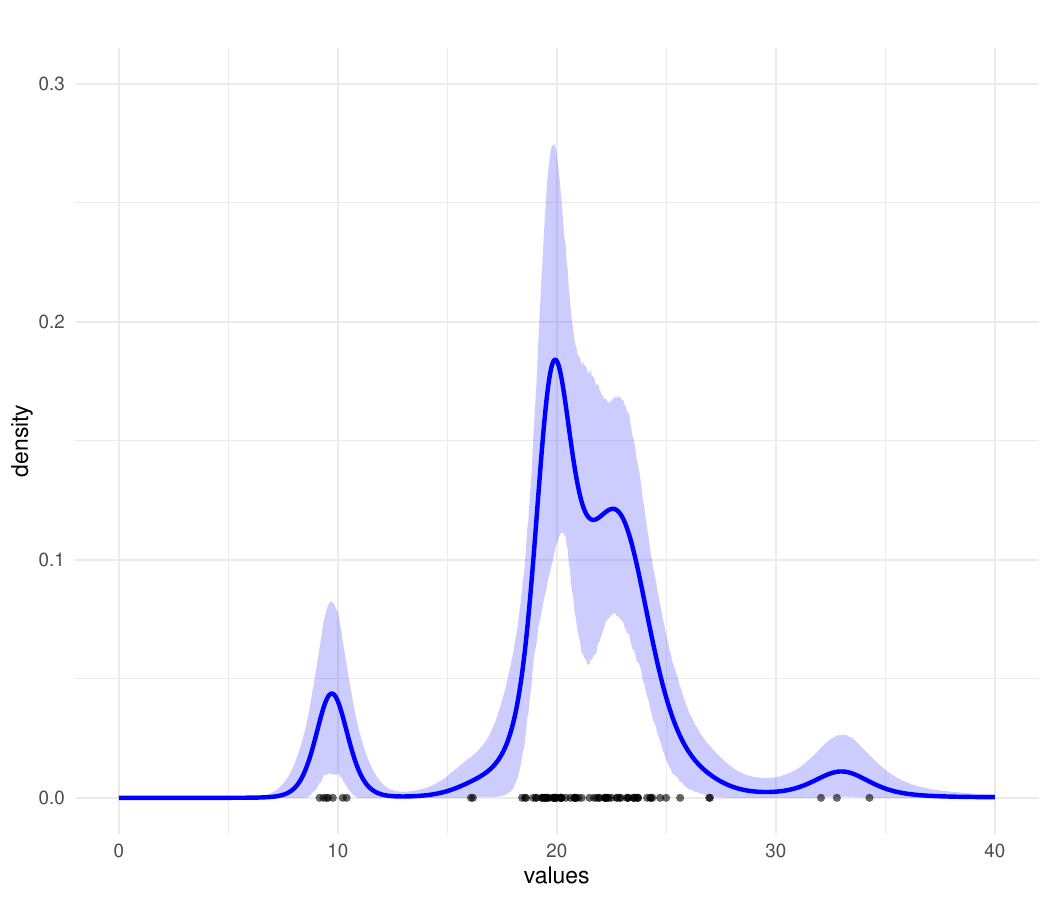}}\\
\subfigure[]{\includegraphics[width=5cm]{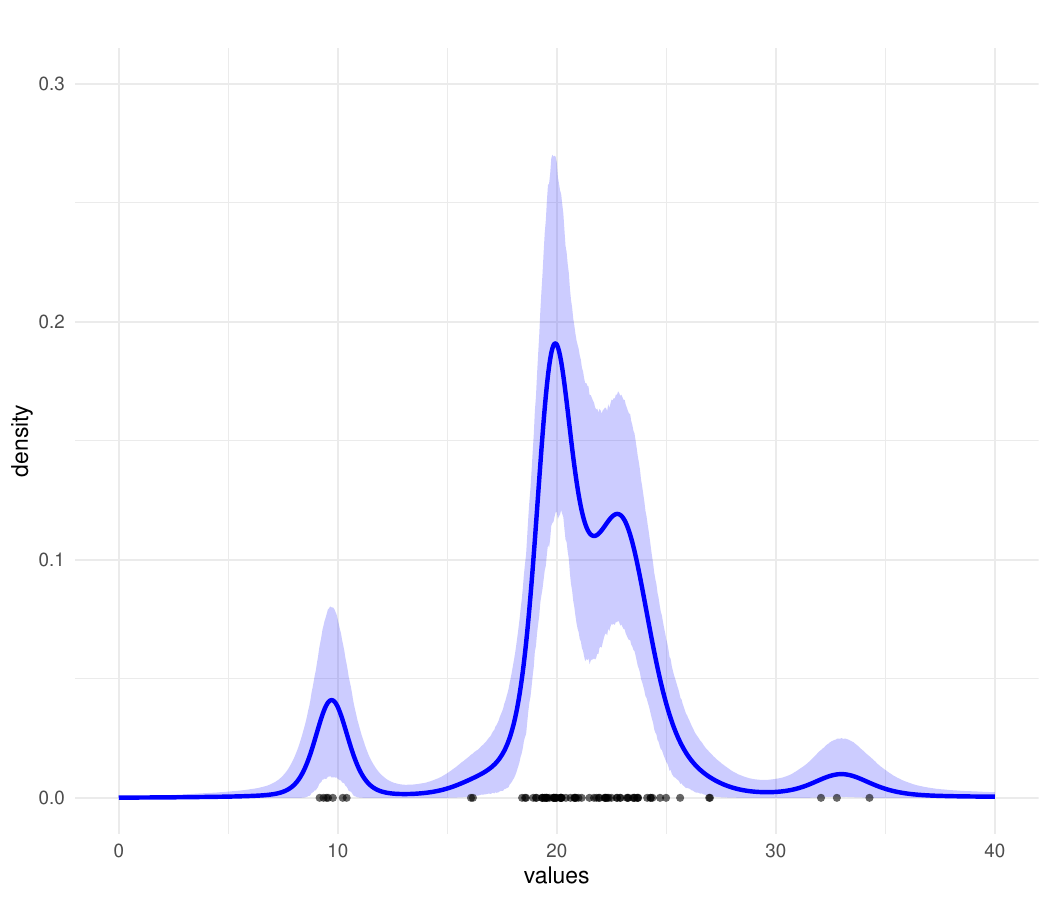}}
\subfigure[]{\includegraphics[width=5cm]{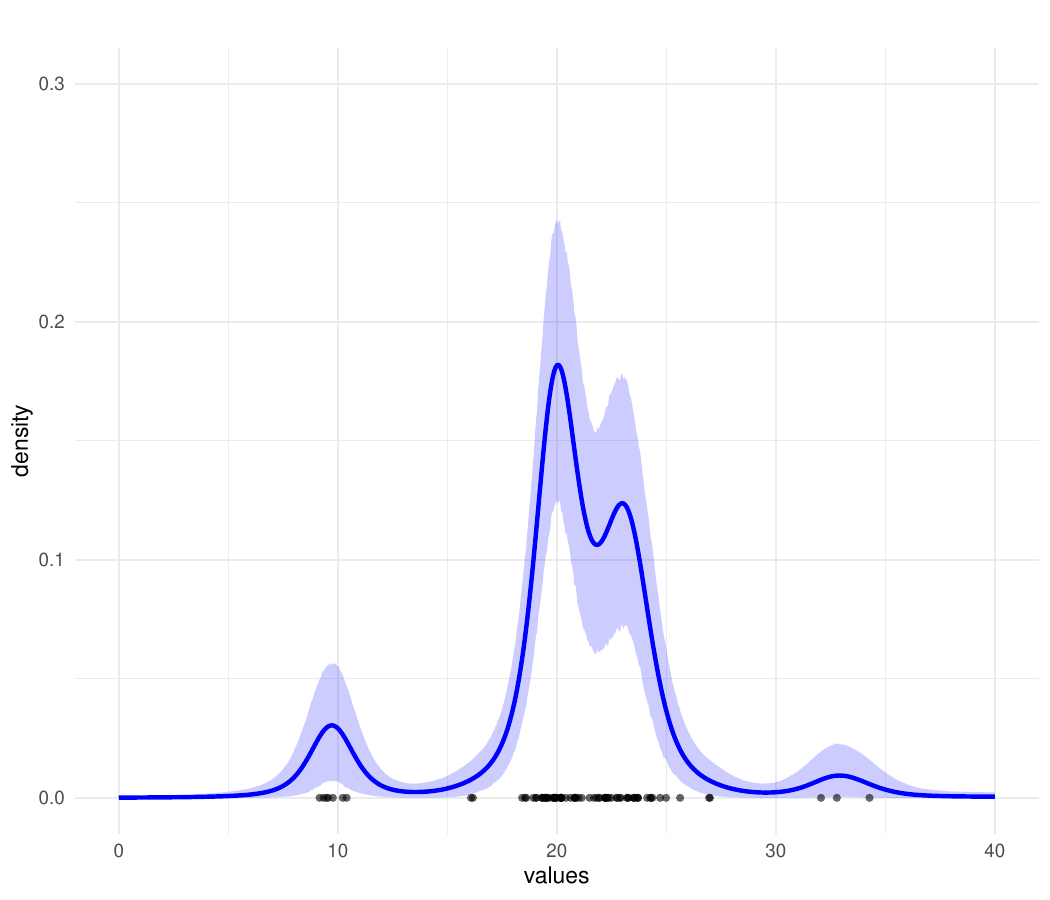}}
\subfigure[]{\includegraphics[width=5cm]{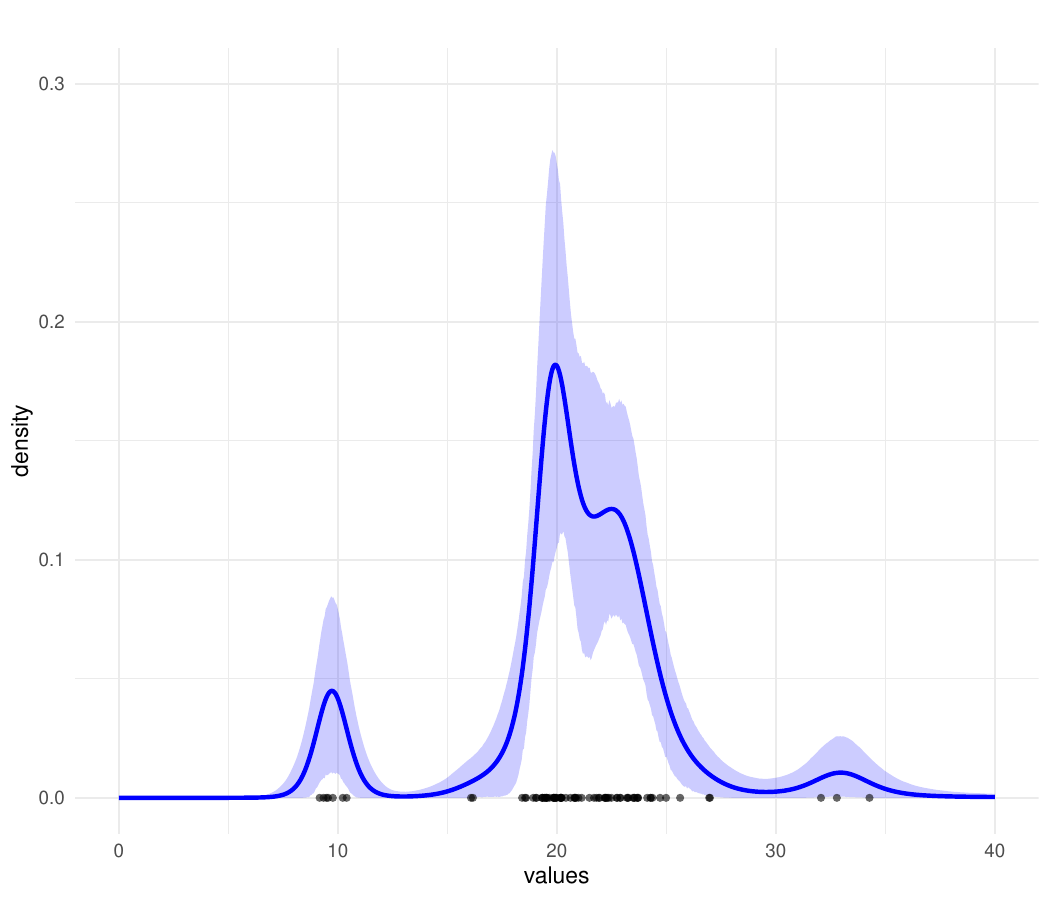}}\\
\subfigure[]{\includegraphics[width=5cm]{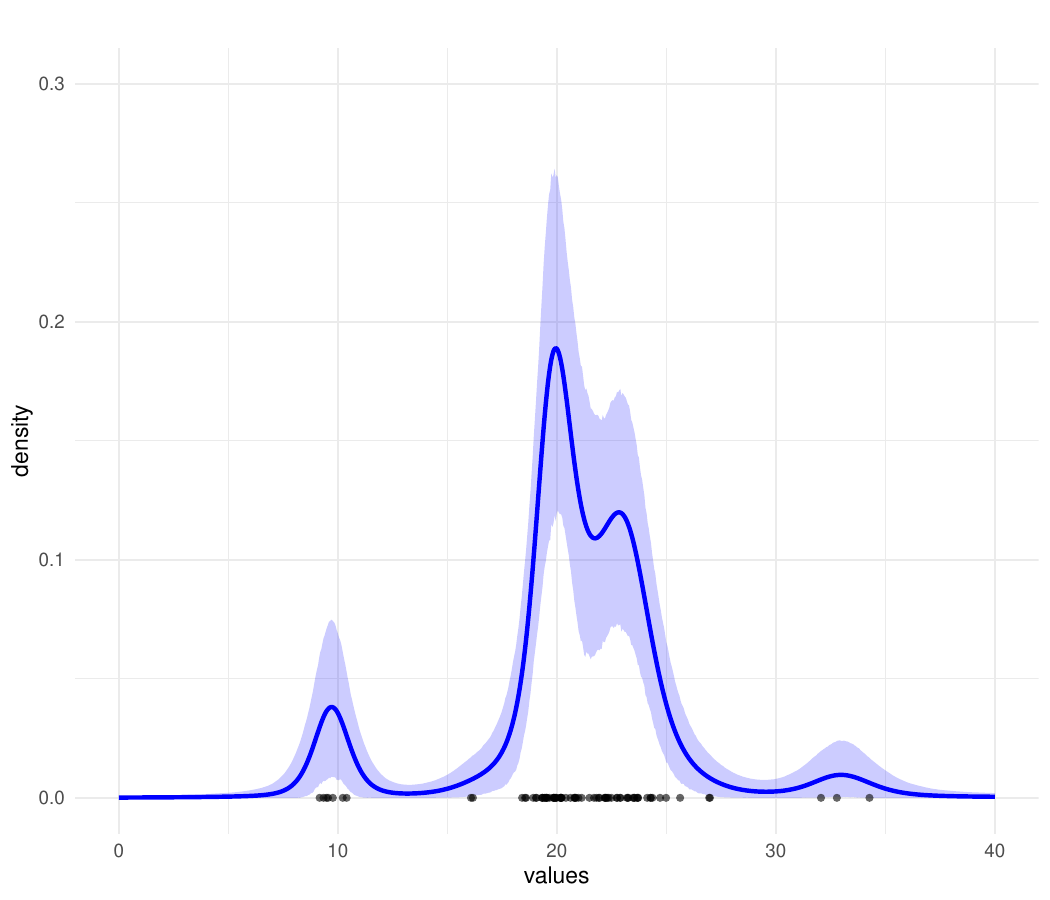}}
\subfigure[]{\includegraphics[width=5cm]{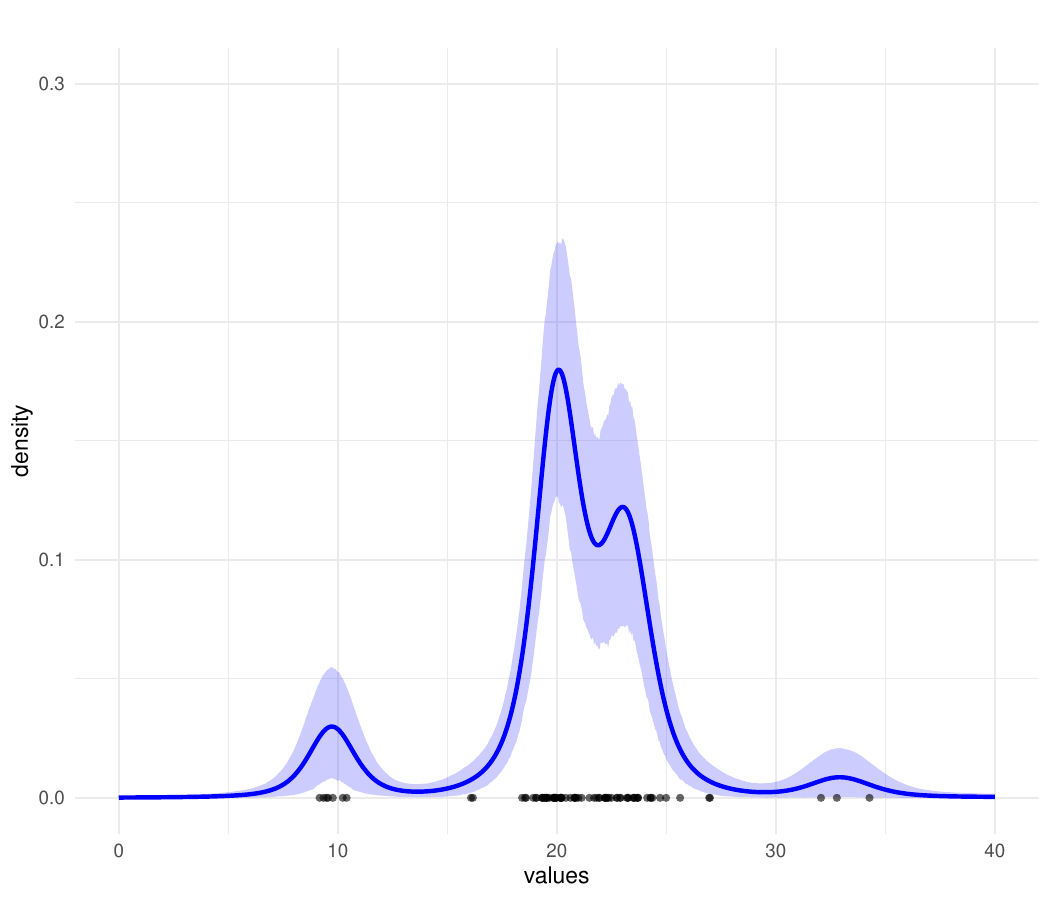}}
\subfigure[]{\includegraphics[width=5cm]{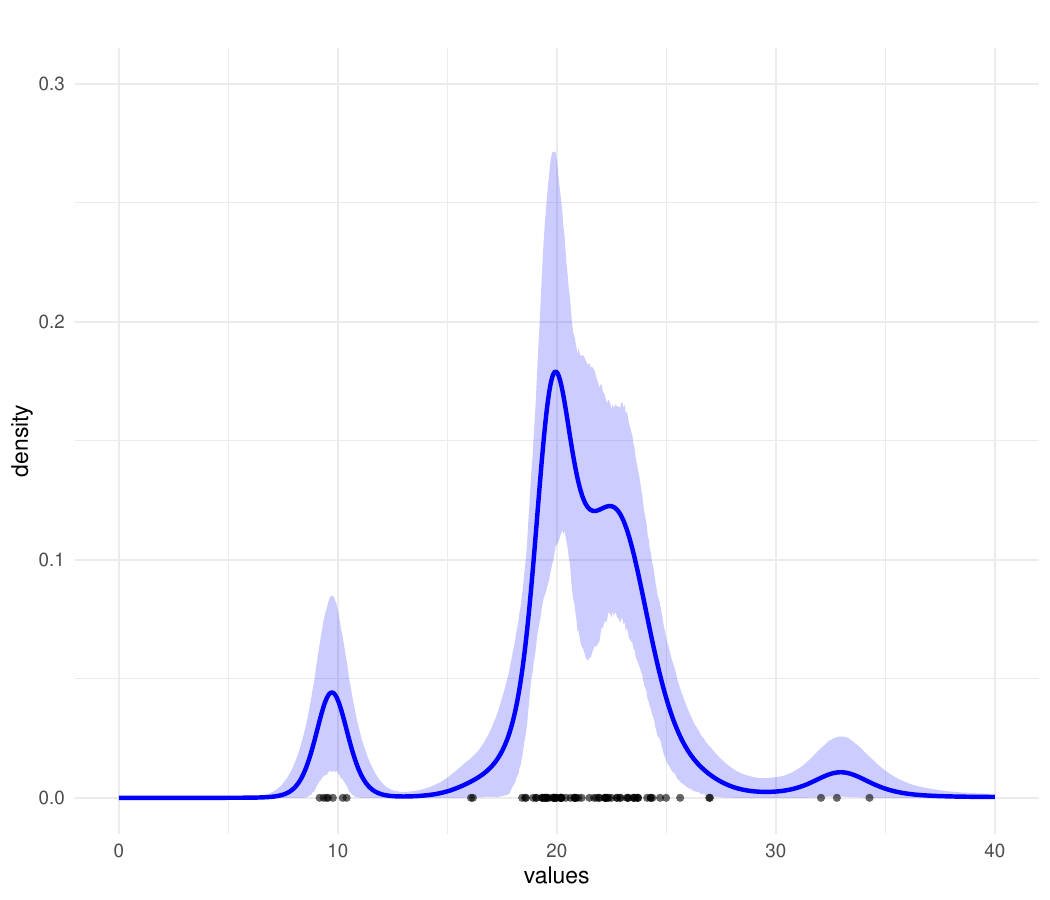}}
\caption{\label{galaxy} 
Galaxy data: Posterior mean and $95\%$ pointwise credible bands for the density. 
Panels (a), (d), and (g) correspond to the parsimonious DSBAS model with $n = 4$, $5$, and $6$, respectively. 
Panels (b), (e), and (h) correspond to the general DSBAS model with $n = 4$, $5$, and $6$, respectively. 
Panels (c), (f), and (i) correspond to the parametric approximation to the DPM model with $L = 8$, $16$, and $32$, respectively.}
\end{figure}

For each model, we ran a total of 220{,}000 MCMC iterations, discarding the first 20{,}000 as burn--in, and then retaining every $10$th iteration to obtain a posterior sample of size $20{,}000$.   Posterior samples were used to estimate the density on a grid of 200 values spanning the range of the observed data.  
Figure~\ref{galaxy} displays the posterior mean and $95\%$ pointwise HPD credible bands for the density under both models.

Across all versions of the DSBAS and DPM models, inferences were nearly identical, suggesting that the proposed method can both incorporate prior information and provide accurate density estimation. The similarity of posterior densities across different values of $n$ and $H$ indicates that, for this dataset, a relatively small number of support points is sufficient to capture the main distributional features. Both the DSBAS and DPM approaches allocate mixture components efficiently to the regions of high density, and the use of informative but weakly regularizing priors on component means and variances further stabilizes inference.  

The WAIC (LPML) values for the parsimonious DSBAS model were $420.02$ ($-210.24$), $422.67$ ($-211.62$), and $423.11$ ($-211.68$), for $n = 4$, $5$, and $6$, respectively.   
For the general DSBAS model  WAIC (LPML) values were $421.96$ ($-211.04$), $425.30$ ($-212.71$), and $424.99$ ($-212.54$), for $n = 4$, $5$, and $6$, respectively.  
For the DPM model, the WAIC (LPML) values were $422.30$ ($-211.42$), $422.77$ ($-211.62$), and $423.21$ ($-211.84$) for $L = 8$, $16$, and $32$, 
respectively. 

The WAIC and LPML values show only negligible differences across the three model classes, indicating that the DSBAS construction achieves a level of model fit comparable to that of the DPM. The parsimonious DSBAS prior, particularly with $n=4$, yielded the best overall predictive performance, although the differences relative to the DPM are minor. This is noteworthy because the DSBAS prior has a fixed and interpretable structure induced by the sequential barycenter array, while also allowing the incorporation of prior beliefs about the marginal mean through the specification of the measures $H_{j,2l-1}$, a feature unavailable in the standard stick--breaking DPM formulation. From a complexity perspective, the parsimonious DSBAS with $n=4$ employs the same number of support points as the DPM with $L=8$, whereas the general DSBAS introduces substantially more support points without delivering improved predictive performance. These findings highlight the advantage of the parsimonious DSBAS in balancing parsimony, interpretability, and predictive adequacy these data.

\subsection{Linear model}

\subsubsection{Simulated data}  
We first illustrate the behavior of the proposed model under a hard constraint on the mean using a linear model and a simulated dataset.  
We generated $n=200$ observations from
\[
y_i = \beta_0 + \beta_1 x_{i1} + \beta_2 x_{i2} + \epsilon_i,
\]
where $\beta_0 = 3$, $\beta_1 = 1$, $\beta_2 = -1$, $x_{i1} \overset{i.i.d.}{\sim} \mathrm{Bern}(0.5)$ is a binary predictor, 
$x_{i2} \overset{i.i.d.}{\sim} N(0,1)$ is a continuous predictor, and the errors are independent draws from a symmetric mixture of normals with mean zero:
\[
\epsilon_i \overset{i.i.d.}{\sim} 0.5 \, N(\cdot \mid -\mu, \sigma^2) + 0.5 \, N(\cdot \mid \mu, \sigma^2),
\]
with $\mu = 1$ and $\sigma = 0.25$.  
The errors $\epsilon_i$ were generated using equally spaced quantiles of the mixture distribution, so they approximate their expected order statistics and represent a ``perfectly representative'' sample.

We fitted the two versions of the DSBAS mixture of normals model for the error distribution, assuming Zellner's $g$-prior \citep{zellner1986gprior} for the regression coefficients, with $g = 100$.  
To avoid identification problems, the DSBA mixture of normals models were defined so that the distribution of the mean is degenerate at zero, i.e., 
$H_{1,1}(\cdot) \equiv \delta_{0}(\cdot)$.  
We set $n = 4$, $H_{j,2\cdot l-1} \sim \mathcal{N}(0, 3)$ for $j = 2, \ldots, n$, 
$m_2 = n$, $\boldsymbol{\alpha}_{m_2} = \mathbf{1}_{m_2}$, and assumed 
$$\phi_j \mid a_\phi, b_\phi \overset{i.i.d.}{\sim} \mathrm{Inverse\mbox{-}Gamma}(1/2, 0.01/2).$$

\begin{figure}
\centering
\subfigure[]{\includegraphics[width=7cm]{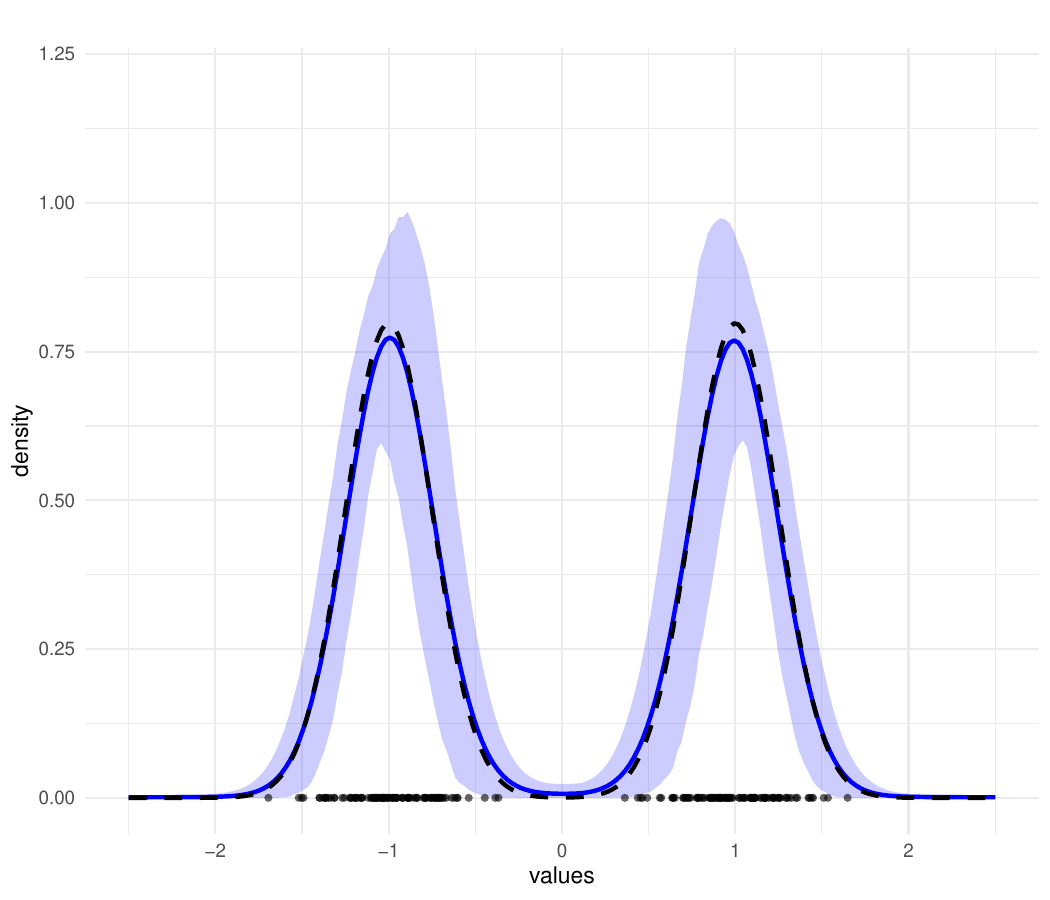}}
\subfigure[]{\includegraphics[width=7cm]{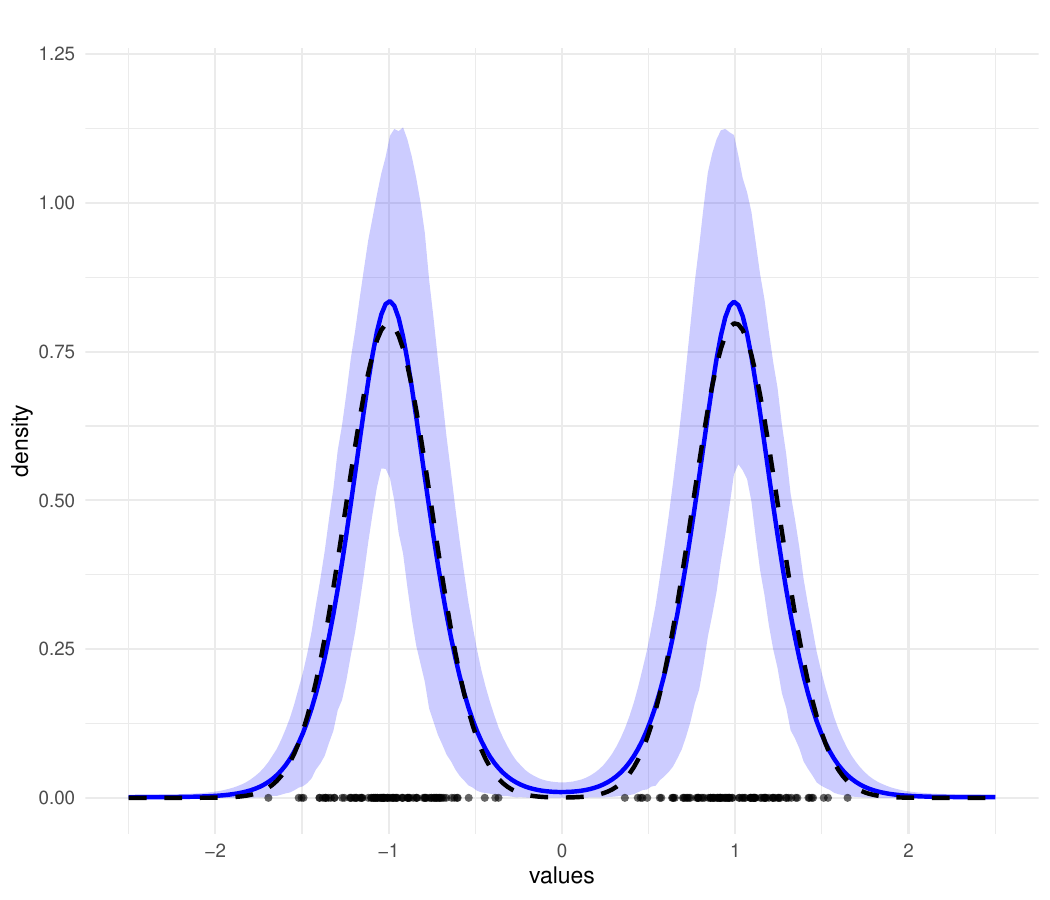}}
\caption{\label{lmsimul} Simulated data: Posterior mean (solid line) and $95\%$ point-wise HPD credible band for the error density. The true density is shown as a dotted line. Panel (a) and (b) correspond to the parsimonious and general DSBAS model, respectively.}
\end{figure}

We ran a total of 220{,}000 MCMC iterations, discarding the first 20{,}000 as burn--in, and then retaining every 10th iteration to obtain a posterior sample of size 20{,}000.  
Posterior samples were used to estimate the error density on a grid of 200 values spanning the range of the observed residuals from an ordinary least squares fit.  
Figure~\ref{lmsimul} shows the estimated error distribution under the DSBAS mixture of normals model.  
The density estimate closely matched the true error distribution across its support, with the truth always lying inside the $95\%$ highest posterior density (HPD) credible bands.  
The posterior means (standard deviations) for $(\beta_0, \beta_1, \beta_2)$ under the parsimonious DSBAS model were $3.0008 \ (0.0766)$, $0.9978 \ (0.0360)$, and $-0.9950 \ (0.0178)$, respectively. The posterior means (standard deviations) for $(\beta_0, \beta_1, \beta_2)$ under the general DSBAS model were $3.0021 \ (0.0780)$, $0.9966 \ (0.0367)$, and $-0.9930 \ (0.0172)$, respectively.

\subsubsection{Australian Institute of Sport data}  
As a second example, we use data from the Australian Institute of Sport (AIS) \citep{cook;weisberg;94}, which have been frequently analyzed in the context of skewed error distributions.  
Specifically, we study the error distribution in a regression model for the lean body mass index (LBM), defined as total body weight minus body fat.  
These and other biomedical variables were collected for 202 athletes (100 females and 102 males) at the AIS.  
A full description of the dataset can be found in \cite{cook;weisberg;94}.  
Enhanced athletic performance is known to be associated with higher LBM.  
Here we study the relationship between the LBM of the AIS \emph{male} athletes and their height (HT) and weight (WT) using the model:
\[
LBM_i = \beta_0 + \beta_1 \times HT_i + \beta_2 \times WT_i + \epsilon_i,
\]
with the error distribution modeled using the two DSBA mixture of normals prior as in the simulated data example. We assumed Zellner's $g$-prior for the regression coefficients \citep{zellner1986gprior}, with $g = 100$, and set $n = 4$, $m_2 = n$, $\boldsymbol{\alpha}_{m_2} = \mathbf{1}_{m_2}$,
$H_{1,1}(\cdot) \equiv \delta_{0}(\cdot)$, and $H_{j,2\cdot l-1} \sim \mathcal{N}(0, 6)$ for $j = 2, \ldots, n$.  
We assumed $$\phi_j \mid a_\phi, b_\phi \overset{i.i.d.}{\sim} \mathrm{Inverse\mbox{-}Gamma}(3/2, 1.5/2),$$ and used the same MCMC settings as in the simulation.

\begin{figure}
\centering
\subfigure[]{\includegraphics[width=7cm]{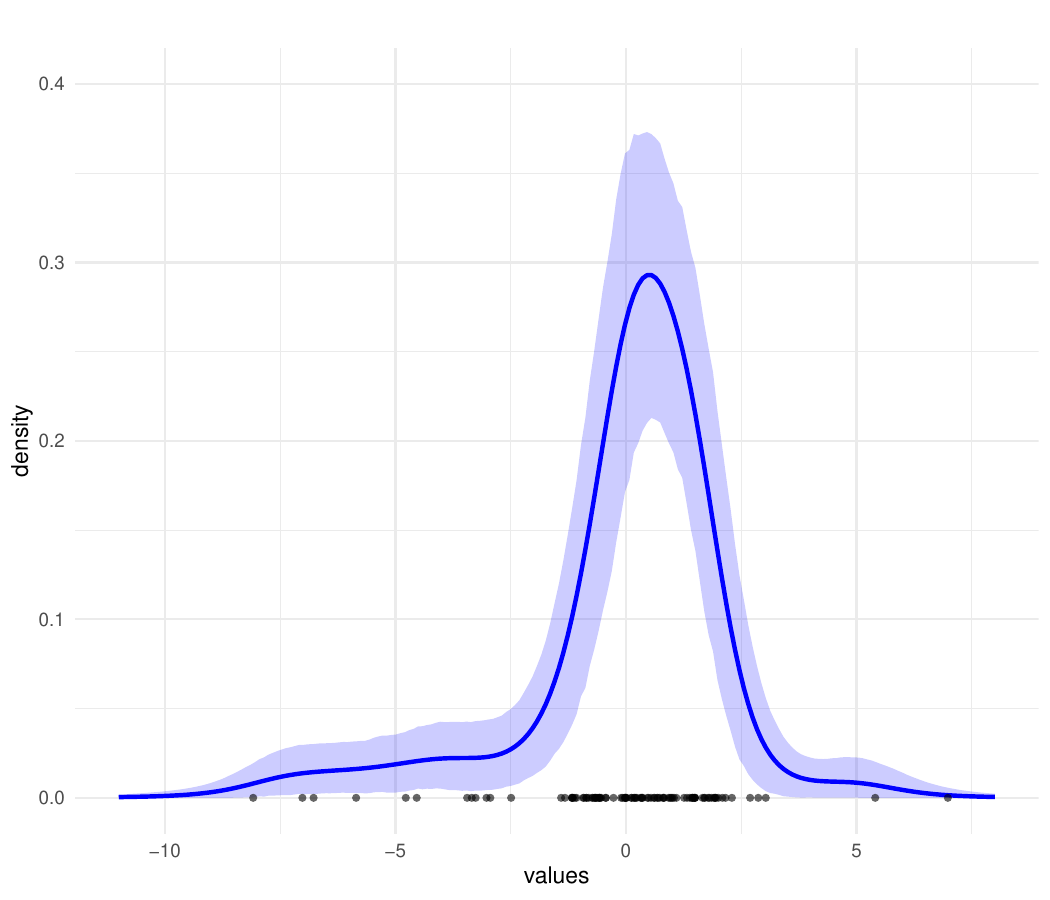}}
\subfigure[]{\includegraphics[width=7cm]{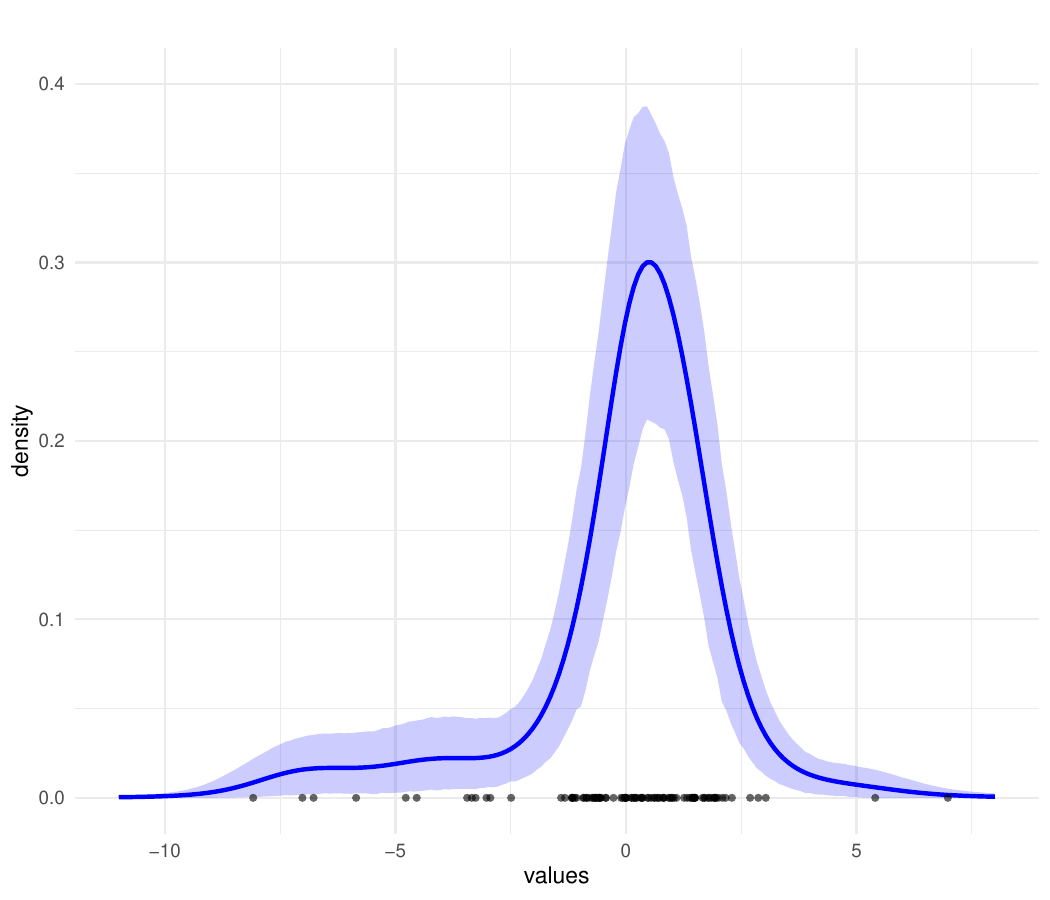}}
\caption{\label{australian} Australian data: Posterior mean (solid line) and $95\%$ point-wise HPD credible band for the error density. Panel (a) and (b) correspond to the parsimonious and general DSBAS model, respectively.}
\end{figure}

Figure~\ref{australian} displays the posterior estimate of the error density.  
There is strong evidence against normality and an apparent skewness in the residual distribution.  
We compared the DSBAS mixture models with the skew--generalized--normal (SGN), skew--normal--$t$ (SNT), and skew--$t$--normal (STN) regression models of \cite{arellano;castro;genton;gomez;2008} using the LPML.  
As reported by those authors, the LPML values for the SGN, SNT, and STN models were $-319.76$, $-349.58$, and $-345.14$, respectively, while the DSBAS mixture model achieved a substantially higher (better) LPML of $-211.15$ and $-212.52$ for the parsimonious and general version, respectively.

\subsection{Hierarchical model}

The Chilean education system is subject to several performance evaluations regularly at the student, school and teacher level. Together with the national voucher system, a national evaluation of student performance was conceived that would provide parents with necessary information to make decisions about schools. Since 1988, the chilean system for measuring the educational quality (Sistema de Medicici\'on de la Calidad de la Educaci\'on, SIMCE) test regularly all the students in a certain grade level in primary and secondary school on an annual basis. Its design and administration is the responsibility of the  Ministry of Education's curriculum and evaluation unit and its main goals are to generate reliable indicators of the improvement of the quality and equity of the education. The SIMCE instruments have been designed to assess the achievement of fundamental goals and minimal contents of the curricular frame in different areas of knowledge.

The SIMCE tests evaluate students in three areas: Spanish, Mathematics and Science. Until 2005 the test was alternatingly given to 4th and 8th graders in primary school (9 and 13 years old, respectively), and 2nd graders in secondary school (16 years old). Since 2006, 4th graders are evaluated every year. We consider a sample from the data obtained in the 2004 SIMCE math test applied to 8th graders. During the test, examinees had to answer 47 questions or items in 90 minutes. Only one of the items was an open question. The remaining 46 were multiple choice items, each of them with 4 possible options. Every item had only one correct answer.

We consider a sample of $N=500$ subjects and the first $J=10$ questions of the test. We define binary response variables $Y_{ij}$ indicating whether the $i$th student answer  correctly the $j$th question ($Y_{ij}=1$) or not ($Y_{ij}=0$),  $i=1,\ldots,N$, $j=1,\ldots,J$. We consider the Rasch model \cite[see, e.g.,][and references therein]{sanmartin;jara;rolin;mouchrat;2011}, given by
\begin{eqnarray}\nonumber
Y_{ij}\mid \lambda_{ij} &\overset{ind}{\sim}&  \mbox{Bernoulli}\,
(\lambda_{ij})\\\nonumber
\lambda_{ij} &=& \frac{\exp\{b_i-\beta_j\}}{1+\exp\{b_i-\beta_j\}},
\end{eqnarray}
where $\beta_j \in \mathbb{R}$ represents the difficulty of the item $j$ and  $b_i$ represents the {\em ability} of subject $i$. The ability parameters
are considered as random effects whereas the difficulty parameters
are interpreted as ``fixed" effects. The classical specification of the model is completed by choosing a probability model for the abilities. A typical assumption in the item-response literatura is to assume that
\begin{eqnarray}\nonumber
b_1,\ldots,b_m \mid \mu_b,\sigma_b^2 \overset{iid}{\sim} \mathcal{N}(\mu_b,\sigma_b^2).
\end{eqnarray}
To avoid identification problems $\mu_b$ must be fixed to zero. This type of restrictions must also be applied when the abilities's distribution is modeled using random probability measures \cite{sanmartin;jara;rolin;mouchrat;2011}. Thus, the distribution of the abilities $b_i$  was modeled using DSBAS mixture of normals. To avoid identification problems, the DSBA mixture of normals models were defined so that the distribution of the mean is degenerate at zero, i.e.,  $H_{1,1}(\cdot) \equiv \delta_{0}(\cdot)$.  We set $H_{j,2l-1}\sim\mathcal{N}(0,1)$ for $j=2,\ldots,n$, $m_2 = n$, $\boldsymbol{\alpha}_{m_2} = \mathbf{1}_{m_2}$, and assumed $\phi_j \,\overset{\text{i.i.d.}}{\sim}\, \mathrm{Inverse\mbox{-}Gamma}(3/2,\,1/2)$. For the difficulty parameters we assumed that $\beta_\ell \,\overset{\text{i.i.d.}}{\sim}\, \mathcal{N}(0,100)$, $\ell=0,\ldots,10$. MCMC settings follow those in the previous sections.

\begin{figure}
\centering
\subfigure[]{\includegraphics[width=7cm]{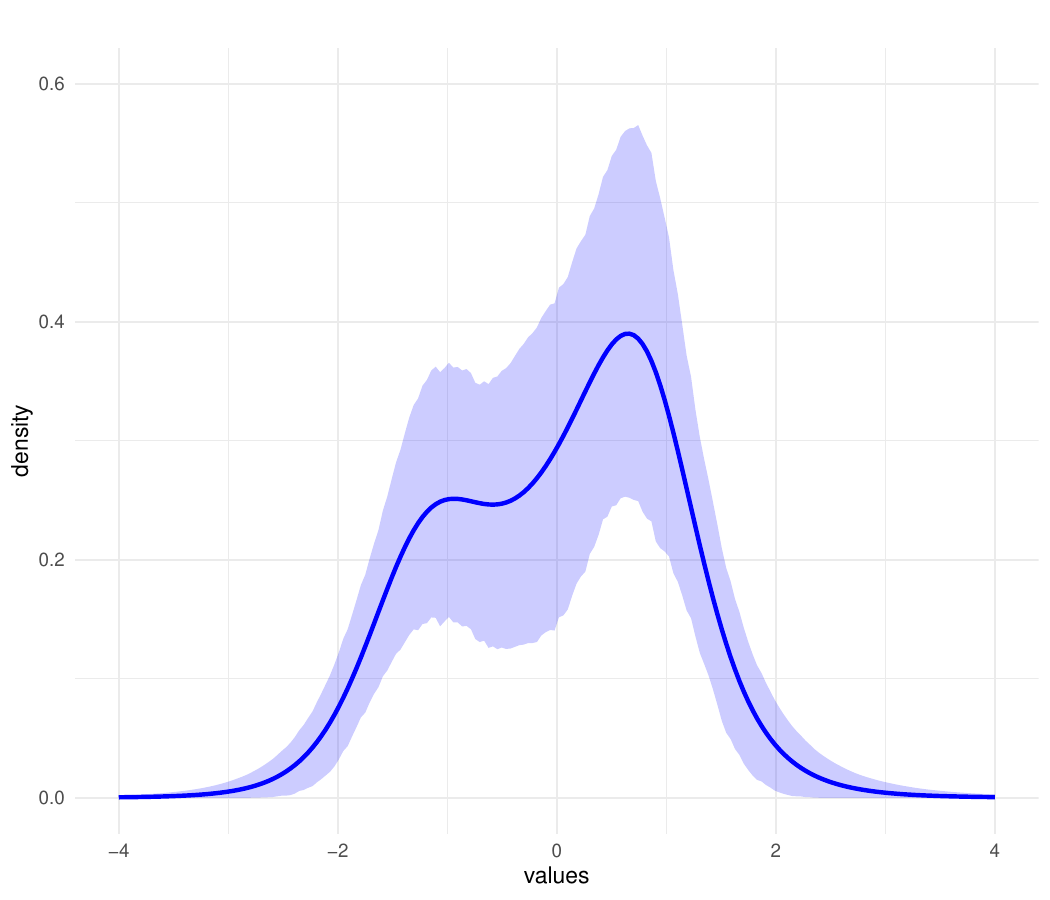}}
\subfigure[]{\includegraphics[width=7cm]{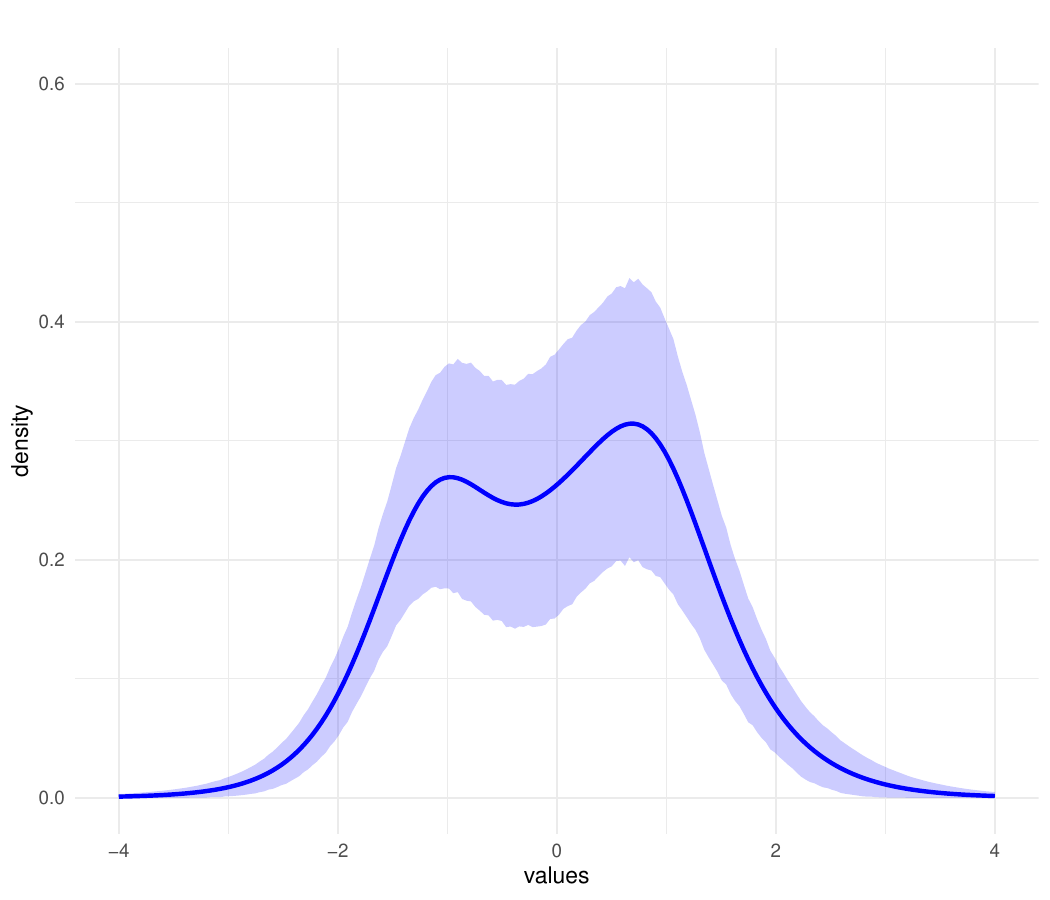}}
\caption{\label{simce} SIMCE  data. Posterior mean (solid line) and $95\%$ point-wise HPD credible band for the abilities density under the parsimonious (panel a) and general (panel b) DSBAS models.}
\end{figure}

Figure~\ref{simce} displays the posterior density of the ability distribution under the parsimonious and independent DSBAS specifications. In both cases the inferred distribution departs from the Gaussian benchmark often assumed in IRT: it is asymmetric and heavier–tailed, indicating marked between–student heterogeneity that item difficulties alone do not absorb. The central tendency and main bulk are stable across priors, while the independent DSBAS yields a slightly wider $95\%$ poin-twise HPD envelope—particularly in the tails—reflecting additional flexibility rather than a shift in location. Local undulations under the full prior resemble mild shoulders but remain within HPD regions compatible with unimodality, offering at most weak evidence for multiple well–separated modes. From a predictive standpoint, a flexible (non-Gaussian) prior primarily affects person-level estimates and calibration at the extremes, mitigating overshrinkage toward the center without materially altering conclusions about item difficulties. Model comparison clearly favors the parsimonious DSBAS. The WAIC (LPML) values were $5434.583$ ($-2719.274$) for the parsimonious model and $5448.279$ ($-2726.857$) for the general model. Since lower WAIC and higher LPML indicate better predictive performance, these differences---$\Delta\text{WAIC}\approx 13.7$ and an LPML improvement of $\approx 7.6$---are meaningful and point to the parsimonious specification as the preferred balance of fit and parsimony; the added flexibility of the general model broadens tail uncertainty but does not translate into predictive gains.

\section{Concluding remarks}\label{section6}
We introduced DSBA priors for density estimation that allow practitioners to encode prior information directly on the mean (or its distribution), while retaining the approximation and support guarantees of the underlying SBA construction. Across simulated and real examples, both the parsimonious and general DSBAS variants delivered fits comparable to widely used BNP baselines, while offering direct and interpretable control over the marginal mean.

The key advantages of the proposal are: (i) transparent control of the mean through the top node of the SBA, allowing for both weak and strong prior information; (ii) a finite, structured set of atoms that scales in a predictable way with the SBA level $n$; (iii) strong support results under different relevant topologies; and (iv) straightforward posterior computation with only a few non-conjugate updates. Practically, the parsimonious specification often strikes the best balance between flexibility and parsimony, whereas the general location--scale variant offers wider tail adaptivity without material loss of interpretability.

Beyond the results established here, several directions remain the subject of ongoing research.  First, we are developing efficient trans-dimensional algorithms (e.g., reversible-jump, slice-RJ, or birth--death samplers) for random $n$ (and, in the independent model, random $m_2$), including data-driven priors on depth and adaptive proposals that preserve the SBA ordering constraints.  
Second, multivariate extensions based on BNP copula priors and other constructions are under investigation. In the former, recent work on grid-uniform copulas \citep{kuschinskiJara2025griduniform} and Bernstein-yett-uniform copulas \citep{kuschinskiWarrJara2024bernsteinYettUniform} provide promising building blocks for this direction.  
Third, we are exploring non-regular SBA random measures, where repeated barycenters are allowed, leading to a reduced number of support points at each level and potential computational and shrinkage benefits.  
Finally, we are studying the induced prior on clustering structures, including the expected number of occupied components, cluster-size distributions, sensitivity to $n$, $\mathcal H_n$, and kernel choices, and comparisons to CRP/PD and NRMI clustering laws. This also involves investigating posterior consistency of the number of clusters and asymptotic behavior under model misspecification.

Although in this paper we focus on a Bayesian implementation of the proposed framework, the methodology is not restricted to the Bayesian paradigm. The discrete SBA representation induces a finite mixture model whose likelihood admits a standard latent variable formulation, which in turn makes NPMLE feasible. In this setting, an EM algorithm can be constructed by introducing allocation variables for the mixture components: the E–step yields the usual posterior responsibilities, while the M–step requires maximizing the expected complete–data log-likelihood with respect to the SBA array. This step is non-trivial because both the atoms and the weights of the mixture are nonlinear functions of the array and must satisfy the structural SBA constraints (monotonicity and inheritance). Developing efficient numerical optimization routines for this constrained M–step—such as projected gradient or Newton-type updates—would complete a full NPMLE implementation. This direction is methodologically appealing and highlights that the SBA framework provides a general representation and estimation mechanism beyond Bayesian inference.

\appendix
\section*{Acknowledgements}
The first author acknowledges support from FONDECYT grant No. 1220907. The second author acknowledges support from a grant Puente 2025 and an Open Seed Fund 2024 from Pontificia Universidad Católica de Chile, and from the National Center for Artificial Intelligence CENIA FB210017, Basal ANID.  Part of this research was carried out during a Visiting Professorship at the Department of Statistics, Brigham Young University. 

\section*{Appendix}
\setcounter{equation}{0}
\renewcommand{\theequation}{A.\arabic{equation}}

\section{Preliminaries}\label{Appendix0}

We assume that  $\Theta$ is a non-degenerate closed interval of the real line, i.e., the entire real line, a closed half-line, or a compact interval with non-empty interior. If \(\Th\) is endowed with the usual topology, then it is Polish. We denote as \(\mPTh\) the set of probability measures on \((\Th,\mBTh)\). If \(A\in \mBTh\) then its Lebesgue measure is denoted simply as \(|A|\). The support of \(G\in\mPTh\) is denoted as \(\supp(G)\).

Recall that a neighborhood basis for the weak topology on \(\mPTh\) can be constructed as follows. Let \(\CobTh\) be the set of continuous and bounded functions on \(\R\). When \(\Th\) is compact then this space coincides with \(\CoTh\). Given \(\Go\in\mPTh\), \(f \in \Cob(\R)\) and \(\eps > 0\) we define the weak neighborhood
\[
    V_0(\Go, f, \eps) := \Lset{G\in \mPTh:\,\, \left|\int_{\Th} f(\th) d(G - \Go)(\th)\right| < \eps}.
\]
The collection of sets of the form 
\[
    V = \bigcap_{i=1}^m V_0(\Go, f_i, \eps_i),
\]
for some \(f_1,\ldots, f_m\in \CobTh\) and \(\eps_1,\ldots, \eps_m > 0\) form a neighborhood basis for \(\Go\). In general, we write
\[
    V(\Go, f_1,\ldots, f_m, \eps_1,\ldots,\eps_m) := \bigcap_{i=1}^m V_0(\Go, f_i, \eps_i).
\]

For \(p\in [1,\infty)\) we denote \(\mPpTh\) the subset of \(\mPTh\) of probability measures with finite \(p\)-th moment, i.e.,
\[
    \mPpTh := \Lset{G\in \mPTh:\,\, \int_{\Th} |\th|^p\, dG(\th) < \infty}.
\]
We can endow \(\mPpTh\) with the subspace topology. However, it is appropriate to endow it with the weak topology. For \(\Go\in \mPpTh\) we define
\[
    V_p(G_0, \eps) := \Lset{G\in \mPpTh:\,\,\left|\int_{\Th} |\th|^p\, d(G - \Go)(\th)\right| < \eps_i }
\]
then a neighborhood basis at \(\Go\) is given by sets of the form
\[
    \bigcap_{i=1}^m V_0(\Go, f_i, \eps_i)\,\,\mbox{and}\,\, V_p(G_0, \eps_p) \cap \bigcap_{i=1}^m V_0(\Go, f_i, \eps_i)
\]
for some \(f_1,\ldots, f_m\in \CobTh\) and \(\eps_1,\ldots, \eps_m > 0\). It is sometimes useful to endow \(\mPpTh\) with the Wasserstein distance of order \(p\). For \(G_1,G_2\in \mPpTh\) we let
\[
    \Gamma(G_1, G_2) := \Lset{\gamma\in \mP(\Th\times\Th):\, A\in\mBTh:\, \gamma(A\times\Th) = G_1(A)\,\,\mbox{and}\,\,\gamma(\Th\times A) = G_2(A)},
\]
denote the set of couplings between \(G_1, G_2\). The Wasserstein distance of order \(p\) between \(G_1\) and \(G_2\) is then defined as
\[
    \Wp(G_1, G_2) = \inf\Lset{\left(\iint_{\Th\times\Th} |\th' - \th|^p\, d\gamma(\th', \th)\right)^{1/p}:\, \gamma\in\Gamma(G_1, G_2)}.
\]
Since \(\Th\) is Polish, the space \((\mPpTh, \Wp)\) is a complete separable metric space  \cite[see, e.g., Theorem~6.18 in][]{villani2008optimal}.  

Unless explicitly stated otherwise, we assume that \(\mPTh\) is endowed with the weak topology and that \(\mPpTh\) is endowed with the metric topology induced by the Wasserstein distance of order \(p\). Furthermore, 
the topological interior and closure of a set are denoted as \(\topint\) and \(\topcl\) respectively, and the convex hull of a set is denoted as \(\cvxhull\).

\setcounter{equation}{0}
\renewcommand{\theequation}{B.\arabic{equation}}

\section{Proofs for the Results of Section 2}\label{AppendixA}

\subsection{Proof of Lemma~\ref{lem:regSBAPartitionSupport}}\label{AppendixA1}

Let \(S = \supp(G)\). We prove the proposition by induction on the level of the SBA. Since \(S\) is a non-degenerate interval, we have that \(S = \cvxhull(S)\). Note that we do not take the closure of the hull. For \(n = 1\) the barycenter \(\mu_{1,1}\) is the expected value and, by standard arguments, \(\mu_{1,1} \in \topint(S)\). It is apparent that \(\tTh_{1,1}\) and \(\tTh_{1,2}\) as defined in~\eqref{eq:regSBA:partitionOfSupport} are intervals with non-empty interior that partition \(S\).

 Suppose now that the statement is true for \(n' \leq n\). It follows that \(\set{\tTh_{n,l}}_{l=1}^{2^n}\) is a partition of \(S\). Let \(l \in \set{1,\ldots, 2^{n}}\) and let \(\mu_{n+1,2\cdot l - 1} = b_{G}(\tTh_{n,l})\). If \(\mu_{n+1,2\cdot l-1}\) is not on the interior of \(\tTh_{n,l}\) it would imply that \(G(\topint(\tTh_{n,l})) = 0\). However, it would follow that \(\topint(\tTh_{n,l}) \subset S^c\) contradicting the induction hypothesis. Therefore, \(\mu_{n+1,2\cdot l-1} \in \topint(\tTh_{n,l})\). Thus \(\tTh_{n+1, 2\cdot l - 1} = \tTh_{n,l}\cap (\mu_{n, 0}, \mu_{n+1,2\cdot l-1}]\) if \(\mu_{n,0} = -\infty\) or \(\tTh_{n+1, 2\cdot l - 1} = \tTh_{n,l}\cap [\mu_{n, 0}, \mu_{n+1,2\cdot l-1}]\) if \(\mu_{n,0} > -\infty\), and \(\tTh_{n+1, 2\cdot l} = \tTh_{n,l}\cap (\mu_{n+1,2\cdot l-1}, \mu_{n, 2^n})\) if \(\mu_{n, 2^n} = +\infty\) or \(\tTh_{j+1, 2\cdot l} = \tTh_{j,l}\cap (\mu_{j+1,2\cdot l-1}, \mu_{n, 2^n}]\) if \(\mu_{n, 2^n} < \infty\), are non-empty and non-degenerate intervals that partition \(\tTh_{n,l}\). It follows that \(\set{\tTh_{n+1,l}}_{l\in 2^{n+1}}\) is a partition of \(S\) and thus the lemma follows.
\hfill$\square$

\subsection{Proof of Lemma~\ref{lem:regSBAIntervalReduction}}\label{AppendixA2}

  Let \(S\) be the support of \(G\). It follows from \(G\in\mPbTh\) that \(S\) is a compact interval and, from Lemma~\ref{lem:regSBAPartitionSupport}, that the sequence of intervals \(\set{\tTh_{n,l(n)}}_{n\in\N}\) is comprised of non-degenerate intervals with \(\tTh_{n,l(n)}\subset S\). To simplify the notation, we write \(\tThn := \tTh_{n, l(n)}\). Define
    \[
        \tThinf = \bigcap_{n\in\N} \tTh_{n}.
    \]
 Then, \(\tThinf\) is a possibly degenerate interval. If \(\topint(\tThinf) = \emptyset\), then \(|\tThinf| = 0\) and, by continuity from above as \(S\) is compact, we conclude that
    \[
        \lim_{n\to\infty}\, |\tTh_{n}| = 0,
    \]
    proving the statement in this case. Hence, assume that \(\topint(\tThinf) \neq \emptyset\). If \(G(\topint(\tThinf)) = 0\), then no point in \(\topint(\tThinf)\) can belong to \(S\). Since \(\tThinf\subset S\), we must have that \(G(\topint(\tThinf)) > 0\) whence \(G(\tThinf) > 0\). In this case, the restriction \(\Ginf\) of \(G\) to \(\tThinf\) is well defined and \(|\tThinf| > 0\) as \(\topint(\tThinf)\neq \emptyset\). We define 
    \[
        \ainf := \inf\, \tThinf\quad\mbox{and}\quad \binf := \sup\, \tThinf,
    \]
    which are both finite as \(S\) is compact, and the barycenter
    \[
        \mu_{\infty} = b_{G}(\tThinf) = \int_{\tThinf} \th d\Ginf(\th).
    \]
    Since \(G(\topint(\tThinf)) > 0\) we must have that
    \[
        \delta_{\infty} := \frac{\mu_{\infty} - \ainf}{|\tThinf|},
    \]
    satisfies \(\delta_{\infty} \in (0, 1)\).
    
    From Lemma~\ref{lem:regSBAPartitionSupport}, each interval \(\tTh_{n}\) has non-empty interior and thus \(|\tTh_{n}| > 0\) for every \(n\). Since \(|\tThinf| > 0\) by assumption, we can choose \(\eps > 0\), such that
    \[
        \eps (1 + \eps)^2 < \min\set{\delta_{\infty},\, 1 - \delta_{\infty}}.
    \]
    Since \(\tThinf\subset S\) and \(|S| < \infty\) by compactness then, by continuity from above, we can find \(n_0\in\N\) such that for all \(n > n_0\) we have that
    \[
        |\tThinf| \leq |\tTh_{n}| < (1+ \eps) |\tThinf|\quad\mbox{and}\quad G(\tThinf) \leq G(\tTh_{n}) < (1+ \eps) G(\tThinf).
    \]
    Fix \(n > n_0\) and let \(\Gn\) be the restriction of \(G\) to \(\tThn\). To simplify the notation, let
    \[
        a_n := \inf\, \tThn\quad\mbox{and}\quad b_n := \sup\, \tThn,
    \]
    which are finite by compactness of \(S\). Since \(\tThinf\) is nondegenerate, we must have that \(a_n \leq \ainf < \binf \leq b_n\). We will show that the barycenter \(\mu_n\) of \(\Gn\) must lie on \(\tThinf\). This leads to a contradiction that yields our claim. 

    First observe that
    \[
        \mu_n - \ainf = \int_{\tThn} (\th - \ainf) d\Gn(\th) = \int_{\tThn\setminus\tThinf} (\th - \ainf) d\Gn(\th) + \int_{\tThinf} (\th - \ainf) d\Gn(\th).
    \]
    The first term in the right-hand side can be bounded above as
    \begin{align*}
        \left|\int_{\tThn\setminus\tThinf} (\th - \ainf) d\Gn(\th)\right| &\leq (b_n - \ainf)\Gn(\tThn\setminus\tThinf), \\
        &\leq |\tThn| \Gn(\tThn\setminus\tThinf), \\
        &< (1+\eps) |\tThinf|  \Gn(\tThn\setminus\tThinf).
    \end{align*}
    Note that
    \[
        \Gn(\tThn\setminus\tThinf) = \frac{G(\tThn) -G(\tThinf)}{G(\tThn)} < \eps \frac{G(\tThinf)}{G(\tThn)} \leq \eps,
    \]
    whence
    \[
        \left|\int_{\tThn\setminus\tThinf} (\th - \ainf) d\Gn(\th)\right| < \eps (1+\eps) |\tThinf|.
    \]
    The second term can be bounded below as
    \[
        \int_{\tThinf} (\th - \ainf) d\Gn(\th) = \frac{G(\tThinf)}{G(\tThn)} \int_{\tThinf} (\th - \ainf) d\Ginf(\th) > \frac{1}{1 + \eps} \frac{\mu_{\infty} - \ainf}{|\tThinf|} |\tThinf|.
    \]
    Remark then that
    \[
        \mu_n - \ainf > \left(\delta_{\infty} - \eps(1+\eps)^2\right)\frac{ |\tThinf| }{1 + \eps} > 0,
    \]
    where the positivity follows from our choice for \(\eps\). Now, remark that the same arguments yield
    \[
        \binf - \mu_n = \int_{\tThn\setminus\tThinf} (\binf - \th) d\Gn(\th) + \int_{\tThinf} (\binf - \th) d\Gn(\th).
    \]
    The first integral can be bounded above exactly in the same way as
    \[
        \left|\int_{\tThn\setminus\tThinf} (\binf - \th) d\Gn(\th)\right| \leq (\binf - a_n)\Gn(\tThn\setminus\tThinf) \leq |\tThn| \Gn(\tThn\setminus\tThinf) < \eps (1+\eps)|\tThinf|.
    \]
    The second can be bounded below similarly as
    \begin{eqnarray}\nonumber
        \int_{\tThinf} (\binf - \th) d\Gn(\th) &=& \frac{G(\tThinf)}{G(\tThn)} \int_{\tThinf} (\binf - \th) d\Ginf(\th),\\\nonumber
         &>& \frac{1}{1 + \eps} \frac{\binf - \mu_n}{|\tThinf|} |\tThinf|,\\\nonumber 
         &=& \frac{1}{1 + \eps} (1 - \delta_{\infty}) |\tThinf|.
        \end{eqnarray}
    Therefore,
    \[
        \binf - \mu_n > \left(1 - \delta_{\infty} - \eps(1+\eps)^2\right)\frac{ |\tThinf| }{1 + \eps} > 0,
    \]
    where the positivity also follows from our choice of \(\eps\). We conclude that
    \[
        \inf\, \tThinf < \mu_n < \sup\, \tThinf,
    \]
    whence we must have that \(\mu_n \in \tThinf\). We conclude that \(\topint(\tThinf) = \emptyset\) whence \(|\tThinf| = 0\). In particular, the same arguments as before imply that in this case
    \[
        \lim_{n\to\infty}\,\, |\tThn| = 0.
    \]
    
To prove the second statement, remark that if the claim is false, then there exists a not necessarily decreasing sequence \(\set{\tTh_{n,l(n)}}_{n\in\N}\) such that \(|\tTh_{n,l(n)}| \not\to 0\). We now show how to extract a decreasing subsequence as follows. By construction of the SBA, for \(n = 1\) at least one of the intervals \(\tTh_{1, 1}, \tTh_{1, 2}\) must contain an infinite number of terms of the sequence \(\set{\tTh_{n, l(n)}}_{n > 1}\). We let \(\tTh_{1,k(1)}\) be the first term of the subsequence. Now suppose that we have constructed the sequence \(\tTh_{1,k(1)} \supset \ldots \supset \tTh_{n, k(n)}\) in this manner. Then one interval \(\tTh_{n+1,l}\) for \(l\in \set{1,\ldots, 2^{n+1}}\) must contain an infinite number of terms of the sequence \(\set{\tTh_{n',l(n')}}_{n' > n}\). Hence, we can select the \(n+1\)-th term of the sequence in this manner. Therefore, the sequence \(\set{\tTh_{n,k(n)}}_{n\in\N}\) is decreasing and \(|\tTh_{n,k(n)}| \not\to 0\). However, our previous result shows that this cannot be the case. This proves the claim.
\hfill$\square$

\subsection{Proof of Lemma~\ref{lem:regSBADenseWeakStar}}\label{AppendixA3}
 Let \(\Go\in \mPTh\). We will prove that for any \(f_1,\ldots, f_m \in \CobTh\) and \(\eps_1,\ldots,\eps_m > 0\) there exists a probability measure \(G\in\mPbTh\), an \(\eps > 0\), and a collection \(f'_1,\ldots, f'_m\in \CobTh\)  with \(|f'_1|,\ldots, |f'_m| \leq 1\), such that
 \[
    V(G, f_1',\ldots, f_m', \eps,\ldots, \eps) \subset 
     V(\Go, f_1,\ldots, f_m, \eps_1,\ldots, \eps_m).
 \]
 Let \(\So = \supp(\Go)\) denote the support of \(\Go\). Let \(B,\eps > 0\) be such that \(|f_i| \leq B\) and \(\eps < \eps_i/2\) for \(i\in\set{1,\ldots, m}\). Since \(\Th\) is Polish, all probability measures in \(\mPTh\) are tight by Prokhorov's theorem. Therefore, there exists a compact set \(K\subset \Th\) such that
    \[
        \Go(\Th\setminus K) < \frac{1}{3B}\eps.
    \]
    Since \(\Th\) is convex by assumption, we may assume without loss that \(K\) is a non-degenerate interval. Let \(G(\cdot) = (1-\pist) \Go|_K(\cdot) +\pist \mathcal{U}_K(\cdot)\), for some \(\pist \in (0, \eps / 3 B]\). Then, \(\supp(G) = K\) and 
    \begin{align*}
        \left|\int_{\Th} f_i(\th) d(G-\Go)(\th)\right| &\leq \left|\int_{K} f_i(\th) d(G-\Go)(\th)\right| + \left|\int_{\Th\setminus K} f_i(\th) d\Go(\th)\right|,\\
        &< \pist \left|\int_{K} f_i(\th) d\Go(\th)\right| + \frac{\pist}{|K|} \left|\int_{K} f_i(\th) d\th\right| + \frac{\eps}{3B}\sup_{\th\in \Th}|f_i(\th)|, \\
        &< \frac{\eps}{3B} \left|\int_{K} f_i(\th) d\Go(\th)\right| + \frac{\eps}{3B}\sup_{\th\in \Th}|f_i(\th)| + \frac{\eps}{3},\\
        &< \frac{\eps}{3B}\sup_{x\in \Th}\, |f_i(\th)| + \frac{2}{3}\eps, \\
        &< \eps,\\
        &< \eps_i,
    \end{align*}
from where it follows that \(G\in V(\Go,f_1,\ldots,f_m,\eps_1,\ldots,\eps_m)\). Now, let \(f'_i = f_i / B\) and let \(\eps' > 0\) be such that \(\eps' < \eps_i/2B\) for \(i\in\set{1,\ldots, m}\). Then, the above shows that if
    \[
        G' \in V(G, f'_1,\ldots, f'_m, \eps',\ldots,\eps'),
    \]
    then,
    \begin{align*}
        \left|\int_{\Th} f'_i(x) d(G'-\Go)(x)\right| &\leq B\left|\int_{\Th} f_i(x) d(G'-G)(x)\right| + B\left|\int_{\Th} f_i(x) d(G-\Go)(x)\right|, \\
        &< \frac{1}{2}\eps_i + \frac{1}{2}\eps, \\
        &< \eps_i,
    \end{align*}
    for \(i\in\set{1,\ldots, m}\), which proves the lemma.
 \hfill$\square$

\subsection{Proof of Lemma~\ref{lem:regSBADenseWasserstein}}\label{AppendixA4}
    
 By Theorem~6.19 in \cite{villani2008optimal}, the set of atomic measures is dense on \((\mPpTh, \Wp)\). Let
    \[
        \bGo(\cdot) = \sum_{i=1}^n \pi_i \delta_{\th_i}(\cdot),
    \]
    be an atomic measure such that \(\Wp(\bGo, \Go) < \eps/2\). We assume without loss that \(\pi_1,\ldots,\pi_n > 0\). If \(n=1\), then it suffices to define
    \[
        \Th_1 = \left(\th_1 - \frac{1}{2^p}\eps^p, \th_1 + \frac{1}{2^p}\eps^p\right)
    \]
    and 
    \[
        G(\cdot) = \mathcal{U}_{\Th_1}(\cdot).
    \]
    Remark that \(\supp(G) = \topcl(\Th_1)\) and is thus a compact interval. By possibly shrinking \(\eps\) we may assume that \(\supp(G) \subsetneq \Th\). Following the proof of Theorem~6.19  in \cite{villani2008optimal}, let \(\th_0\in \Th\setminus\supp(G)\) and define the map \(T:\Th\to\Th\) as
    \[
        T(\Th_1) = \th_1\quad\mbox{and}\quad T(\Th\setminus\Th_1) = \set{\th_0}.
    \]
    Then,
    \[
        \int_{\Th} |\th - T(\th)|^p\, dG(\th) \leq \frac{1}{2^p}\eps^p.
    \]
    Since the pushforward measure of $G$ through $T$ is precisely \(\bGo\), i.e.,   
    \(T_{\sharp} G = \bGo\), we conclude that \(\Wp(\bGo, G) < \eps/2\) whence \(\Wp(\Go,G) < \eps\), as we wanted to show. If \(n > 1\), then we may assume that
    \[
        \th_1 < \ldots < \th_n.
    \]
    This allow us to define the maximum separation
    \[
        \Delta = \max\set{|\th_{i'} -\th_i|:\,\, i',i\in\set{1,\ldots, n},\, i'\neq i},
    \]
    and the intervals
    \[
        \Th_i = \begin{cases}
            \left[\th_1, \th_1 + \frac{1}{2}(\th_2 - \th_1)\right], & i = 1, \\
            \left(\th_i - \frac{1}{2}(\th_i - \th_{i-1}), \th_l + \frac{1}{2}(\th_{i+1} - \th_i)\right], & i\in\set{2, n-1},\\
            \left(\th_n - \frac{1}{2}(\th_n - \th_{n-1}), \th_n\right], & i = n.
        \end{cases}
    \]
    It is apparent that the intervals \(\set{\Th_i:\,\, i\in\set{1,\ldots, n}}\) are disjoint and
    \[
        S:= \topcl\left(\bigcup_{i=1}^n \Th_i\right) = [\th_1, \th_n].
    \]
    Now define
    \[
        G(\cdot) = \sum_{i=1}^n \pi_i ((1-\pist) \delta_{\th_i}(\cdot) + \pist \mathcal{U}_{\Th_i}(\cdot),
    \]
    for \(\pist \in (0 ,1)\) such that
    \[
        \pist < \frac{\Delta^{-p}}{2^p} \eps^p.
    \]
    Then, \(\supp(G) = S\) and is thus compact. If \(S\subsetneq \Th\), then we let \(\th_0\in \Th\setminus S\) and we define the map \(T:\Th\to\Th\) as
    \[
        i\in\set{1,\ldots, n}:\,\, T(\Th_i) = \th_i\quad\mbox{and}\quad T(\Th\setminus\supp(G)) = \set{\th_0}.
    \]
    In this case,
    \begin{align*}
        \int_{\Th} |\th - T(\th)|^p\, dG(\th) &= \sum_{i=1}^n \int_{\Th_i} |\th - \th_i|^p\, dG(\th), \\
        &= \sum_{i=1}^n \pist \pi_i |\th - \th_i|^p, \\
        &= \pist \Delta^p, \\
        &< \frac{1}{2^p}\eps^p.
    \end{align*}
  As before, we have that \(T_{\sharp} G = \bGo\). Therefore, we conclude that \(\Wp(\bGo, G) < \eps/2\) whence \(\Wp(\Go,G) < \eps\). If \(S = \Th\), then it is apparent that \(\Th\setminus\supp(G) = \emptyset\) and thus the same map yields the desired claim, which proves the lemma.
 \hfill$\square$

\setcounter{equation}{0}
\renewcommand{\theequation}{C.\arabic{equation}}
\section{Proofs for the Results of Section 3}\label{AppendixB}

\subsection{Proof of Theorem~\ref{theorem1}}\label{AppendixB1}

The proof is based on the induction principle on the cardinality of the support of $G$. First, suppose that $G$ is a degenerated probability measure, giving unit mass to the set $\{\theta^*_1\}$, that is $G(\cdot)=\delta_{\theta_1}(\cdot)$. By Definition~\ref{def1}, it follows that
\begin{eqnarray}\nonumber
\mu_{1,1} &=& b_{G}(\Theta) = \int_{\Theta} \theta dG(\theta) = \theta^*_1. 
\end{eqnarray}
In a similar manner,
\begin{eqnarray}\nonumber
\mu_{2,1} &=& b_G(a, \mu_{1,1}],\\
\nonumber &=& \frac{\int_{(a,\mu_{1,1}]}\theta dG(\theta) }{G(a,\mu_{1,1}]},\\
\nonumber &=& \frac{\int_{(a,\theta^*_1]}\theta dG(\theta) }{G(a,\theta^*_1]},\\
\nonumber &=& \theta^*_1, 
\end{eqnarray}
and, by Definition~\ref{def1}, $m_{2,3} = \theta^*_1$. A repeated application of Definition~\ref{def1} and Definition~\ref{def2}, shows that
$\mu_{n,2\cdot l-1}=\theta^*_1$, for every $n\ge 1$ and $l \in \{1,2,\ldots,2^{n-1}\}$. It follows that
\begin{eqnarray}\nonumber
G^{(n)}(\cdot) =\sum_{l=1}^{2^n}  G(\Theta_j)  \delta_{\theta^*_1}(\cdot)=\delta_{\theta^*_1}(\cdot)=G(\cdot),
\end{eqnarray}
for every $n\ge 1$.

Now assume that if $G$ has support on a set of $k-1$ distinct elements $\{\theta^*_1,\ldots,\theta^*_{k-1}\} \in \Theta^{k-1}$, then $G^{(n)}(\cdot)=G(\cdot)$, for every $n\ge k-1$. Suppose now that
$G$ is supported on a set of $k$ distinct elements $\{\theta^*_1,\ldots,\theta^*_k\} \in \Theta^{k}$. Consider the 2nd SBA level decomposition of $G$,
\begin{eqnarray}\nonumber
G(\cdot) = G\left((a,\mu_{1,1}]\right) G\left(\cdot\mid (a,\mu_{1,1}]\right) + G\left( (\mu_{1,1},b)\right) G\left(\cdot\mid (\mu_{1,1},b)\right),
\end{eqnarray}
where, as before, $G\left(\cdot\mid (a,\mu_{1,1}]\right)$ and $G\left(\cdot\mid (\mu_{1,1},b)\right)$ is the restriction of $G$ to
the set $(a,\mu_{1,1}]$ and $(\mu_{1,1}, b)$, respectively, and $\mu_{1,1}=\int_{\Theta} \theta G(d\theta)$. Now, it is straightforward to see that there
can be at most $k-1$ elements in the support of $G$ greater than or equal to $\mu_{1,1}$. Thus,  $G\left(\cdot\mid (a,\mu_{1,1}]\right)$ and $G\left(\cdot\mid (\mu_{1,1},b)\right)$
have support on a set of at most $k-1$ distinct elements. It follows, by the induction assumption, that
\begin{eqnarray}\nonumber
G\left(\cdot\mid (a,\mu_{1,1}]\right)=G^{(n)}\left(\cdot\mid (a,\mu_{1,1}]\right),
\end{eqnarray}
and
\begin{eqnarray}\nonumber
G\left(\cdot\mid (\mu_{1,1},b)\right)=G^{(n)}\left(\cdot\mid (\mu_{1,1},b)\right),
\end{eqnarray}
for every $n \ge k-1$. It follows that, for every $n \ge k$,
\begin{eqnarray}\nonumber
G(\cdot) &=& G\left((a,\mu_{1,1}]\right) G\left(\cdot\mid (a,\mu_{1,1}]\right) + G\left( (\mu_{1,1},b)\right) G\left(\cdot\mid (\mu_{1,1},b)\right),\\\nonumber
&=& G\left((a,\mu_{1,1}]\right) G^{(n)}\left(\cdot\mid (a,\mu_{1,1}]\right) + G\left( (\mu_{1,1},b)\right) G^{(n)}\left(\cdot\mid (\mu_{1,1},b)\right),\\\nonumber
&=& G^{(n)}(\cdot),
\end{eqnarray}
which completes the proof of the theorem.
\hfill$\square$

\subsection{Proof of Theorem~\ref{thm:SBAApproximatingMeasureDenseWeakStar}}\label{AppendixB2}

To prove Theorem~\ref{thm:SBAApproximatingMeasureDenseWeakStar} we need the following auxiliary lemma.

\begin{lemma}\label{lem:regSBAApproximatingMeasureDenseWeakStar}
Let \(G\in \mPbTh\). For any \(f_1,\ldots, f_m \in \CobTh\) and \(\eps_1,\ldots,\eps_m > 0\) there exists \(n\in \N\) such that for the measure
    \[
        \Gbn(\cdot) := \sum_{l=1}^{2^{n}} G(\Th_{n, l}) \delta_{\mu_{n+1, 2\cdot l-1}}(\cdot),
    \]
we have that
    \[
        \Gbn \in V(G, f_1,\ldots, f_m, \eps_1,\ldots, \eps_m).
    \]
\end{lemma}

\noindent {\sc Proof:}  Let \(S = \supp(G)\) and note that, by assumption, it is a compact interval. Let \(\eps > 0\) be such that \(\eps < \eps_i\) for \(i\in\set{1,\ldots, m}\). Since \(S\) is a compact interval, the functions \(f_1,\ldots, f_m\) are uniformly continuous over \(S\). Let \(\delta > 0\) be such  that
    \[
        \th,\th'\in S,\, i\in \set{1,\ldots, n}:\,\, |\th - \th'| < \delta\quad\Rightarrow\quad |f_i(\th) - f_i(\th')| < \eps.
    \]
    By Lemma~\ref{lem:regSBAPartitionSupport} \(G\) has a regular SBA. By Lemma~\ref{lem:regSBAIntervalReduction} we can choose \(n_0\in \N\) sufficiently large so that \(n > n_0\) implies that
    \[
        l\in\set{1,\ldots, 2^n}:\,\, |\tTh_{n, l}| < \frac{1}{2}\delta.
    \]
    Fix \(n > n_0\) and let \(\Gbn\) be as in the statement. Remark that it can be written equivalently as
    \[
        \Gbn(\cdot)  = \sum_{l=1}^{2^{n}} G(\tTh_{n, l}) \delta_{\mu_{n+1, 2\cdot l-1}}(\cdot),
    \]
    where \(\mu_{n+1,2\cdot l-1} \in \tTh_{n,l}\) for \(l\in\set{1,\ldots, 2^n}\). Since \(\supp(\Gbn) \subset S\) we have that
    \begin{align*}
        \int_{\Th} f_i(\th) d(\Gbn-G)(\th) &= \int_{S} f_i(\th) d(\Gbn-G)(\th), \\
        &= \sum_{l=1}^{2^{n}} \int_{\tTh_{n,l}} f_i(\th) d(\Gbn - G)(\th), \\
        &= \sum_{l=1}^{2^{n}}\left(f(\mu_{n+1, 2\cdot l-1})G(\tTh_{n,l}) - \int_{\tTh_{n,l}} f_i(\th) dG(\th)\right),\\
        &= -\sum_{l=1}^{2^{n}}\int_{\tTh_{n,l}} (f_i(\th) - f_i(\mu_{n+1, 2\cdot l-1})) dG(\th),
    \end{align*}
    where we used the fact that \(\Gbn(\tTh_{n,l}) = G(\tTh_{n,l})\). Therefore,
    \begin{align*}
        \left|\int_{\Th} f_i(\th) d(\Gbn - G)(\th)\right| &\leq \sum_{l=1}^{2^{n}} \int_{\tTh_{n,l}} |f_i(\th) - f_{i}(\mu_{n+1,2\cdot l-1})| dG(\th),\\
        &< \eps \sum_{l=1}^{2^{n}}G(\tTh_{n,l}),\\
        &< \eps_i,
    \end{align*}
    for every \(i\in\set{1,\ldots, m}\) proving the claim.
\hfill$\square$

We now provide the proof of Theorem~\ref{thm:SBAApproximatingMeasureDenseWeakStar}. From Lemma~\ref{lem:regSBAApproximatingMeasureDenseWeakStar}, we deduce that if a probability measure is supported on a compact interval, then the sequence of atomic measures induced by its level \(n\) SBA converges to itself as \(n\to\infty\) in the weak topology. By applying Lemma~\ref{lem:regSBADenseWeakStar} we conclude that the set of terms of all such sequences is weakly dense.
\hfill$\square$

\subsection{Proof of Theorem~\ref{thm:SBAApproximatingMeasureDenseWasserstein}}\label{AppendixB3}

To prove the theorem, it suffices to assume that the neighborhoods are balls of radius \(\eps > 0\) centered at \(\Go\). We need the following auxiliary lemma.

\begin{lemma}\label{lem:regSBAApproximatingMeasureDenseWasserstein}
    Let \(p\in[1,\infty)\) and let \(G\in \mPbTh\). For any \(\eps > 0\) there exists \(n\in \N\) such that for the measure defined as
    \[
        \Gbn(\cdot) := \sum_{l=1}^{2^{n}} G(\tTh_{n, l}) \delta_{\mu_{n+1, 2l-1}}(\cdot),
    \]
    we have that
    \[
        \Wp(\Gbn, G) < \eps.
    \]
\end{lemma}
\noindent {\sc Proof:} Let \(S = \supp(G)\) and note that, by assumption, it is a compact interval. Therefore, the function \(f(\th) = \th\) is uniformly continuous on \(S\). Let \(\delta > 0\) be such that
    \[
        \th',\th\in S:\,\, |\th' - \th| < \delta \quad\Rightarrow\quad |f(\th') - f(\th)| < \eps.
    \]
    From Lemma~\ref{lem:regSBAIntervalReduction} we may choose \(n\in \N\) such that
    \[
        l\in\set{1,\ldots, 2^n}:\,\, |\tTh_{n,l}| < \frac{1}{2}\delta. 
    \]
    Consider the coupling
    \[
        \gamma = \sum_{l=1}^{2^n} G|_{\tTh_{n,l}}\otimes \delta_{\mu_{n+1,2\cdot l-1}},
    \]
    such that
    \[
        \iint_{\Th\times\Th} f(\th',\th)\, d\gamma(\th',\th) = \sum_{l=1}^{2^n} \int_{\tTh_{n,l}} f(\th',\mu_{n+1,2\cdot l-1}) dG(\th'),
    \]
    for any \(f\in \Cob(\Th\times\Th)\). Then,
    \begin{align*}
        \iint_{\Th\times\Th} |\th' - \th|^p\, d\gamma(\th',\th) &= \sum_{l=1}^{2^n} \int_{\tTh_{n,l}} |\th' - \mu_{n+1,2\cdot l-1}|^p\, dG(\th'),\\
        &< \eps^p\sum_{l=1}^{2^n} G(\tTh_{n,l}),\\
        &= \eps^p,
    \end{align*}
    where we used the fact that \(\mu_{n+1,2l-1} \in \tTh_{n,l}\). From this inequality the lemma follows.
\hfill$\square$

We now provide the proof of Theorem~\ref{thm:SBAApproximatingMeasureDenseWasserstein}. From Lemma~\ref{lem:regSBAApproximatingMeasureDenseWasserstein} we deduce that if a probability measure is supported on a compact interval, then it belongs to \(\mPpTh\) and the sequence of atomic measures induced by its level \(n\) SBA converges to itself as \(n\to\infty\) in the Wasserstein distance of order \(p\). By applying Lemma~\ref{lem:regSBADenseWasserstein} we conclude that the set of terms of all such sequences is dense in \(\mPpTh\).
\hfill$\square$

\subsection{Proof of Lemma~\ref{lem:DSBAmeasurability}}\label{AppendixB5}

\def\RBC{\mathcal{RBC}}
\def\RBCn{\RBC_n}
\def\BC{\mathcal{BC}}
\def\BCn{\BC_n}
\def\mT{\mathcal{T}}
\def\mTn{\mT_n}
\def\bC{\mathcal{C}}
\def\bCn{\bC_n}

The space \(\bCn\) has a natural metric and thus a metric topology. Let \(\RBCn \subset \bCn\) be the space of level \(n\) barycenters of measures with a {\em regular} level \(n\) SBA endowed with the subspace topology and its Borel \(\sigma\)-algebra; we clearly have that \(\RBCn\subset \BCn\). Under the hypotheses of the lemma, the random barycenters \(\set{\mu_{j,l}}_{j=1,l=0}^{n+1, 2^{j}-1}\) are a well defined random variable taking values in \(\RBCn\). To prove the lemma for \(\mTn:\RBCn \to \mPTh\) it suffices to prove that it is continuous. Similarly, to prove the lemma for \(\mTn:\RBCn \to \mPpTh\) for \(p\in [1,\infty)\), it suffices to show that it is measurable. To prove measurability, it suffices to show that it is measurable with respect to the \(\sigma\)-algebra induced by the weak topology on \(\mPpTh\). To see why this is the case, observe that, as \(\Th\) is Polish and locally compact, but not necessarily compact, for every \(\Go\in \mPpTh\) the map \(G\mapsto W_p(G,\Go)\) is lower semicontinuous with respect to the weak topology on \(\mPpTh\) \citep[][Proposition 7.4]{Santambrogio2015}. Therefore, the closed unit ball
\[
    \set{G\in \mPTh:\,\, W_p(G,\Go) \leq \eps},
\]
is weakly closed, and thus measurable with respect to the \(\sigma\)-algebra generated by the weak topology on \(\mPpTh\). In particular, the \(\sigma\)-algebra generated by the topology induced by the Wasserstein distance of order \(p\) is contained in it. Thus, the measurability follows in both cases from the following lemma.

\begin{lemma}\label{lem:DSBAmeasurability:continuityOfTn}
    The following assertions are true. 
    \begin{enumerate}
        \item[(i)]{The map \(\mTn:\RBCn \to \mPTh\) is continuous.
        }
        \item[(ii)]{Let \(p\in [1,\infty)\) and let \(\mPpTh\) be endowed with the weak topology. Then the map \(\mTn:\RBCn\to \mPpTh\) is continuous.
        }
    \end{enumerate}
\end{lemma}

\noindent {\sc Proof:} The proof of both assertions follows the same argument. Let \(U\subset \mPTh\) be open, and let \(\set{\mu_{0,j,l}}_{j=1,l=0}^{n+1, 2^{j}-1} \in \mTn^{-1}(U)\). Define \(\Go = \mTn(\set{\mu_{0,j,l}}_{j=1,l=0}^{n+1, 2^{j}-1})\). Then \(\Go \in U\) and, as \(U\) is open, there exist \(f_1\in \CobTh\) or \(f_1(\th) = |\th|^p\), \(f_2,\ldots,f_{m} \in \CobTh\), and \(\eps_1,\ldots,\eps_m > 0\) such that
\[
    V(\Go, f_1,\ldots, f_m, \eps_1,\ldots, \eps_m) \subset U. 
\]
Let \(\eps > 0\) be such that \(\eps < \min\set{\eps_i:\, i\in\set{1,\ldots,m}}\). From now on, let \(\set{\mu_{j,l}}_{j=1,l=0}^{n+1, 2^{j}-1} \in \RBCn\) and denote \(G = \mTn(\set{\mu_{j,l}}_{j=1,l=0}^{n+1, 2^{j}-1})\). Observe that 
\begin{align*}
    \int_{\Th} f_i(\th) d(G -\Go)(\th) &= \sum_{l=0}^{2^n - 1} (G(\Th_{n,l}) - \Go(\Th_{n,l})) f_i(\mu_{n+1, 2j-1}) \\
    &\quad +\: \sum_{l=0}^{2^n - 1} \Go(\Th_{n,l}) (f_i(\mu_{n+1, 2l-1}) - f_i(\mu_{0,n+1, 2l-1})),
\end{align*}
for \(i\in\set{1,\ldots, m}\). First, by continuity we can choose \(B > 0\) and \(\delta_f > 0\) such that
\begin{multline*}
    |\mu_{n+1, 2l-1} - \mu_{0,n+1, 2l-1}| < \delta_f\quad\Rightarrow\quad \\
    |f_i(\mu_{n+1, 2l-1}) - f_i(\mu_{0,n+1, 2l-1})| < B,\,\, |f_i(\mu_{0,n+1, 2l-1})| < B,
\end{multline*}
for \(l\in\set{0,\ldots, 2^{n}-1}\) and \(i\in\set{1,\ldots,m}\). Second, since \(f_1,\ldots,f_m \in \CoTh\), we can choose \(\delta_\mu \in (0, \delta_f)\) such that
\[
    |\mu_{n+1, 2l-1} - \mu_{0,n+1, 2l-1}| < \delta_\mu\quad\Rightarrow\quad |f_i(\mu_{n+1, 2l-1}) - f_i(\mu_{0,n+1, 2l-1})| < \frac{1}{2}\eps.
\]
for \(l\in\set{0,\ldots, 2^{n}-1}\) and \(i\in\set{1,\ldots,m}\).

Since the barycenters on \(\RBCn\) correspond to the barycenters of level \(n\) of measures with a regular level \(n\) SBA, we deduce from Proposition~\ref{property1} that the probabilities \(\set{G(\Th_{n,l}):\, l\in\set{0,\ldots, 2^n-1}}\) are continuous functions, in fact rational functions, of the barycenters \(\set{\mu_{j,l}}_{j=1,l=0}^{n+1, 2^{j}-1}\). Therefore, we can find \(\delta_\pi \in (0, \delta_f)\) such that
\begin{multline*}
    j\in\set{1,\ldots, n+1},\, l\in\set{0,\ldots, 2^{j}-1}:\,\, |\mu_{j, 2l} - \mu_{0,j, 2l}| < \delta_\mu\quad\Rightarrow\quad \\
    l\in\set{0,\ldots, 2^n - 1}:\,\, |G(\Th_{n,l}) - \Go(\Th_{n,l})| < \frac{1}{4B}\eps.
\end{multline*}
Consequently, by choosing \(\delta > 0\) such that \(\delta < \min\set{\delta_f,\delta_{\mu}}\) we conclude that
\begin{multline*}
    j\in\set{1,\ldots, n+1},\, l\in\set{0,\ldots, 2^{j}-1}:\,\, |\mu_{j, 2l} - \mu_{0,j, 2l}| < \delta\quad\Rightarrow\quad \\
    i\in\set{1,\ldots,m}:\,\, \left|\int_{\Th} f_i(\th) d(G -\Go)(\th)\right| < \eps_i,
\end{multline*}
or, equivalently,
\begin{multline*}
    \set{j\in\set{1,\ldots, n+1},\, l\in\set{0,\ldots, 2^{j}-1}:\,\, |\mu_{j, 2l} - \mu_{0,j, 2l}| < \delta} \subset \\ \mTn^{-1}(V(\Go, f_1,\ldots, f_m, \eps_1,\ldots, \eps_m)),
\end{multline*}
from where the continuity of \(\mTn\) follows.
\hfill$\square$

\subsection{Proof of Theorem~\ref{thm:dsbaFullSupport}}\label{AppendixB6}

We first prove the following auxiliary lemma.
\begin{lemma}\label{lem:boundingCoupling} Let \(\mX = \set{1,\ldots, n}\) and let \(\pi,\mu\) be two probability measures on \((\mX,2^{\mX})\). We write \(\mu(i) := \mu(\set{i})\). Assume that \(\pi\neq \mu\) and let
    \[
        C := (\pi - \mu)(\set{i\in\mX:\, \pi(i) > \mu(i)})^{-1}.
    \]
    Then,
    \[
        \Pi(i,j) = \min\set{\pi(i), \mu(j)}\delta(i-j) + C(\pi(i)-\mu(i))_{+}(\pi(j) - \mu(j))_{-},
    \]
    is a coupling for \((\pi,\mu)\). Furthermore
    \[
        \max\set{\Pi(i,j):\,\, i,j\in\mX,\, i\neq j} < \max\set{|\pi(i) - \mu(i)|:\,\, i\in \mX}.
    \]
\end{lemma}

\noindent {\sc Proof:}  Let \(I = \set{i\in\mX:\,\, \pi(i) > \mu(i)}\). First, note that
    \begin{align*}
        \sum_{j=1}^n \Pi(i,j) &= \min\set{\pi(i),\mu(i)} + C(\pi(i)-\mu(i))_{+} \sum_{j=1}^n (\pi(j) - \mu(j))_{-}, \\
            &= \min\set{\pi(i),\mu(i)} + C(\pi(i)-\mu(i))_{+} (\mu(I^c) - \pi(I^c)), \\
            &= \min\set{\pi(i),\mu(i)} + C(\pi(i)-\mu(i))_{+} (\pi(I) - \mu(I)).
    \end{align*}
    If \(\pi(i) \leq \mu(i)\), then
    \[
        \sum_{j=1}^n \Pi(i,j) = \pi(i),
    \]
    whereas if \(\pi(i) > \mu(i)\), then
    \[
        \sum_{j=1}^n \Pi(i,j) = \mu(i) + C(\pi(i)-\mu(i)) (\pi(I) - \mu(I)) = \mu(i) + (\pi(i)-\mu(i)) = \pi(i).
    \]
    On the other hand
    \begin{align*}
        \sum_{i=1}^n \Pi(i,j) &= \min\set{\pi(j),\mu(j)} + C  (\pi(j) - \mu(j))_{-}\sum_{i=1}^n(\pi(i)-\mu(i))_{+}, \\
            &= \min\set{\pi(j),\mu(j)} + C  (\pi(j) - \mu(j))_{-}(\pi(I)-\mu(I)).
    \end{align*}
    If \(\pi(j) \leq \mu(j)\), then
    \[
        \sum_{i=1}^n \Pi(i,j) = \pi(j) + C (\mu(j) - \pi(j)) (\pi(I)-\mu(I)) = \mu(j),
    \]
    whereas if \(\pi(j) > \mu(j)\), then
    \[
        \sum_{i=1}^n \Pi(i,j) = \pi(j).
    \]
    Therefore, \(\Pi\) is a coupling as stated. Furthermore, notice that 
    \[
        C = \sum_{i=1}^n (\pi(i) - \mu(i))_{+}.
    \]
    Thus, for \(i \neq j\) we have that
    \begin{align*}
        \Pi(i,j) &= C(\pi(i)-\mu(i))_{+}(\pi(j) - \mu(j))_{-}, \\
        &\leq (\pi(j) - \mu(j))_{-}, \\
        &\leq \max\set{|\pi(j) - \mu(j)|:\,\, j\in\mX},
    \end{align*}
    from where the claim follows.

\hfill$\square$

We now provide and prove an approximation property for the process \(\mbox{DSBA}(n, \mathcal{H}_n)\) for a fixed \(n\).

\begin{lemma}\label{lem:dsbaSamplingApproximationProperty}
    Let \(G \mid n,  \mathcal{H}_n \sim \mbox{DSBA}(n, \mathcal{H}_n)\) and suppose that the hypotheses of Theorem~\ref{thm:dsbaFullSupport} hold. For every \(\Go\in\mPbTh\) the following assertions are true.
    \begin{enumerate}
        \item[(i)] For every \(f_1,\ldots, f_m \in \Cob(\Th)\) and \(\eps_1,\ldots,\eps_m > 0\) we have that
        \[
            \Prob(\set{\omega\in\Omega:\,\, G(\omega)\in V(\Gbon, f_1,\ldots, f_m, \eps_1, \ldots, \eps_m)}) > 0.
        \]
        
        \item[(ii)] For every \(p\in [1,\infty)\) and \(\eps > 0\) we have that
        \[
            \Prob(\set{\omega\in\Omega:\,\, \Wp(G(\omega), \Gbon) < \eps}) > 0.
        \]
        
    \end{enumerate}
\end{lemma}

\noindent {\sc Proof:} Let \(\set{\mu_{0,n+1,l}:\,\, l\in\set{0,\ldots, 2^{n+1}+1}}\) be the barycenters of level \(n+1\) of \(\Go\) and let \(\set{\Th_{0, n,l}:\,\, l\in\set{1,\ldots, 2^{n}}}\) be the corresponding intervals used to define \(\Gbon\). Let \(\So\) be the support of \(\Go\) which, by assumption, is compact. Denote as \(\pist_1,\ldots, \pist_{2^{n}}\) the probabilities obtained as a function of the barycenters
    \[
        \bst_1 < \bst_2 < \ldots < \bst_{2^{n+1} - 2} < \bst_{2^{n+1} - 1},
    \]
    Since the choice of probabilities is unique, the choice
    \[
        l\in\set{1,\ldots, 2^{n+1}-1}:\,\, \bst_{l} = \mu_{0,n+1, l}
    \]
    yields
    \[
        l\in\set{1,\ldots, 2^{n}}:\,\, \pist_{l} = \Go(\Th_{0,n,l}).
    \]
    Furthermore, the hypotheses of Theorem~\ref{thm:dsbaFullSupport} imply that for every choice \(b_1,\ldots, b_{2^{n}-1} \in \Th\) such that \(b_1 < \ldots < b_{2^{n+1}-1}\) and \(\Delta_1,\ldots, \Delta_{2^{n+1} - 1} > 0\) we have that
    \[
        \Prob(\set{\omega\in \Omega:\,\, |\mu_{n+1, l}(\omega) - b_l| < \Delta_l,\, l\in\set{1,\ldots, 2^{n+1}-1}}) > 0.
    \]
    We now prove the statements.
    
    \begin{enumerate}
        \item[(i)] Let \(\eps > 0\) be such that \(\eps < \eps_i\) for \(i\in\set{1,\ldots, m}\) and let \(B > 0\) be such that \(|f_i| < B\) for \(i\in\set{1,\ldots, m}\). Using the fact that \(\mu_{0,n+1, 2l-1} \in \topint(\Th_{0,n, l})\) we can find \(\Delta_b > 0\) such that
        \[
            l\in\set{1,\ldots, 2^{n-1}}:\,\, |b - \mu_{0,n+1, 2l-1}| < \Delta_b\,\,\Rightarrow\,\, b\in \Th_{0,n,l}.
        \]
        Let \(\delta_f > 0\) be, such that
        \begin{multline*}
            l\in\set{1,\ldots, 2^{n-1}},\, i\in\set{1,\ldots, m}:\,\, |b - \mu_{0,n+1, 2l-1}| < \delta_f \\
            \,\,\Rightarrow\,\, |f_i(b) - f_i(\mu_{0,n+1, 2l-1})| < \frac{1}{2}\eps. 
        \end{multline*}
        Furthermore, there exists \(\delta_{\pi} > 0\), such that
        \[
            \max_{l\in\set{1,\ldots, 2^{n+1}-1}}\,\, |\bst_{l} - \mu_{0, n+1, l}| < \delta_\pi\,\,\Rightarrow\,\, \max_{l\in\set{1,\ldots, 2^{n}}}\,\,|\pist_{l} - \Go(\Th_{0, n,l})| < \frac{1}{2B}\eps.
        \]
        Finally, let \(\delta > 0\) be such that \(\delta < \min\set{\Delta_b, \delta_f,\delta_{\pi}}\) and define
        \[
            \Omega^{\star} := \set{\omega\in\Omega:\,\, |\mu_{n+1,l}(\omega) - \mu_{0, n+1,l}| < \delta,\,\, l\in\set{1,\ldots, 2^{n+1}-1}},
        \]
        which, by hypothesis, has positive probability. Then, for \(\omega \in \Omega\) we have that, for any \(i\in\set{1,\ldots, m}\), 
        \begin{align*}
            \int_{\Th} f_i(\th) d(G(\omega) - \Gbon)(\th) &= \sum_{l=1}^{2^{n}} \int_{\Th_{0, n,l}} f_i(\th) d(G(\omega) - \Gbon)(\th), \\
            &= \sum_{l=1}^{2^{n}} (w_l(\omega) f_i(\mu_{n+1,2l-1}(\omega)) - \Gbon(\Th_{0,n,l})f_i(\mu_{0,n+1,2l-1})), \\
            &= \sum_{l=1}^{2^{n}} (w_l(\omega) - \Gbon(\Th_{0,n,l})) f_i(\mu_{n+1,2l-1}(\omega)), \\
            &\quad+\: \sum_{l=1}^{2^{n}} \Gbon(\Th_{0,n,l})(f_i(\mu_{n+1,2l-1}(\omega)) - f_i(\mu_{0,n+1,2l-1})).
        \end{align*}
        The first sum in the right-hand side can be bounded as
        \[
            \left|\sum_{l=1}^{2^{n}} (w_l(\omega) - \Gbon(\Th_{0,n,l})) f_i(\mu_{n+1,2l-1}(\omega))\right| \leq B \sum_{l=1}^{2^{n-1}} |w_l(\omega) - \Gbon(\Th_{0,n,l})| < \frac{1}{2}\eps.
        \]
        The second sum can be bounded as
        \begin{multline*}
            \left|\sum_{l=1}^{2^{n-1}} \Gbon(\Th_{0,n,l})(f_i(\mu_{n+1,2l-1}(\omega)) - f_i(\mu_{0,n,2l-1}))\right| \\
            \leq
            \sum_{l=1}^{2^{n-1}} \Gbon(\Th_{0,n,l})|f_i(\mu_{n+1,2l-1}(\omega)) - f_i(\mu_{0,n,2l-1})|, \\
            \leq \frac{1}{2}\eps \sum_{l=1}^{2^{n-1}} \Gbon(\Th_{0,n,l}) = \frac{1}{2}\eps.
        \end{multline*}
        Therefore, for every \(\omega\in\Omega\) we have that,
        \[
            G(\omega) \in V(\Gbon, f_1,\ldots, f_m, \eps_1,\ldots, \eps_m),
        \]
        proving the claim.

        \item[(ii)] Let \(\So\) be the support of \(\Go\in\mPbTh\) and let \(h(\th) = |\th|^p\). Since \(\So\) is compact, \(h\) is uniformly continuous and we may choose \(\delta_p > 0\), such that
        \[
            \th',\th \in \So:\,\, |\th' - \th| < \delta_p\,\,\Rightarrow\,\, |h(\th') - h(\th)| < \frac{1}{2}\eps.
        \]
        Define \(\Delta_b > 0\) exactly as before, and let \(\delta_\pi > 0\), such that 
        \[
            \max_{l\in\set{1,\ldots, 2^{n+1}-1}}\,\, |\bst_{l} - \mu_{n+1, l}| < \delta_\pi\,\,\Rightarrow\,\, \max_{l\in\set{1,\ldots, 2^n}}\,\,|\pist_{l} - \Go(\Th_{0,n,l})| < \frac{2^{1 - n-p}}{(2^n - 1)|\So|^{p}} \eps^p.
        \]
        Finally, let \(\delta > 0\) be such that \(\delta < \min\set{\Delta_b,\delta_p,\delta_{\pi}}\) and define
        \[
            \Omega^{\star} := \set{\omega\in\Omega:\,\, |\mu_{n+1, l}(\omega) - \mu_{0,n,l}| < \delta,\,\, l\in\set{1,\ldots, 2^{n+1}-1}},
        \]
        which, by hypothesis, has positive probability. Then, for \(\omega \in \Omega\) we define first the auxiliary measure
        \[
            \bGwn(\cdot) := \sum_{l=1}^{2^{n}} \Go(\Th_{0, n, l}) \delta_{\mu_{n+1,2l-1}(\omega)}(\cdot).
        \]
        By using the coupling
        \[
            \gamma(\omega)(\cdot) = \sum_{k,l=1}^{2^{n-1}} \Pi_{k,l}(\omega)(\delta_{\mu_{n+1,2l-1}(\omega)}\otimes \delta_{\mu_{n+1,2l-1}(\omega)})(\cdot)
        \]
        where the \(2^n\times 2^n\) matrix \(\Pi \geq 0\) is as defined in Lemma~\ref{lem:boundingCoupling}, we deduce that
        \begin{multline*}
            \iint |\th' - \th|^p\, d\gamma(\omega)(\th',\th) \\
            = \sum_{l',l=1}^{2^n} \Pi_{l',l}(\omega) |\mu_{n+1, 2l'-1}(\omega) - \mu_{n+1, 2l-1}(\omega)|^p = \sum_{l'\neq l} \Pi_{l',l}(\omega) |\mu_{n+1, 2l'- 1} - \mu_{n+1, 2l-1}|^p,
            \\< |\So|^p 2^{n-1} (2^n - 1) \max\set{|w_{l}(\omega) - \Go(\Th_{0,n,l})|:\, l\in\set{1,\ldots, 2^{n-1}}} < \frac{1}{2^p}\eps^p,
        \end{multline*}
        whence
        \[
            \Wp(G(\omega), \bGwn) < \frac{1}{2}\eps.
        \]
        Now, consider the coupling
        \[
            \gamma(\omega)(\cdot) = \sum_{l=1}^{2^n} \Go(\Th_{0,n,l}) (\delta_{\mu_{n+1,2l-1}(\omega)}\otimes \delta_{\mu_{0,n+1,2l-1}})(\cdot),
        \]
        between \(\bGwn\) and \(\Gbon\). Then
        \begin{align*}
            \iint |\th' - \th|^p\, d\gamma(\omega)(\th',\th) &= \sum_{l=1}^{2^{n}} \Go(\Th_{0,n,l}) |\mu_{n+1,2l-1}(\omega) - \mu_{0,n+1,2l-1}|^p,\\
            &< \frac{1}{2^p}\eps^p\sum_{l=1}^{2^{n}} \Go(\Th_{0,n,l}), \\
            &= \frac{1}{2^p}\eps^p.
        \end{align*}
        Therefore,
        \[
            \Wp(\bGwn, \Gbon) < \frac{1}{2}\eps,
        \]
        whence
        \[
            \Wp(G(\omega), \Gbon) < \Wp(G(\omega), \bar{G}(\omega)) + \Wp(\bar{G}(\omega), \Gbon) < \eps,
        \]
        proving the claim. 
        
    \end{enumerate}
\hfill$\square$

To prove the theorem, we need to be able to select \(n\) at random. To achieve this, we need to embed all processes on the same space. For \(n \in\N\) denote as \((\Omega_n,\mathcal{A}_n,\Prob_n)\) the probability space on which \(M^{(n)} : \Omega_n \to \RBCn\) is well defined. First, let
\[
    \Omega_{\infty} := \bigsqcup_{n=1}^{\infty} \Omega_n,
\]
and denote as \(i_n :\Omega_n \to \Omega_{\infty}\) the canonical inclusion. We endow this space with the \(\sigma\)-algebra \(\mathcal{A}_{\infty}\) generated by the canonical inclusions \(\set{\iota_n}_{n=1}^{\infty}\). Let \(\pi_N\) be a probability measure on \((\N, 2^{\N})\). We endow \(\Omega_{\infty}\) with the probability measure
\[
    \Prob(A) = \sum_{n\in \N} \pi_N(\set{n}) \Prob_n(i_n^{-1}(A)).
\]
It is apparent that the random variable \(N : \Omega_{\infty} \to \N\) given by
\[
    N(\omega_{\infty}) = N((n,\omega_n)) = n,
\]
has law
\[
    \Prob(\set{\omega_{\infty}\in\Omega_{\infty}:\,\, N(\omega_{\infty}) = n}) = \pi(\set{n}). 
\]
Furthermore, for any \(A_n\in\mathcal{A}_n\) we have that
\[
    \Prob(\set{(n,\omega_n)\in\Omega_{\infty}:\, \omega_n\in A_n}) = \pi(\set{n}) \Prob_n(A_n).
\]
Using the notation from Lemma~\ref{lem:DSBAmeasurability}, let
\[
    \RBC_{\infty} := \bigsqcup_{n=1}^{\infty} \RBCn.
\]
Let \(\iota_n : \RBCn \to \RBC_{\infty}\) be the canonical inclusion. The natural topology in the above space is the weakest topology for which the family of inclusions \(\set{\iota_n}_{n=1}^{\infty}\) is continuous. We endow \(\RBC_{\infty}\) with this topology and the corresponding Borel \(\sigma\)-algebra. Then the map \(M : \Omega_{\infty}\mapsto \RBC_{\infty}\) given by
\[
    M(n,\omega_n) = M^{(n)}(\omega_n),
\]
is measurable, as for every \(n\in \N\) we have that the map \(M\circ i_n : \Omega_n \to \RBC_{\infty}\) is such that
\[
    (M\circ i_n)(\omega_n) = M^{(n)}(\omega_n),
\]
and thus measurable by Lemma~\ref{lem:DSBAmeasurability}. Note that for any \(A\in \Bor(\RBC_{\infty})\) we have
\begin{align*}
    \Prob(\set{\omega_{\infty}\in\Omega_{\infty}:\,\, M(\omega_{\infty}) \in A}) &= \sum_{n\in\N}\pi(\set{n}) \Prob_n(\set{\omega_{n}\in\Omega_{n}:\,\, (M\circ i_n)(\omega_{n}) \in A})\\
    &= \sum_{n\in\N}\pi(\set{n}) \Prob_n(\set{\omega_{n}\in\Omega_{n}:\,\, M^{(n)}(\omega_{n}) \in \iota_n^{-1}(A)})\\
    &= \sum_{n\in\N}\pi(\set{n}) \Prob_n((M^{(n)})^{-1}(\iota_n^{-1}(A))).
\end{align*}
as desired. Therefore, the random truncated barycentric array \(M\) is a well defined random process on \(\RBC_{\infty}\). From now on, we simply denote as \((\Omega,\mathcal{A}, \Prob)\) the space on which \(M\) is defined. 

Define \(\mT_{\infty}:\RBC_{\infty}\to \mPTh\) as
\[
    \mT_{\infty}(n,\set{\mu_{j,l}}_{j=1,l=0}^{n+1, 2^j - 1}) = \mT_n(\set{\mu_{j,l}}_{j=1,l=0}^{n+1, 2^j - 1}). 
\]
Since \(\mT_{\infty} \circ \iota_n = \mTn\) for every \(n\in\N\) we conclude that \(\mT_\infty\) is measurable. Remark that the same arguments hold if we let \(\mT_{\infty} : \RBC_{\infty} \to \mPpTh\) for \(p\in [1,\infty)\). Thus, the resulting process can be represented as
\[
    G = \mT_{\infty}(M) = \sum_{n=1}^{\infty}\ind\set{N = n} G_n
\]
where \(G_n \sim \mbox{DSBA}(n, \mathcal{H}_n)\).

We can now provide the proof of the main result. It is apparent that the hypotheses of Lemma~\ref{lem:dsbaSamplingApproximationProperty} hold. We prove both statements separately. 
\begin{enumerate}
    \item[(i)] By Lemma~\ref{lem:regSBADenseWeakStar}, there exists \(\bGo\in\mPbTh\) and an \(n_0\in \N\) such that \(\bGbon\) satisfies
    \[
        \bGbon \in V(\Go, f_1,\ldots, f_m, \eps_1, \ldots, \eps_m).
    \]
    Furthermore, we may choose \(f'_1,\ldots, f'_{m'} \in \Cob(\Th)\) and \(\eps' > 0\) such that
    \[
        V(\bGbon, f'_1,\ldots, f'_{m'}, \eps', \ldots, \eps') \subset V(\Go, f_1,\ldots, f_m, \eps_1, \ldots, \eps_m).
    \]
    Note that
    \begin{multline*}
        \set{\omega\in\Omega:\,\, G(\omega)\in V(\bGbon, f'_1,\ldots, f'_{m'}, \eps', \ldots, \eps')} \supseteq \\
        \set{\omega\in\Omega:\,\, G(\omega)\in V(\bGbon, f'_1,\ldots, f'_{m'}, \eps', \ldots, \eps')}\cap\set{\omega\in\Omega:\,\, n(\omega) = n_0}.
    \end{multline*}
    Since the hypotheses imply the hypotheses of Lemma~\ref{lem:dsbaSamplingApproximationProperty} for \(n_0\), we conclude that
    \begin{align*}
        0 &< \Prob(\set{\omega\in\Omega:\,\, n(\omega) = n_0,\, G(\omega)\in V(\bGbon, f'_1,\ldots, f'_{m'}, \eps', \ldots, \eps')})\\
        &< \Prob(\set{\omega\in\Omega:\,\, G(\omega)\in V(\bGbon, f'_1,\ldots, f'_{m'}, \eps', \ldots, \eps')})\\
        &\leq \Prob(\set{\omega\in\Omega:\,\, G(\omega) \in V(\Go, f_1,\ldots, f_m, \eps_1, \ldots, \eps_m)})
    \end{align*}
    proving the statement.
    
    \item[(ii)] Similarly, by Lemma~\ref{lem:regSBADenseWasserstein} there also exists \(\bGo\in\mPbTh\) and an \(n_0\in \N\) such that \(\bGbon\) satisfies
    \[
        \Wp(\bGbon,\Go) < \eps. 
    \]
    A similar argument as before shows that
    \begin{multline*}
        \set{\omega\in\Omega:\,\, \Wp(G(\omega),\bGbon) < \eps} \supseteq \\
        \hspace{1cm} \set{\omega\in\Omega:\,\, \Wp(G(\omega),\bGbon) < \eps}\cap \set{\omega\in\Omega:\,\, n(\omega) =n_0}.
     \end{multline*}
    Since the hypotheses imply the hypotheses of Lemma~\ref{lem:dsbaSamplingApproximationProperty} for \(n_0\), the same arguments as before yield
    \begin{align*}
        0 &< \Prob(\set{\omega\in\Omega:\,\, n(\omega) = n_0,\, \Wp(G(\omega),\bGbon) < \eps}), \\
        &\leq \Prob(\set{\omega\in\Omega:\,\, \Wp(G(\omega),\bGbon) < \eps}, \\
        &\leq \Prob(\set{\omega\in\Omega:\,\, \Wp(G(\omega),\Go) < \eps}),
    \end{align*}
    proving the claim.
    
\end{enumerate}
\hfill$\square$

\setcounter{equation}{0}
\renewcommand{\theequation}{D.\arabic{equation}}
\section{Proofs for the Results of Section 4}\label{AppendixC}

We introduce some preliminary notation. We denote the \(\ell_{\infty}\)-norm in \(\R^m\) as
\[
    x\in \mathbb{R}^m:\,\, \nrmInf{x} = \max\set{|x_1|,\ldots, |x_m|},
\]
and the simplex in \(\R^m\) as \(\Delta^{m-1}\). If \(S\subset \R^m\), then its convex hull is the set
\[
    \cvxhull(S) := \left\{\sum_{i = 1}^k \pi_i x_i:\,\, x_1,\ldots, x_k\in \R^m,\,\, \pi \in \Delta^{k-1}\right\}.
\]
To simplify the notation, we denote \(\Gam = \Th\times\Phi\). Then, every \(\gam\in \Gam\) can be represented as \(\gam = (\th, \phi)\). In this case, 
\[
    |\gam - \gam'| := \max\set{|\th - \th'|,\, |\phi - \phi'|}.
\]
To prove the theorem for \(\mP(\Gam)\) it suffices to prove that for any \(\Go\in\mP(\Gam)\), any \(f_1,\ldots, f_m\in \Cob(\Gam)\), and any \(\eps_1,\ldots, \eps_m > 0\), we have that
\begin{equation}\label{eq:mixing:statementToProve}
    \Prob(\set{\omega\in\Omega:\,\, G(\omega)\in V(\Go,f_1,\ldots,f_m,\eps_1,\ldots, \eps_m)} > 0.
\end{equation}
Similarly, to prove the theorem for \(\mPp(\Gam)\) endowed with the weak topology it suffices to prove that~\eqref{eq:mixing:statementToProve} holds for any \(\Go\in\mP(\Gam)\), any choice \(f_1,\ldots, f_m\in \Cob(\Gam)\), or
\[
    f_i(\gamma) = |\theta|^p\,\,\mbox{or}\,\, f_i(\gamma) = |\phi|^p,
\]
and any \(\eps_1,\ldots, \eps_m > 0\); we emphasize that this {\em does not} yield the result for \(\mPp(\Gam)\) endowed with the topology induced by the Wasserstein distance of order \(p\). Therefore, instead of distinguishing between both cases, we will simply prove~\eqref{eq:mixing:statementToProve} under the assumption that some functions may be polynomials, and being clear in which cases the proof strategy must be adapted according to this fact; in particular, our arguments do not require uniform bounds on \(f_1,\ldots, f_m\). Finally, by choosing \(\eps > 0\) such that \(\eps < \eps_i\) for \(i\in\set{1,\ldots, m}\) we deduce that
\[
    V(\Go,f_1,\ldots,f_m, \eps,\ldots, \eps) \subset V(\Go,f_1,\ldots,f_m,\eps_1,\ldots, \eps_m)
\]
whence we can reduce our arguments to the case \(\eps_1 = \ldots = \eps_m = \eps\). Finally, it will be useful to define the function \(F:\Gam\to \R^m\) as
\[
    F(\gam) = \begin{bmatrix}
        f_1(\gam) \\ \vdots \\ f_m(\gam)
    \end{bmatrix}.
\]
It is apparent that \(F\) is continuous and that
\[
    V(\Go,f_1,\ldots,f_m,\eps,\ldots, \eps) = \left\{G\in\mPp(\Gamma):\,  \LnrmInf{\int_{\Gam} F(\gam) d(G-\Go)(\gam)} < \eps\right\}.
\]

\subsection{Proof of Theorem~\ref{thm:dsbagFullSupport}}\label{AppendixC1}

In this case we perform a reduction argument that allows us to assume without loss that \(\Go\) is compactly supported. 

\begin{lemma}\label{lem:mixing:restrictionToSquare}
    There exist compact intervals \(I_{\Th}\subset \Th\) and \(I_{\Phi}\subset \Phi\), such that there exists \(\bGo \in \mP(\Th\times\Phi)\) such that \(\supp(\bGo) = I_{\Th}\times I_{\Phi}\) and 
    \[
        \bGo \in V(\Go, f_1,\ldots, f_m, \eps,\ldots, \eps).
    \]
\end{lemma}

\noindent {\sc Proof:} 
    Let \(\eps_0 > 0\) be such that \(\eps_0 < \min\set{\eps, 1/2}\). If \(\Go\in \mP(\Gam)\) then \(f_i\in \Cob(\Gam)\) and it is apparent that we can choose a compact \(K_i\subset \Gamma\), such that
    \[
        \int_{\Gam\setminus K} |f(\gam)|\, d\Go(\gam) < \frac{1}{3}\eps_0. 
    \]
    If \(\Go\in\mPp(\Gam)\) and \(f_i\in \Cob(\Gam)\) the same argument holds. If \(f_i(\gam) = |\th|^p\) or \(f_i(\gam) = |\phi|^p\), then from \(f_i \geq 0\) and 
    \[
        \int_{\Gam} f_i(\gam)\, d\Go(\gam) < \infty,
    \]
    we conclude that the same argument holds. Let \(B > 0\) be such that
    \[
        i\in\set{1,\ldots, m}:\,\, \int_{\Gam} |f_i(\gam)|\, d\Go(\gam) \leq B.
    \]
    Since \(\Gam\) is Polish, we can find a compact \(K_0\subset \Gam\) such that
    \[
        \Go(\Gam\setminus K_0) < \frac{1}{6\max\set{1, B}}\eps_0.
    \]
    Therefore, after choosing \(K_0, K_1,\ldots, K_m\) in this manner, we can find compact intervals \(\ITh\subset \Th\) and \(\IPh\subset \Phi\) such that
    \[
        \bigcup_{i=0}^m K_i \subset \ITh\times\IPh. 
    \]
    Since \(f_1,\ldots, f_m\) are bounded on \(\ITh\times\IPh\), we can find \(C > 0\) such that
    \[
        i\in\set{1,\ldots, m}:\,\, |f_i(\gam)| \leq C.
    \]
    Let \(\pi \in (0 ,1)\) be such that
    \[
        1 - \pi < \frac{1}{C|\ITh\times\IPh|},
    \]
    and define
    \[
        \bGo = (1-\pi) \mathcal{U}_{\ITh\times\IPh} + \pi \Go|_{\ITh\times\IPh}.
    \]
    Then,
    \begin{align*}
        \left|\int_{\Gam} f_i(\gam) d(\bGo - \Go)(\gam)\right| &< \int_{\Gam\setminus \ITh\times\IPh} |f_i(\gam)|\, d\Go(\gam) + (1-\pi) \int_{\ITh\times\IPh} |f_i(\gamma)|\, d\gamma, \\
        &\quad +\: \pi\frac{\Go(\Gam\setminus \ITh\times\IPh)}{\Go(\ITh\times\IPh)} \int_{\ITh\times\IPh} |f_i(\gam)| \, d\Go(\gam),\\
        &< \frac{1}{3}\eps_0 + (1-\pi) C |\ITh\times\IPh| + \pi \frac{\eps_0}{3B(1-\eps_0)} B,\\
        &< \frac{1}{3}\eps_0 + (1-\pi) C |\ITh\times\IPh| + \frac{2\eps_0}{6},\\
        &=\eps_0,\\
        & < \eps,
    \end{align*}
    where we used the fact that
    \[
        \frac{\eps_0}{1- \eps_0} < 2\eps_0,
    \]
    for our choice of \(\eps_0\). From this statement the lemma follows.
\hfill$\square$

As a consequence of the lemma, we can assume without loss that \(\supp(\Go) = \ITh\times\IPh\) for compact interval \(\ITh\subset \Th\) and \(\IPh\in\Phi\) compact intervals. In this case, we can choose \(B > 0\), such that
\[
    \gam \in \ITh\times\IPh:\,\, \nrmInf{F(\gam)} \leq B.
\]
Furthermore, \(F\) is uniformly continuous on \(\ITh\times\IPh\) and there exists \(\delta_{F} > 0\), such that
\[
    \gam', \gam \in\ITh\times\IPh:\,\, |\gam' - \gam| < \delta_{F} \,\,\Rightarrow\,\, \nrmInf{F(\gam') - F(\gam)} < \frac{1}{3}\eps.
\]
Note that \(\set{(\bphi-\delta_{F}/2,\bphi+\delta_{F}/2)}_{\bphi\in\IPh}\) is an open cover for \(\IPh\) from which we can extract a finite cover \(\set{(\bphi_k-\delta_{F}/2,\bphi_k+\delta_{F}/2)}_{k=1}^K\). From this finite cover, we can construct a partition \(\set{\Phi_k}_{k=1}^{K}\) of \(\IPh\), such that \(\bphi_k \in \Phi_k\) and \(|\Phi_k| < \delta_{F}\) for \(k\in\set{1,\ldots, K}\).

If we denote as \(G_{0,1}\in \mPTh\) the marginal
\[
    A\in\mathcal{B}(\Th):\,\, G_{0,1}(A) := \Go(A\times\Phi),
\]
then, it is apparent that \(\supp(\Go|_{\Th}) = \ITh\). In particular, it has a regular SBA and there is \(n_0 \in \N\), such that, for \(N_0 = 2^{n_0}\), the level \(n_0+1\) SBA approximation
\[
    G_{0,1}^{(n)}(\cdot) = \sum_{l=1}^{N_0} G_{0,1}(\Th_{n_0,l}) \delta_{\mu_{0,n_0+1,2l-1}}(\cdot), 
\]
satisfies
\[
    l\in\set{1,\ldots, N_0 - 1}:\,\, |\Th_{n_0,l}| < \delta_F.
\]
Note that \(\set{\Th_{n_0,l}\times\Phi_k}_{l=1,k=1}^{N_0,K}\) is a partition of \(\ITh\times\IPh\), such that
\[
    \gam', \gam\in \Th_{n_0,l}\times\Phi_k\,\,\Rightarrow\,\, \nrmInf{F(\gam') - F(\gam)} < \frac{1}{3}\eps.
\]
The same arguments in the proofs of Lemma~\ref{lem:dsbaSamplingApproximationProperty} and Theorem~\ref{thm:dsbaFullSupport} allow us to prove the existence of \(\delta^* > 0\), such that the set
\[
    \Omega_{\th} := \set{\omega\in\Omega:\,\,  m_1(\omega) = n_0,\,\, |\mu_{n_0+1,l}(\omega) - \mu_{0,n_0+1,l}| < \delta^*,\, l\in\set{1,\ldots, 2N_0 - 1}},
\]
has positive measure, and such that for any \(\omega\in \Omega_{\th}\), we have that
\[
    l\in \set{1,\ldots, N_0}:\,\, |w_{n_0,l}^{\th}(\omega) - \Go^1(\Th_{n_0,l})| < \frac{2}{3 B K N_0}\eps\,\,\mbox{and}\,\, \th_{l}(\omega) \in \Th_{n_0,l}.
\]
We may choose \(\delta_{\phi} > 0\), such that 
\[
    \phi\in \IPh,\, k\in\set{1,\ldots, K}:\,\, |\phi - \bphi_k| < \delta_{\phi}\,\,\Rightarrow\,\, \phi \in \Phi_k.
\]
Define the events
\[
    \Omega_{\phi} := \set{\omega\in\Omega:\,\, m_2(\omega) = K,\,\, |\phi_k(\omega) - \bphi_k| < \delta_{\phi},\, k\in\set{1,\ldots, K}},
\]
and
\[
    \Omega_{w^\phi} := \bigcap_{l=1}^{N_0}\Lset{\omega\in\Omega:\,\, m_2(\omega) = K,\,\, \left|w^{\phi}_{l,k} - \frac{\Go(\Th_{n_0,l}\times \Phi_{k})}{G_{0, 1}(\Th_{n_0,l})}\right| < \frac{2\eps}{3BKN_0},\, k\in\set{1,\ldots, K}}.
\]
Then, for \(\omega\in \Omega_{\th}\cap\Omega_{\phi}\cap\Omega_{w^{\phi}}\), we have that for any \(l',l\in\set{1,\ldots, N_0}\), and \(k',k\in\set{1,\ldots, K}\) it holds that
\[
    (\th_{l'}(\omega), \phi_{k'}(\omega)) \in \Th_{n_0,l}\times \Phi_k\,\,\Leftrightarrow\,\, l' = l\,\,\mbox{and}\,\, k' = k,
\]
and that
\begin{align*}
    |w^{\th}_{n_0,l} w^{\phi}_k - \Go(\Th_{n_0,l}\times\Phi_k)| &< |w^{\th}_{n_0,l} - G_{0,1}(\Th_{n_0,l})| + |G_{0,1}(\Th_{n_0,l}) w^{\phi}_k - \Go(\Th_{n_0,l}\times \Phi_k)|,\\
    &< \frac{1}{3 B K N_0}\eps + \frac{1}{3 BKN_0}\eps,\\
    &= \frac{2}{3B K N_0}\eps.
\end{align*}
Therefore, for any \(\omega\in \Omega_{\th}\cap\Omega_{\phi}\cap\Omega_{w^{\phi}}\), we have that
\[
    G(\omega)(\cdot) = \sum_{l=1}^{N_0}\sum_{k=1}^K w_{n_0,l}^{\th}(\omega) w_k^{\phi}(\omega)\delta_{(\th_l(\omega),\phi_k(\omega))}(\cdot),
\]
and it follows that
\begin{align*}
    \LnrmInf{\int_{\Gam} F(\gam)\, d(G(\omega) - \Go)(\gam)} &= \LnrmInf{\int_{\ITh\times\IPh} F(\gam)\, d(G(\omega) - \Go)(\gam)}, \\
    &\leq \sum_{l=1}^{N_0}\sum_{k=1}^{K}\LnrmInf{\int_{\Th_{n_0,l}\times\Phi_{k}} F(\gam)\, d(G(\omega) - \Go)(\gam)}.
\end{align*}
For each term in the sum, we have that
{\small
\begin{align*}
    \LnrmInf{\int_{\Th_{n_0,l}\times\Phi_{k}} F(\gam)\, d(G(\omega) - \Go)(\gam)}
    &\leq \LnrmInf{(w_{n_0,l}(\omega)w^{\phi}_{l,k}(\omega) - \Go(\Th_{n_0,l}\times\Phi_k))F(\th_l(\omega), \phi_k(\omega))}, \\ 
    &\,\, +\: \LnrmInf{\int_{\Th_{n_0,l}\times\Phi_{k}} (F(\th,\phi) - F(\th_l(\omega), \phi_k(\omega)))\, d\Go(\th,\phi)}, \\
    &< \frac{2}{3 B K N_0} \eps B + \frac{1}{3}\eps G(\Th_{n_0,l}\times\Phi_k),\\
    & = \frac{2}{3 K N_0}\eps + \frac{1}{3}\eps G(\Th_{n_0,l}\times\Phi_k).
\end{align*}}Therefore,
\[
    \LnrmInf{\int_{\Th\times\Phi} F(\th,\phi)\, d(G(\omega) - \Go)(\th,\phi)} < \frac{2}{3}\eps + \frac{1}{3}\eps = \eps.
\]
Consequently,
\[
    0 < \Prob(\Omega_{\th}\cap \Omega_{\phi} \cap \Omega_{w^\phi})) \leq \Prob(\set{\omega\in\Omega:\,\, G(\omega)\in V(\Go,f_1,\ldots, f_m,\eps,\ldots,\eps}),
\]
and the theorem follows.
\hfill$\square$

\subsection{Proof of Theorem~\ref{thm:dsbapFullSupport}}\label{AppendixC2}

Since \(\Gam\) is Polish, measures of finite support are dense in \(\mPp(\Gam)\). It is apparent that
\[
    \int_{\Gam} F(\gam)\, d\Go(\gam) \in \topcl(\cvxhull(\set{F(\gam):\,\, \gam\in \Gam})).
\]
Let \(n_0 > \lceil\log_2(m)\rceil\) and let \(N_0 = 2^{n_0}\). An application of  Carathéodory's theorem \citep{Schneider2013} shows that
\[
    \int_{\Gam} F(\gam)\, d\Go(\gam) \in \Lset{\sum_{l=1}^{N_0} \pi_l F(\gam_l):\,\, \pi \in \Delta^{N_0 - 1},\, \gam_1,\ldots, \gam_{N_0}\subset \Gam}.
\]
Therefore, there exists \(\bpi\in S_{N_0}\) and \(\set{\bgam_l}_{l=1}^{N_0}\subset \Gam\), such that
\[
    \int_{\Gam} F(\gam)\, d\Go(\gam) = \sum_{l=1}^{N_0} \bpi_l F(\bgam_l).
\]
Since the map \(\Delta^{N_0-1}\times \Gam\times \ldots \Gam\mapsto \R^m\) given by 
\[
    (\pi, \gam_1,\ldots, \gam_{N_0})\mapsto \sum_{l=1}^{N_0}\pi_l F(\gam_l),
\]
is continuous, there exists \(\delta > 0\), such that
\[
    l\in\set{1,\ldots, N_0}:\,\, \max\set{|\pi_l - \bpi_l|, |\gam_l - \bgam_l|} < \delta\,\,
    \Rightarrow\,\, \LnrmInf{\sum_{l=1}^{N_0}\pi_l F(\gam_l) - \int_{\Gam} F(\gam)\, d\Go(\gam)} < \frac{1}{3}\eps. 
\]
Therefore, without loss, we may select \(\pi^* \in \Delta^{N_0 - 1}\) and \(\set{\gam^*_l}_{l=1}^{N_0}\), such that \(\pi_1^*,\ldots, \pi_{N_0}^* > 0\), \(\th_1^* < \ldots < \th_{N_0}^*\), and
\[
    \LnrmInf{\sum_{l=1}^{N_0}\pi_l^* F(\gam_l^*) - \int_{\Gam} F(\gam)\, d\Go(\gam)} < \frac{1}{3}\eps.
\]
Since \(F\) is continuous and \(\set{\gam_1^*,\ldots, \gam_m^*}\) is discrete, there exists \(B > 0\), such that
\[
    l\in \set{1,\ldots, N_0}:\,\, \nrmInf{F(\gam_l^*)} \leq B.
\]
Remark that this holds even when some function in \(f_1,\ldots, f_m\) is a polynomial. Let \(\bGo\in \mP(\Gam)\) be
\[
    \bGo(\cdot) = \sum_{i=1}^{N_0} \pi^*_l \delta_{\gam_l^*}(\cdot),
\]
and let \(\bG_{0,1}\in \mPTh\) be its marginal
\[
    \bG_{0,1}(\cdot) = \sum_{i=1}^{N_0} \pi^*_l \delta_{\th_{l}^*}(\cdot).
\]
Remark that \(\bG_{0,1}\) coincides with its own level \(n_0\) SBA. In particular, its level \(n_0\) SBA is regular. Let \(\set{\bmu_{0,j,l}}_{j=1,l=1}^{n_0, 2^{n_0}-1}\) denote its barycenters. Then the same arguments in the proofs of Lemma~\ref{lem:dsbaSamplingApproximationProperty} and Theorem~\ref{thm:dsbaFullSupport} allow us to prove the existence of \(\delta^* > 0\), such that the set
\[
    \Omega_{\th} := \set{\omega\in\Omega:\,\,  n(\omega) = n_0,\,\, |\mu_{n_0+1,l}(\omega) - \bmu_{0,n_0+1,l}| < \delta^*,\, l\in\set{1,\ldots, 2N_0- 1}},
\]
has positive measure, and such that for any \(\omega\in \Omega_{\th}\), we have that
\[
    l\in \set{1,\ldots, N_0}:\,\, |w_{n_0,l}(\omega) - \pi^*_l| < \frac{1}{3N_0 B}\eps.
\]
Furthermore, by possibly shrinking \(\delta^*\), we may further assume that
\[
    l\in\set{1,\ldots, N_0}:\,\, |\gam-\gam_l^*| < \delta^*\,\,\Rightarrow\,\, \nrmInf{F(\gam) - F(\gam_l^*)} < \frac{1}{3}\eps.
\]
Let
\[
    \Omega_{\phi} := \set{\omega\in\Omega:\,\, n(\omega) = n_0,\,\, |\phi_l - \phi_l^*| < \delta_{\phi},\, l\in\set{1,\ldots, N_0}}.
\]
which, by hypothesis, has positive measure. Then, by independence of \(\Omega_{\th}\) and \(\Omega_{\phi}\), we conclude that
\[
    \Prob(\Omega_{\phi}\cap\Omega_{\th}) = \Prob(\Omega_{\phi})\Prob(\Omega_{\th}) > 0.
\]
If \(\omega\in\Omega_{\th}\cap\Omega_{\phi}\), then 
\begin{align*}
    \LnrmInf{\int_{\Gam} F(\gam)\, d(G(\omega) - \Go)(\gam)} &\leq \LnrmInf{\int_{\Gam} F(\gam)\, d(\bGo - \Go)(\gam)}, \\
    &\quad +\: \LnrmInf{\int_{\Gam} F(\gam)\, d(G(\omega) - \bGo)(\gam)},\\
    &< \frac{1}{3}\eps + \LnrmInf{\sum_{l=1}^{N_0} (w_{n_0,l} - \pi_l^*)F(\gam_l^*)}, \\
    &\quad +\: \LnrmInf{\sum_{l=1}^{N_0} w_{n_0,l} (F(\gam_l) - F(\gam_l^*))},\\
    &< \frac{1}{3}\eps + \frac{1}{3N_0 B}\eps  N_0 B + \frac{1}{3}\eps, \\
    &=\eps.
\end{align*}
Therefore,
\[
    0 < \Prob(\Omega_{\phi}\cap\Omega_{\th}) \leq \Prob(\set{\omega\in\Omega:\,\, G(\omega)\in V(\Go, f_1,\ldots, f_m, \eps,\ldots,\eps)}),
\]
and from this inequality the theorem follows.
\hfill$\square$

\subsection{Proof of Theorem~\ref{thm:hellinger:support}}\label{AppendixC3}

Before proceeding with the proof, we require some preparatory results. We introduce some auxiliary notation. We assume that \(\mY \in \set{\R, \R_+, [0, 1]}\). Remark that in all cases \(\mY\) is Polish and the restriction of the Lebesgue measure to \(\mY\) is well defined. It will be useful to consider the set \(\mYs\subset \mY\) defined as
\[
    \mYs = \begin{cases}
        \R, & \mY = \R,\\
        (0, \infty), & \mY = \R_+,\\
        (0, 1), & \mY = [0, 1].
    \end{cases}
\]
This set will play the role of the interior of \(\mY\). To simplify the notation, from now on we write \(\Gam = \Th\times\Phi\) whence every \(\gam\in \Gam\) has the representation \(\gam = (\th, \phi)\). The space \(\mP(\Gam)\) is assumed to be endowed with the weak topology, and the Borel \(\sigma\)-algebra.

In Section~\ref{sec:hellinger:densities} we introduce some preliminary results about the spaces of densities with respect to the Lebesgue measure on \(\mY\). Then, in Section~\ref{sec:hellinger:mixtureMap} we define the mixture map on a suitable subset of \(\mP(\Gam)\) for each of the kernels of interest. In Section~\ref{sec:hellinger:mixtureMapRange} we show that in all the cases of interest the mixture map is able to approximate any density in \(\mY\) to an arbitrary accuracy. In Section~\ref{sec:hellinger:mixingMapMeasurable} we show that the mixture map is measurable, and thus induces a proper probability measure on \(\mDY\). Finally, in Section~\ref{sec:hellinger:supportOfMixtures} we prove that the support of this induced measure is \(\mDY\). 

\subsubsection{The space of densities on \(\mY\)}
\label{sec:hellinger:densities}

We denote as \(\mDY\) be the space of probability density functions with respect to the Lebesgue measure on \(\mY\) and we denote as \(\mDoY\) the subset of continuous probability density functions with respect to the Lebesgue measure. Furthermore, we denote as \(\mDocY\) the densities in \(\mDoY\) with compact support, and we define
\[
    \mDsY := \set{f\in \mDocY:\,\, \supp(f)\in \mYs}.
\]
From now on, we assume that \(\mDY\) is endowed with the Hellinger distance
\[
    f_1,f_2\in \mDY:\,\, \Hd(f_1, f_2) = \left(\frac{1}{2}\int_{\mY}(\sqrt{f_1(y)} - \sqrt{f_2(y)})^2\, dy\right)^{1/2},
\]
with the metric topology, and with the Borel \(\sigma\)-algebra.

Since the square-root is H\"older continuous with exponent \(1/2\), we have the bound
\[
    \Hd(f_1, f_2) \leq \left(\frac{1}{2}\int_{\mY}|f_1(y) - f_2(y)|\, dy\right)^{1/2}.
\]
Conversely, 
\[
    \int_{\mY} |f_1(y) - f_2(y)|\, dy = \int_{\mY} |\sqrt{f_1(y)} - \sqrt{f_2(y)}|(\sqrt{f_1(y)} + \sqrt{f_2(y)})\, dy \leq 2\sqrt{2} \Hd(f_1, f_2).
\]
From the above inequality the following proposition follows.

\begin{proposition}\label{prop:hellinger:densityOfCoc}
    The inclusions \(\mDsY \subset \mDocY \subset \mDoY \subset \mDY\) are dense.
\end{proposition}

For \(\eps > 0\) and \(f_0\in \mDY\) we denote as \(B_{\Hd}(f_0, \eps)\) and \(\bar{B}_{\Hd}(f_0, \eps)\) the open and closed ball of center \(f_0\) and radius \(r\) respectively. For \(f_1,f_2\in\mDY\) the Bhattacharyya coefficient is defined as
\[
    BC(f_1,f_2) = \int_{\mY} \sqrt{f_1(y) f_2(y)}\, dy.
\]
It is apparent that \(\Hd^2 = 1 - BC\). This allows us to bound \(\Hd^2\) as follows. Let \(K\subset \mY\) be compact. Then
\begin{align*}
    \Hd^2(f_1, f_2) &= \int_{\mY\setminus K} \sqrt{f_1(y)}(\sqrt{f_1(y)} - \sqrt{f_2(y)})\, dy \\
    &\quad +\: \int_{K} \sqrt{f_1(y)}(\sqrt{f_1(y)} - \sqrt{f_2(y)})\, dy,\\
    &\leq \left(\int_{\mY\setminus K} f_1(y)\, dy\right)^{1/2}\left(\int_{\mY\setminus K} (\sqrt{f_1(y)} - \sqrt{f_2(y)})^2\, dy\right)^{1/2} \\
    &\quad +\: \left(\int_K f_1(y)\, dy\right)^{1/2}\left(\int_{K} (\sqrt{f_1(y)} - \sqrt{f_2(y)})^2\, dy\right)^{1/2},\\
    &\leq 2\left(\int_{\mY\setminus K} f_1(y)\, dy\right)^{1/2} + \left(\int_{K} |f_1(y) - f_2(y)|\, dy \right)^{1/2}.
\end{align*}
This implies the following proposition for all our choice of \(\mY\).

\begin{proposition}\label{prop:hellinger:compactIntervalBound}
    Let \(f_0\in\mDY\) and let \(\eps > 0\). Then there exists \(K\subset \mYs\) compact such that for any \(f\in\mDY\), we have that
    \[
        \Hd^2(f_0, f) \leq \frac{1}{2}\eps^2 + \left(\int_{K} |f(y) - f_0(y)|\, dy \right)^{1/2}.
    \]
\end{proposition}

\subsubsection{The mixture map}
\label{sec:hellinger:mixtureMap}

To associate an element of \(\mDY\) to every measure \(G\in\mP(\Gam)\) we use a kernel. For \(\mY = \R\) we use the Gaussian kernel, for \(\mY = \R_+\) we use the Gamma kernel, and for \(\mY = [0, 1]\) we use the Beta kernel. Remark that in all cases \(\Phi = \R_+\) and \(\Th = \mY\). This allows us to identify any \(f\in\mDY\) as a probability density on \(\Th\) with respect to the Lebesgue measure and viceversa. However, our notation will differentiate these two sets to preserve the conceptual difference between the spaces \(\mY\) and \(\Th\). Therefore, we denote as \(\Ths = \mYs\) and we define similarly \(\Phis = (0, \infty)\). The identification between \(\Th\) and \(\mY\) allows us to define \(\mDs(\Th)\) analogously.

It is apparent that for some choices of \(\gam\) the resulting function \(y\mapsto k(y|\,\gam)\) is not an element of \(\mDY\). As a concrete example, for the Gaussian kernel \(k(y|\, \th, 0) \equiv 0\), for the Gamma kernel \(k(y|\, \th, 0) = 0\) for any \(y > 0\), and for the Beta kernel \(k(y|\, \th, 0)\) is simply undefined. Therefore, we define the set \(\Gam_k^0 \subset \Gam\) as
\[
    \Gam_k^0 := \begin{cases}
        \Th\times \set{0}, & \Th = \R,\\
        \Th\times \set{0}, & \Th = \R_+,\\
        \Th\times \set{0}, \cup \set{0, 1}\times \R_+ & \Th = [0, 1].
    \end{cases}
\]
and let \(\Gam_k^* = \Gam\setminus \Gam_k^0\). Remark that \(\Gam_k^0\) is a measurable subset of \(\Gam\) and that
\[
    f\in \mDsY,\, \phi \in \Phis:\,\, \supp(f) \times \set{\phi} \subset \Gam_k^*.
\]
From now on, we refer to the kernel simply as \(k\) as we consider a single kernel for each choice of \(\mY\). This leads to the following proposition.

\begin{proposition}\label{prop:hellinger:continuityKernel}
    Let \(K_{\phi}\subset \Phis\) be a compact, let \(K_y \in \mYs\) be compact, and let \(K_{\th} \subset \Ths\) be compact. Then the map \((y,\th,\phi) \mapsto k(y|\, \th,\phi)\) restricted to \(K_y\times K_{\th}\times K_{\phi}\) is continuous and
    \[
        (\th,\phi) \in K_{\th}\times K_{\phi}:\,\, \int_{\mY} k(y|\, \th,\phi)\, dy = 1.
    \]
\end{proposition}

We define the set
\[
    D_k := \set{G\in\mP(\Gam):\,\, G(\Gam_k^0) = 0}.
\]
Remark that in each case \(D_k\) is a measurable subset of \(\mP(\Gam)\). Furthermore, it is apparent that
\[
    G\in D_k :\,\, T_k G(y) = \int k(y|\, \gam)\, dG(\gam),
\]
does define an element of \(\mDY\). Therefore, the map \(T_k : D_k \mapsto \mDY\) is well defined. We call this map the {\em mixture map}. Furthermore, we have the following proposition.

\begin{proposition}
    The set \(D_k\) is a measurable subset of \(\mP(\Gam)\). The processes DSBASp and DSBASg assign probability zero to \(D_k^c\).
\end{proposition}

Therefore, we can restrict ourselves to \(D_k\) endowed with the subspace topology, and the Borel \(\sigma\)-algebra. The above proposition states that there exist a measurable set \(\Omega_k\subset \Omega\) of full measure such that \(G(\omega)\in D_k\) for \(\omega\in \Omega_k\). From now on, we study the restriction of the processes DSBASp and DSBASg to this set. 

From now on, we let
\[
    \im(T_k) := \set{T_k G:\,\, G\in D_k}.
\]
Finally, remark that for any \(f\in \mDs(\Th)\) and \(\phi\in\Phis\) the mixing measure \(G\) given by
\[
    A\in \mathcal{B}(\Th),\, B\in \mathcal{B}(\Phi):\,\, G(A\times B) = \left(\int_A f(\th)\, d\th\right) \delta_{\phi}(B),
\]
is in \(D_k\).

\subsubsection{The range of the mixture map}
\label{sec:hellinger:mixtureMapRange}

We now prove that in all cases the mixture map that we defined is expressive, being able to approximate any density in \(\mY\) to an arbitrary accuracy. 

\begin{lemma}\label{lem:hellinger:mixtureRangeDense}
    The range of \(T_k\) is dense in \(\mDY\).
\end{lemma}

 \noindent{\sc Proof of  Lemma~\ref{lem:hellinger:mixtureRangeDense}:} From Proposition~\ref{prop:hellinger:densityOfCoc} it suffices to show that for any \(f_0\in\mDsY\) and \(\eps > 0\) there exists \(G\in D_k\) such that \(\Hd(f_0, T_kG) < \eps\).
    \begin{enumerate}
        \item{For \(\phi\in \R_+\) let
        \[
            g_\phi(\th) = \sqrt{\frac{\phi^2}{2\pi}} e^{-\frac{1}{2}\phi^2 \th^2},
        \]
        define
        \[
            f_0^{\phi}(y) = \int_{\R} f_0(\th) k(y|\, \th, \phi)\, d\th = (f_0\ast g_{\phi})(y),
        \]
        and let \(f_0^\phi := f_0 \ast g_\phi\) where \(\ast\) denotes the convolution. It is apparent that \(f_0^{\phi}\in \mD(\R)\). A classical result states that
        \[
            \lim_{\phi\to \infty} \int |f_0^{\phi}(y) - f_0(y)|\, dy = 0.
        \]
        Therefore, there exists \(\phi\adj > 0\) such that
        \[
            \Hd(f_0^{\phi\adj}, f_0) < \eps.
        \]
        If we define
        \[
            A\in\mB(\R),\, B\in\mB(\R_+):\,\,  G(A\times B) = \left(\int_{A} f_0(\th)d\th \right)\delta_{\phi\adj}(B),
        \]
        we deduce that
        \begin{align*}
            f_0^{\phi}(y) &= \int_{\R} f_0(\th) g_{\phi\adj}(y - \th)\, d\th,\\
                &= \int_{\R}  k(y|\, \th,\phi\adj)\, f_0(\th) d\th,\\
                &= \iint_{\Th\times\Phi} k(y|\, \th,\phi)\, dG(\th,\phi),\\
                &= T_k G(y).
        \end{align*}
        It is apparent that \(G\in D_k\) as desired.
        }
        \item{Let \(K\in \mYs\) be as in Proposition~\ref{prop:hellinger:compactIntervalBound}. We use the auxiliary Lemma~\ref{lem:hellinger:auxiliaryGamma}. Let \(\phi > 3\) and define
        \[
            f_0^{\phi}(y) = \int_0^{\infty} f_0(\th) k(y|\,\th,\phi)\, d\phi.
        \]
        Then \(f_0^{\phi} \in \mDY\). Since \(K\subset \mYs\) we can use Lemma~\ref{lem:hellinger:auxiliaryGamma} once again to select \(\phi\adj\) sufficiently large so that
        \[
            \sup_{y\in K}\, |f_0^{\phi}(y) - f_0(y)| < \frac{1}{4|K|}\eps^4,
        \]
        holds. In this case, we conclude that
        \[
            \Hd^2(f_0^{\phi}, f_0) < \frac{1}{2}\eps^2 + \left(\frac{1}{4|K|} \eps^2 |K|\right)^{1/2} = \frac{1}{2}\eps^2 + \frac{1}{2}\eps^2 = \eps^2,
        \]
        whence
        \[
           \Hd(f_0^{\phi\adj}, f_0) < \eps.
        \]
        Therefore, by defining
        \[
            A\in\mB(\R),\, B\in\mB(\R_+):\,\,  G(A\times B) = \left(\int_{A} f_0(\th)d\th \right)\delta_{\phi\adj}(B),
        \]
        we deduce that
        \[
            f_0^{\phi\adj}(y) = \int_{\R_+} f_0(\th) k(y|\, \th,\phi\adj)\, d\th = \iint_{\Th\times\Phi} k(y|\, \th,\phi)\, dG(\th,\phi) = T_k G(y).
        \]
        Since \(\phi\adj > 0\) and \(f_0\in \mDsY\) it is clear that \(G\in D_k\).
        }
        \item{Let \(n\in\N\) and define \(h_0^n \in C^0([0, 1])\) as
        \[
            h_0^n(y) = \sum_{m=0}^n f_0(m/n) b_{m,n}(y),
        \]
        where \(b_{m,n}\) is the \(m\)-th Bernstein polynomial of degree \(n\) defined as
        \[
            b_{m,n}(y) = {n \choose m} y^m (1-y)^{n - m}.
        \]
        Remark that the terms for \(m = 0\) and \(m = n\) do not contribute to the sum as \(f_0(0) = f_0(1) = 0\) for \(f_0\in\mDsY\). A classical result states that
        \[
            \lim_{n\to\infty}\, \sup_{y\in [0, 1]}\, |f_0(y) - h_0^n(y)| < \eps.
        \]
        However, \(h_0^n \notin \mD([0, 1])\) as
        \[
            I_{n} := \int_0^1 h_0^n(y)\, dy = \frac{1}{n+1}\sum_{m=0}^n f_0(m/n).
        \]
        Therefore, define \(f_0^n = \frac{1}{I_n} h_0^n\). From
        \[
            f_0(y) - f_0^n(y) = f_0(y) - h^n_0(y) + \frac{I_n - 1}{I_n} h_0^n(y),
        \]
        it follows that
        \begin{align*}
            \int_0^1 |f_0(y) - f_0^n(y)|\, dy &\leq \int_0^1 |f_0(y) - h^n_0(y)|\, dy + |I_n - 1|,\\
            &\leq |I_n - 1| + \sup_{y\in[0, 1]}\, |f_0(y) - h^n_0(y)|.
        \end{align*}
        Thus, we can choose \(n\) sufficiently large so that \(H(f_0^n, f_0) < \eps\). Finally, by defining \(G\) as
        \[
            A\in \mB([0, 1]),\, B\in \mB(\R_+):\, G(A\times B) = \sum_{m=0}^n \frac{f_0(m/n)}{I_n} \delta_{m/n}(A)\delta_{n}(B),
        \]
        it follows that
        \begin{align*}
            f_0^n(y) &= \sum_{m=0}^n \frac{f_0(m/n)}{I_n} {n \choose m} y^m (1-y)^{n - m}, \\
            &= \sum_{m=0}^n \frac{f_0(m/n)}{I_n} k(y|\, (m+1)/(n + 2), n + 2), \\
            &= \iint_{\Th\times\Phi} k(y|\, \th, \phi)\, dG(\th,\phi),\\
            &= T_kG(y).
        \end{align*}
        Since \(f_0(0) = f_0(1) = 0\) we conclude that \(G\in D_k\).
        }
    \end{enumerate}
\hfill$\square$

We now prove the auxiliary approximation lemma for the Gamma kernel.

\begin{lemma}\label{lem:hellinger:auxiliaryGamma}
    Let \(\mY = \R_+\) and let \(f \in \mDsY\). For \(\phi > 3\), define
    \[
        f^{\phi}(y) = \int_0^{\infty} f(\th) k(y|\, \th,\phi)\, d\th,
    \]
    where \(k\) is the Gamma kernel. Then, \(f^{\phi}\in \mDY\) and for any \(K\subset \mYs\) compact, we have that
    \[
        \lim_{\phi\to\infty}\, \sup_{y \in K}\, |f^{\phi}(y) - f(y)| = 0.
    \]
\end{lemma}

\noindent{\sc Proof of Lemma~\ref{lem:hellinger:auxiliaryGamma}:} A direct application of Tonelli's theorem shows that \(f^\phi\in \mDY\). Let \(y > 0\) and \(\phi > 3\). In this case, the function
    \[
        \th \mapsto \left(\frac{\phi}{\th}\right)^{\phi} e^{-\frac{\phi}{\th} y},
    \]
    is integrable on \(\R_+\). Let \(m\in\set{0,1,2}\). Observe that
    \begin{align*}
        \int_0^{\infty} \th^{m} k(y|\, \th,\phi)\, d\th = \frac{y^{\phi-1}}{\Gamma(\phi)}\int_0^{\infty} \th^m \left(\frac{\phi}{\th}\right)^{\phi} e^{-\frac{\phi}{\th} y}\, d\th.
    \end{align*}
    By using the change of variables \(z = \phi y / \th\), we deduce that
    \begin{align*}
        \int_0^{\infty} \left(\frac{\phi}{\th}\right)^{\phi} e^{-\frac{\phi}{\th} y} f(\th)\, d\th &= \int_0^{\infty}\left(\frac{\phi y}{z}\right)^m \left(\frac{z}{y}\right)^{\phi}  e^{-z} \frac{\phi y}{z} \frac{dz}{z},\\
        &=  \frac{\phi^{1 +m}}{y^{\phi - 1 - m}}\int_0^{\infty} z^{\phi - 2 - m} e^{-z}\, dz,\\
        &= \frac{\phi^{1 + m}}{y^{\phi - 1 - m}}\Gamma(\phi - 1 - m).
    \end{align*}
    By letting \(m = 0\) we define
    \[
        I_{\phi} = \int_0^{\infty} k(y|\, \th,\phi)\, d\th = \frac{\phi \Gamma(\phi - 1)}{\Gamma(\phi)} = \frac{\phi}{\phi - 1},
    \]
    whence \(I_{\phi} \to 0\) as \(\phi\to\infty\) uniformly in \(y\). By letting \(m = 1\), we define
    \[
        \mu_{\phi} = \frac{1}{I_{\phi}}\int_0^{\infty} \th k(y|\, \th,\phi)\, d\th = \frac{\phi - 1}{\phi}\frac{\phi^2\Gamma(\phi-2)}{\Gamma(\phi)} y = \frac{1}{1 - 2\phi^{-1}} y,
    \]
    In particular,
    \[
        y - \mu_{\phi} = -y \frac{2\phi^{-1}}{1 - 2\phi^{-1}} = -\frac{2y}{\phi - 2},
    \]
    and \(\mu_{\phi}\to y\) as \(\phi\to\infty\). Now, remark that for \(m = 2\), we obtain
    \[
        \frac{1}{I_{\phi}}\int_0^{\infty} \th^2 k(y|\, \th,\phi)\, d\th = \frac{\phi - 1}{\phi}\frac{\phi^3\Gamma(\phi-3)}{\Gamma(\phi)} y^2 = \frac{1}{(1-2\phi^{-1})(1- 3\phi^{-1})} y^2.
    \]
    Therefore,
    \begin{align*}
        \sigma_{\phi}^2 &= \left((1- 2\phi^{-1}) - (1-3\phi^{-1})\right) \frac{y^2}{(1-2\phi^{-1})^2(1-3\phi^{-2})},\\
        &= \frac{\phi^{-1}}{(1-2\phi^{-1})^2(1-3\phi^{-2})} y^2,
    \end{align*}
    whence \(\sigma_{\phi}^2 \to 0\) as \(\phi\to\infty\).

    Let \(K\subset \R_+\) be compact and such that \(0\notin K\), and let \(\eps > 0\). For any \(\delta > 0\), such that
    \[
        \delta < \inf_{y\in K}\, y,
    \]
    we have
    \begin{align*}
        f^{\phi}(y) - f(y) &= \int_0^{\infty} (f(\th) - f(y))\, k(y|\, \th,\phi)\, d\th - (1-I_\phi) f(y),\\
        &= \int_{|\th - y| < \delta} (f(\th) - f(y))\, k(y|\, \th,\phi)\, d\th, \\
        &\quad +\:\int_{|\th - y| \geq \delta} (f(\th) - f(y))\, k(y|\, \th,\phi)\, d\th - (1-I_\phi) f(y).
    \end{align*}
     Since \(f\) is compactly supported, it is uniformly continuous, and by possibly shrinking \(\delta\) we must have that
    \[
        |\th - y| < \delta\,\,\Rightarrow\,\, |f(\th) - f(y)| < \frac{1}{2}\eps.
    \]
    Importantly, this choice for \(\delta\) is independent of \(y\). Therefore
    \[
        \left|\int_{|\th - y| < \delta/2} (f(\th) - f(y))\, k(y|\, \th,\phi)\, d\th  \right| < \frac{1}{2}\eps I_{\phi}.
    \]
    Since
    \[
        |y - \mu_{\phi}| < \frac{2}{\phi - 2}\left(\sup_{y\in K}\, y\right),
    \]
    we may choose \(\phi\) sufficiently large, and independent of \(y\), such that \(|\mu_{\phi} - y| < \delta/2\) for any \(y\in K\). In this case,
    \[
        |\th - \mu_{\phi}| = |\th - y| - |y - \mu_{\phi}| > \delta - \delta/2 = \delta/2.
    \]
    Therefore,
    \begin{align*}
        \left| \int_{|\th - y| \geq \delta} (f(\th) - f(y))\, k(y|\, \th,\phi)\, d\th \right| &= \left|\int_{|\th - y| \geq \delta} \frac{f(\th) - f(y)}{(\th - \mu_{\phi})^2}\, (\th - \mu_{\phi})^2 k(y|\, \th,\phi)\, d\th\right|,\\
        &\leq \left(\frac{4I_{\phi}}{\delta}\sup_{y\in\R_+} f(y)\right)\left( \frac{1}{I_{\phi}} \int_{|\th - y| \geq \delta} (\th - \mu_{\phi})^2 k(y|\, \th,\phi)\, d\th\right),\\
        &\leq \left(\frac{4I_{\phi}}{\delta}\sup_{y\in\R_+} f(y)\right) \sigma_{\phi}^2.
    \end{align*}
    Thus, it follows that,
    \[
        \sup_{y \in K}\, |f^{\phi}(y) - f(y)| < I_{\phi} \left(\frac{1}{2}\eps + \frac{4\sigma_{\phi}^2}{\delta}\left(\sup_{y\in\R_+} f(y)\right)\right) + |1-I_{\phi}| \left(\sup_{y\in \R_+}\, f(y)\right).
    \]
    Since
    \[
        \sigma_{\phi}^2 \leq \frac{\phi^{-1}}{(1-2\phi^{-1})^2(1-3\phi^{-2})}  \left(\sup_{y\in K}\, y^2\right),
    \]
    and \(\delta\) is independent of \(y\), we can select \(\phi\) sufficiently large such that
    \[
        \sup_{y \in K}\, |f^{\phi}(y) - f(y)| < \eps,
    \]
    as we wanted to prove.
\hfill$\square$

\subsubsection{Measurability of the mixture map}
\label{sec:hellinger:mixingMapMeasurable}

Lemma~\ref{lem:hellinger:mixtureRangeDense} together with the fact that the processes DSBASp and DSBASg have full support on \(D_k\) suggests that the process 
\[
    T_k G :\Omega \to \mDY,
\]
has full support on \(\mDY\). However, to make this conclusion rigorous, it is necessary to prove that the above map is measurable.

To prove this we first need the following technical result showing that the Borel \(\sigma\)-algebra in \(\mDY\) is countably generated by neighborhoods of elements in \(\mDsY\).

\begin{lemma}\label{lem:hellinger:sigmaAlgebraCountablyGenerated}
    There exists a countable subset \(\mF\subset \mDsY\) such that the collection
    \[
        \mathcal{C}_H := \set{B_{\Hd}(f, r):\,\, f\in \mF,\, r \in \mathbb{Q}_+},
    \]
    generates the Borel \(\sigma\)-algebra on \(\mDY\).
\end{lemma}

\noindent{\sc Proof of Lemma~\ref{lem:hellinger:sigmaAlgebraCountablyGenerated}:} We provide the main elements of the proof and leave some of the details to the reader. It is apparent that in all cases the space \(\mY\) is \(\sigma\)-compact and there exists a monotone increasing family \(\set{K_n}_{n\in \N}\) of compact subsets of \(\mYs\) such that
    \[
        \mYs = \bigcup_{n\in\N}\, K_n.
    \]
    Define
    \[
        \mDs(K_n):= \set{f\in \mDsY:\,\, \supp(f)\subset K_n}.
    \]
    Then,
    \[
        \mDsY := \bigcup_{n\in\N} \mDs(K_n),
    \]
    whence it suffices to show that \(\mDs(K_n)\) is separable for each \(n\). Remark that the natural injection 
    \[
        \iota_n : \mDs(K_n) \to C^0(K_n),
    \]
    is an isometry {\em with respect to the uniform norm}. Since the space \(C^0(K_n)\) is separable in the uniform norm, so is \(\iota_n(\mDs(K_n))\). Hence, \(\mDs(K_n)\) is separable {\em in the uniform norm}. Let \(\mF_n\) be the corresponding countable dense subset. We now prove that \(\mF_n\) is also dense {\em in the Hellinger distance}. For any \(f_0\in \mDs(K_n) \) and \(\eps > 0\) there exists \(f \in \mF_n\) such that
    \[
        \sup_{y\in K_n}\,\, |f(y) - f_0(y)| < \frac{2}{|K_n|} \eps^2.
    \]
    However, this implies that 
    \[
        \Hd(f, f_0) \leq \left(\frac{1}{2}\int_{K_n} |f(y) - f_0(y)|\, dy\right)^{1/2} < \eps.
    \]
    Therefore, \(\mF_n\) is a countable dense subset of \(\mDs(K_n)\) in the Hellinger distance. We conclude that \(\mDsY\) is separable in the Hellinger distance. In fact, 
    \[
        \mF = \bigcup_{n\in\N}\, \mF_n,
    \]
    is a countable dense subset of \(\mDsY\). By Proposition~\ref{prop:hellinger:densityOfCoc} it is also dense in \(\mDY\). Therefore, for any \(f\in \mDY\) and \(r > 0\),
    \[
        B_{\Hd}(f, r) := \bigcup\set{B_{\Hd}(f', r'):\,\, B_{\Hd}(f', r')r'}.
    \]
    Since \(B_{\Hd}(f, r)\) can be represented as the countable union of element in \(\mathcal{C}_H\) the lemma follows.
\hfill$\square$

Therefore, it suffices to show that for any \(f_0\in \mDsY\) and \(r > 0\), the set
\[
    A_{f_0, r} := \set{G\in D_k:\,\, T_k G \in \bar{B}_{\Hd}(f_0, r)},
\]
is measurable. To do this, we first represent the closed balls as level sets of a suitable function. From now on, we fix \(f_0 \in \mF\) and \(r > 0\). Remark that
\[
    \bar{B}_{\Hd}(f_0, r) := \set{ f\in \mDY:\,\, u_0(f) \geq \alpha},
\]
for \(u_0:\mDY \to \R\) given by
\[
    u_0(f) := \int_{\mY} \sqrt{f_0(y)} \sqrt{f(y)}\, dy,
\]
and \(\alpha = 1 - r^2\). Remark then that
\[
    A_{f_0, r} = (u_0\circ T_k)^{-1}([\alpha, \infty)).
\]
Consequently, to prove that \(A_{f_0, r}\) is measurable, it suffices to show that \(u_0\circ T_k : D_k \to \R\) is measurable. 

To our knowledge, this cannot be proved directly. For this reason, we regularize the function \(u_0\) as follows. Fix \(\pi \in (0, 1)\) and fix a \(f^*\in \mDoY\) such that \(\supp(f^*) = \mY\). Define
\[
    f_s = \pi f_0 + (1-\pi) f^*. 
\]
Then, \(f_s \in \mDoY\) and \(\supp(f_s) = \mY\). Thus, for any \(\delta > 0\), we define \(\eta_{\delta} : \mY\times \R_+ \to \R\) as
\[
    \eta_{\delta}(y, t) = \sqrt{f_0(y)} \sqrt{(1 - \delta) t + \delta f_s(y)},
\]
and we define the regularized function \(\uod:\mDY \to \R\) as
\[
    \uod(f) := \int \eta_\delta(y, f(y))\, dy.
\]
It is apparent that \(\eta_{\delta}\) is continuous, and that for any \(y\in \mY\) the map \(t\mapsto \eta_{\delta}(y, t)\) is differentiable and concave. Furthermore, \(\uod\) is continuous, as for \(f_1,f_2\in \mDY\) we have that
\begin{align}\label{eq:hellinger:regFunctionalsLipschitz}
\begin{split}
    |\uod(f_1) - \uod(f_2)| &\leq \int_{\mY} \sqrt{f_0(y)}|\sqrt{(1-\delta) f_1(y) + \delta f_s(y)}, \\
    &\quad -\:  \sqrt{(1-\delta) f_2(y) + f_s(y)}|\, dy, \\
    &\leq (1-\delta)\int_{\mY} \sqrt{f_0(y)}|f_1(y) - f_2(y)|^{1/2}\, dy,\\
    &\leq (1-\delta) \Hd(f_1, f_2).
\end{split}
\end{align}

We have the following lemma.

\begin{lemma}\label{lem:hellinger:regFunctionalsMeasurable}
    For any \(\delta > 0\) the function \(\uod \circ T_k : D_k \to \R\) is measurable. As a consequence \(u_0\circ T_k\) is measurable. 
\end{lemma}

\noindent{\sc Proof of Lemma~\ref{lem:hellinger:regFunctionalsMeasurable}:} To prove the lemma we will prove that \((\uod \circ T_k)^{-1}([\alpha,\infty))\) is weakly closed and thus measurable. Let
    \[
        U := \set{G\in D_k:\,\, (\uod\circ T_k)(G) < \alpha}.
    \]
    Let \(\tilde{G} \in U\) and denote \(\tilde{f}\in T_k \tilde{G}\). We first show that we can assume without loss that \(\tilde{f}\) is continuous. Let 
    \[
        \eps = \alpha - \uod(\tilde{f}).
    \]
    The constructions in the proof of Lemma~\ref{lem:hellinger:mixtureRangeDense} show that there exists \(\bar{G}\in D_k\) such that \(\bar{f} := T_k \bar{G}\) is in \(\mDoY\) and such that
    \[
        \Hd(\tilde{f}, \bar{f}) < \frac{1}{4\sqrt{2}}\eps.
    \]
    In particular, this implies that
    \[
        \int_{\mY} |\tilde{f}(y) - \bar{f}(y)|\, dy < \frac{1}{2}\eps.
    \]
    From~\eqref{eq:hellinger:regFunctionalsLipschitz} we deduce that 
    \[
        \uod(\bar{f}) \leq |\uod(\bar{f}) - \uod(\tilde{f})| + \uod(\tilde{f}) <  \frac{1}{2}(1-\delta) \eps + \uod(\tilde{f}) < \alpha.
    \]
    Therefore, \(\bar{f} \in U\). Hence, we will construct a weakly open neighborhood of \(\bar{f}\) contained in \(U\) that also contains \(\tilde{f}\). This will prove the lemma.
    
    By a direct computation, we have
    \[
        \partial_t \eta_{\delta}(y, t) = \frac{1-\delta}{2}\sqrt{\frac{f_0(y)}{(1-\delta)t + \delta f_s(y)}},
    \]
    and
    \[
        |\partial_t \eta_{\delta}(y, t)| \leq \frac{1-\delta}{2\sqrt{\delta}}\sqrt{\frac{f_0(y)}{f_s(y)}} < \frac{1-\delta}{2\sqrt{\delta \pi}}.
    \]
    Remark that the upper bound is independent of \(y\) and \(t\). Since \(\eta_{\delta}\) is smooth and concave, we have that for any \(y\in \mY\),
    \[
        t,s\in \R_+:\,\, \eta_{\delta}(y, s) \leq \eta_{\delta}(y, t) + \partial_t\eta_{\delta}(y, t)(s -t).
    \]
    Consequently, the above inequality and the uniform boundedness of the derivative implies that for any \(f\in \mDY\),
    \[
        \uod(f) \leq \uod(\bar{f}) + \int_{\mY} \partial_t \eta_{\delta}(y, \bar{f}(y))(f(y) - \bar{f}(y))\, dy.
    \]
    Therefore, let
    \[
        \alpha^* := \int_{\mY} \partial_t \eta_{\delta}(y, \bar{f}(y))\bar{f}(y)\, dy - \frac{1}{4}\frac{\eps (1-\delta)}{\sqrt{\pi\delta}},
    \]
    and define
    \[
        V := \Lset{G\in D_k:\,\,  \int_{\mY} \partial_t \eta_{\delta}(y, \bar{f})T G(y)\, dy < \alpha^*}.
    \]
    It is apparent that \(V\subset U\). We now prove that \(V\) is weakly open. Since \(\partial_t \eta_{\delta} \geq 0\), by Tonelli's theorem we deduce that
    \begin{align*}
        \int_{\mY} \partial_t \eta_{\delta}(y, \bar{f}(y))T G(y)\, dy &= \int_{\mY}\int_{\Gam} \partial_t \eta_{\delta}(y, \bar{f}(y)) k(y|\, \gam)\, dG(\gam) dy,\\
        &= \int_{\Gam} \left(\int_{\mY} \partial_t \eta_{\delta}(y, \bar{f}(y)) k(y|\, \gam)\, dy\right) dG(\gam).
    \end{align*}
    Define now \(h : \Gam^*_k \to \R\) as
    \[
        h(\gam) := \int_{\mY} \partial_t \eta_{\delta}(y, \bar{f}(y)) k(y|\, \gam)\, dy = \frac{1-\delta}{2} \int_{K_y} \sqrt{\frac{f_0(y)}{(1-\delta)\bar{f}(y) + \delta f_s(y)}} k(y|\, \gam)\, dy,
    \]
    where \(K_y = \supp(f_0)\) is such that \(K_y \subset \mYs\). Remark that for any \(\gam \in \Gam_k^*\), we have that
    \[
        |h(\gam)| \leq \frac{1-\delta}{2\sqrt{\delta\pi}},
    \]
    whence \(h\) is bounded. For any \(\gam_0\in \Gam_k^*\) we can find a sufficiently small compact neighborhood \(K_{\th}\in \Ths\) of \(\th_0\) and \(K_{\phi}\in \Phis\) of \(\phi_0\). Since \(y\mapsto \partial_t \eta_\delta(y, \bar{f}(y))\) is continuous, as \(f_0, f_s\) and \(\bar{f}\) are continuous, we deduce from Proposition~\ref{prop:hellinger:continuityKernel} that \(h\in \Cob(\Gam_k^*)\). Consequently, \(V\) is weakly open. Finally, remark that
    \begin{align*}
        \int_{\mY} \partial_t \eta_{\delta}(y, \bar{f}(y)) T_k \tilde{G}(y)\, dy &\leq \frac{1-\delta}{2\sqrt{\delta\pi}} \int_{\mY}|\tilde{f}(y) - \bar{f}(y)|\, dy +  \int_{\mY} \partial_t \eta_{\delta}(y, \bar{f}(y)) T_k \bar{G}(y)\, dy \\
        &< \frac{1}{2}\eps\frac{1-\delta}{2\sqrt{\delta\pi}} + \alpha^* - \frac{1}{2}\eps\frac{1-\delta}{2\sqrt{\delta\pi}}\\
        &< \alpha^*
    \end{align*}
    from where \(\tilde{G}\in V\). This proves the first part of the lemma.

    To prove the second, it suffices to prove that \(\uod\circ T_k\to u_0\circ T_k\) pointwise. We prove this as follows. Let \(G\in D_k\) and let \(f = T_k G\). Then
    \begin{align*}
        |(u_0\circ T_k)(G) - (\uod\circ T_k)(G)| &= |u_0(f) - \uod(f)|,\\
        &\leq \int_{\mY} \sqrt{f_0(y)}|\sqrt{f(y)} - \sqrt{(1-\delta) f(y) + \delta f_s(y)}|\, dy,\\
        &\leq \sqrt{\delta} \int_{\mY} \sqrt{f_0(y)}|f(y) - f_s(y)|^{1/2}\, dy,\\
        &\xrightarrow{\delta\to 0} 0.
    \end{align*}

\hfill$\square$

This result allows us to prove the following theorem.

\begin{theorem}\label{thm:hellinger:mixtureMapMeasurable}
    The mixture map \(T_k : D_k \to \mDY\) is measurable.
\end{theorem}

\noindent{\sc Proof of Theorem~\ref{thm:hellinger:mixtureMapMeasurable}:}
    From Lemma~\ref{lem:hellinger:sigmaAlgebraCountablyGenerated} it suffices to consider for \(f_0 \in \mF\) and \(r \in \mathbb{Q}_+\) the set 
    \[
        T_k^{-1}(B_{\Hd}(f_0, r)) = \int{G\in D_k:\,\, (u_0\circ T_k)(G) \geq 1 - r^2} = (u_0\circ T_k)^{-1}([1-r^2, \infty)).
    \]
    We conclude from Lemma~\ref{lem:hellinger:regFunctionalsMeasurable} that \(T_k^{-1}(B_{\Hd}(f_0, r))\) is measurable. The theorem then follows.
\hfill$\square$

\subsubsection{Support of the mixtures induced by the processes DSBASp and DSBASg}
\label{sec:hellinger:supportOfMixtures}

Given \(f_0\in \mDY\) and \(\eps > 0\) we want to show that
\[
    \Prob(\set{\omega\in \Omega^*:\,\, T_k G(\omega) \in B_{\Hd}(f_0, \eps)}) = \Prob(\set{\omega \in \Omega^*:\,\, G(\omega) \in T_k^{-1}(B_{\Hd}(f_0,\eps))}) > 0.
\]
Remark that the above is well defined as \(T_k^{-1}(B_{\Hd}(f_0,\eps))\) is measurable. Our arguments rely heavily on the following decomposition. As it follows from elementary algebraic manipulations, we omit its proof for brevity.

\begin{proposition}\label{prop:hellinger:supportDecomposition}
    Let \(f\in \mDsY\) and fix \(\phi \in \Phis\). Define
    \[
        f^{\phi}(y) = \int_{\Th} f(\th) k(y|\, \th, \phi)\, d\th,
    \]
    and let \(K\subset \Th\) be measurable. Then for any partition \(\set{A_m}_{m=1}^{M}\) of \(K\) into measurable sets and any selection \(\th_m\in A_m\) for which the collection \(\set{|A_m| f(\th_m)}_{m=1}^M\) is not identically zero, the function
    \[
        f_M^\phi(y) = \frac{1}{I_M}\sum_{m=1}^M |A_m| f(\th_m) k(y|\, \th_m, \phi),
    \]
    where
    \[
        I_M = \sum_{m=1}^M |A_m| f(\th_m),
    \]
    is in \(\mDY\) and
    satisfies
    \begin{align*}
        f^{\phi}(y) - f_M^{\phi}(y) &= \int_{\mY\setminus K} f(\th) k(y|\,\th,\phi)\, d\th + \sum_{m=1}^M \int_{A_m} (f(\th) - f(\th_m)) k(y|\, \th,\phi)\, d\th \\
        &\quad +\: \sum_{m=1}^M f(\th_m) \int_{A_m} (k(y|\, \th,\phi) - k(y|\, \th_m,\phi))\, d\th\\
        &\quad +\: \left(1 - \frac{1}{I_M}\right) \sum_{m=1}^M |A_m| f(\th_m) k(y|\, \th_m, \phi).
    \end{align*}
\end{proposition}

The assumption \(f\in\mDsY\) implies that
\[
    \th\in \supp(f):\,\, \int_{\mY} k(y|\, \th,\phi)\, dy = 1.
\]
Therefore, the mixing
\begin{equation}\label{eq:hellinger:discreteMixing}
    G_M^\phi(\cdot) = \frac{1}{I_M}\sum_{m = 1}^M |A_m| f(\th_m) (\delta_{\th_m}\otimes\delta_{\phi})(\cdot),
\end{equation}
is in \(D_k\) and \(f_M^\phi = T_k G_M^\phi\). We now show that, in all cases that we consider and under suitable conditions, we can control the approximation error in each term of the decomposition in Proposition~\ref{prop:hellinger:supportDecomposition}.

\begin{lemma}\label{lem:hellinger:discreteApproximation}
    Let \(f\in \mDsY\) and let \(\phi \in \Phis\). Then for every \(\eps > 0\) there exists \(M \in \N\) such that
    \[
        \Hd(f^\phi, f_M^{\phi}) < \eps.
    \]
    where \(f^\phi\) and \(f_M^\phi\) are as in Proposition~\ref{prop:hellinger:supportDecomposition}.
\end{lemma}

\noindent{\sc Proof of Lemma \ref{lem:hellinger:discreteApproximation}:}
    From Proposition~\ref{prop:hellinger:compactIntervalBound} we can choose \(K_y\subset \mYs\) compact such that
    \[
        \Hd(f^{\phi}, f)^2 \leq \frac{1}{2}\eps^2 + \left(\int_{K_y} |f^{\phi}(y) - f(y)|\, dy\right)^{1/2}.
    \]
    We will use Proposition~\ref{prop:hellinger:supportDecomposition} to control the difference \(f^\phi - f_M^{\phi}\) over \(K_y\). To do this, we will use Proposition~\ref{prop:hellinger:continuityKernel}. First, let \(K_{\th}\subset \Ths\) be such that
    \[
        \int_{\Th\setminus K_{\th}} f(\th)\, d\th < \frac{1}{32}\eps^4.
    \]
    Since for all the cases that we consider \(\Ths\) is convex, by taking the convex envelope of \(K_{\th}\) we can assume that \(K_{\th}\) is a compact interval. Therefore, it suffices to consider its partition into \(M\) intervals of equal length. Since \(f\) is uniformly continuous on \(K_{\th}\) we may select \(M\) sufficiently large so that
    \[
        m\in\set{1,\ldots, M}:\,\, \th',\th\in A_m\,\,\Rightarrow\,\, |f(\th') - f(\th)| < \frac{1}{32 |K_{\th}|}\eps^4.
    \]
    In this case, by choosing \(\th_m \in A_m\) so that the collection \(f(\th_1),\ldots, f(\th_M)\) is not identically zero, we obtain the bounds
    \[
        \left|\int_{K_{\th}} f(\th)\, d\th - I_M\right| = \sum_{m=1}^M \int_{A_m} |f(\th) - f(\th_m)|\,d\th < \frac{1}{32|K_{\th}|}\eps^4 \sum_{m=1}^M |A_m| < \frac{1}{32}\eps^4,
    \]
    and
    \begin{align*}
        \left|\sum_{m=1}^M \int_{A_m} (f(\th) - f(\th_m)) k(y|\, \th,\phi)\, d\th\right| &< \frac{1}{32|K_{\th}|}\eps^4 \sum_{m=1}^M\int_{A_m} k(y|\, \th,\phi)\,d\th, \\
        &= \frac{1}{32|K_{\th}|}\eps^4 \int_{K_{\th}} k(y|\, \th,\phi)\,d\th.
    \end{align*}
    Remark that the first condition implies that
    \[
        |I_M - 1| < \frac{1}{32}\eps^4 + \int_{\Th\setminus K_{\th}} f(\th)\, d\th < \frac{1}{16}\eps^4.
    \]
    If we let \(K_\phi = \set{\phi}\), then \(K_y, K_\th\) and \(K_{\phi}\) satisfy the hypotheses of Proposition~\ref{prop:hellinger:continuityKernel}. Therefore, we can choose \(\delta > 0\), such that
    \[
        y\in K_y,\,\, \th',\th\in K_{\th}:\,\, |\th'-\th| < \delta\,\,\Rightarrow\,\, |k(y|\, \th', \phi) - k(y|\,\th,\phi)| < \frac{1}{16|K_{\th}| |K_y| B}\eps,
    \]
    where \(B > 0\) is an upper bound for \(|f|\). Thus, for \(y\in K_y\) we have that
    \begin{align*}
        \left|\sum_{m=1}^M f(\th_m) \int_{A_m} (k(y|\,\th,\phi) - k(y|\,\th_m,\phi))\, d\th\right| &\leq \sum_{m=1}^M f(\th_m) \int_{A_m} |k(y|\,\th,\phi) - k(y|\,\th_m,\phi)|\, d\th, \\
        &<\sum_{m=1}^M B \frac{1}{16|K_{\th}||K_y| B} \eps^4 |A_m|,\\
        &= \frac{1}{16|K_y|} \eps^4.
    \end{align*}
    Consequently, for any \(y\in K_y\) we have that
    \begin{align*}
        |f^\phi(y) - f_M^\phi(y)| &< \int_{\mY\setminus K_{\th}} f(\th) k(y|\,\th,\phi)\, d\th + \frac{1}{32|K_{\th}|}\eps^4 \int_{K_{\th}} k(y|\, \th,\phi)\,d\th \\
        &\qquad +\: \frac{1}{16|K_y|} \eps^4 + \frac{|{I_M} - 1|}{I_M}\sum_{m=1}^M |A_m| f(\th_m) k(y|\,\th_m, \phi).
    \end{align*}
    Since \(f\in \mDsY\) we must have for any \(\th\in \supp(f)\) that
    \[
        \int_{K_y} k(y|\, \th,\phi)\, dy \leq \int_{\mY} k(y|\, \th,\phi)\, dy = 1.
    \]
    We deduce from this that
    \begin{align*}
        \int_{K_y} |f^\phi(y) - f_M^\phi(y)|\, dy &< \int_{\mY\setminus K_{\th}} f(\th) \left(\int_{K_y} k(y|\,\th,\phi)\, dy\right)\, d\th \\
        &\quad +\: \frac{1}{32|K_{\th}|}\eps^4 \int_{K_{\th}} \left(\int_{K_y} k(y|\, \th,\phi)\, dy\right)\,d\th \\
        &\qquad +\: \frac{1}{16} \eps^4 |K_y| +  \frac{|{I_M} - 1|}{I_M}\sum_{m=1}^M |A_m| f(\th_m) \int_{K_y} k(y|\,\th_m, \phi)\,dy,\\
        &\leq \int_{\Th\setminus K_{\th}} f(\th) \, d\th + \frac{1}{16|K_{\th}|}\eps^4 \int_{K_{\th}}\,d\th \\
        &\qquad +\: \frac{1}{16|K_y|} \eps^4 |K_y| + |{I_M} - 1| \frac{1}{I_M}\sum_{m=1}^M |A_m| f(\th_m),\\
        &< \frac{1}{32}\eps^4 + \frac{1}{16}\eps^4 + \frac{1}{16}\eps^4 + \frac{1}{16}\eps^4,\\
        &< \frac{1}{4}\eps^4,
    \end{align*}
    whence
    \[
        \Hd(f^{\phi}, f_M^{\phi}) < \eps
    \]
    as we wanted to prove.
\hfill$\square$

We now prove the last auxiliary result.

\begin{lemma}\label{lem:hellinger:positiveProbability}
    Suppose that the hypotheses of Theorem~\ref{thm:dsbagFullSupport} hold for the DSBASp and the hypotheses of Theorem~\ref{thm:dsbapFullSupport} hold for the DSBASg. Consider the same hypotheses of Lemma~\ref{lem:hellinger:discreteApproximation} and let \(f^\phi_M\) be as in Proposition~\ref{prop:hellinger:supportDecomposition} for \(M = 2^{n_0 - 1}\) for some \(n_0 \in \N\). Then for each of the processes DSBASp and DSBASg and every \(\eps > 0\) there exists an event \(\Omega^\star \subset \Omega^*\) of positive measure such that
    \[
        \omega\in \Omega^{\star}:\,\, \Hd(G(\omega), f^\phi_M) < \eps.
    \]
\end{lemma}

\noindent{\sc Proof of Lemma~\ref{lem:hellinger:positiveProbability}:}
    Since \(f^\phi_M = T_k G_M^\phi\) with \(G_M^\phi\) as in~\eqref{eq:hellinger:discreteMixing} it suffices to show that we can approximate this mixing measure. We first prove the following auxiliary result. If we define
    \[
        \bar{f}_M^\phi(y) = \sum_{m=1}^M \pi_m k(y|\,\th_m,\phi),
    \]
    for some \(\pi_1,\ldots, \pi_M \geq 0\) with \(\pi_1 + \ldots + \pi_M = 1\), then
    \[
        \int_{\mY} |f_M^\phi(y) - \bar{f}_M^\phi(y)|\, dy \leq \sum_{m=1}^M \left|\pi_m - \frac{1}{I_M} |A_m| f(\th_m)\right|.
    \]
    Hence, we can select \(\pi_1,\ldots,\pi_M > 0\), such that
    \[
        \Hd(f_M^\phi, \bar{f}_M^\phi) < \frac{1}{2}\eps.
    \]
    Using Proposition~\ref{prop:hellinger:compactIntervalBound} we can find a compact \(K_y\subset \mY\), such that
    \[
        \Hd(\bar{f}_M^\phi, f)^2 < \frac{1}{4}\eps^2 + \left(\int_{K_y} |\bar{f}_M^\phi(y) - f(y)|\, dy\right)^{1/2}.
    \]
    Let \(K_{\th}\subset \cvxhull(\supp(f))\) be a compact interval with non-empty interior, such that
    \[
        \th_1,\ldots, \th_m \in \topint(K_{\th}),
    \]
    and let \(K_{\phi}\subset \Phi\) be a compact interval that does not contain \(0\) and contains \(\phi\) on its interior. Then, from Proposition~\ref{prop:hellinger:compactIntervalBound}, the compact sets \(K_y, K_{\th}\) and \(K_\phi\) satisfy the hypotheses of Proposition~\ref{prop:hellinger:continuityKernel}. Then, for any
    \[
        \th'_1,\ldots,\th_M' \in K_{\th}\,\,\mbox{and}\,\, \phi_1',\ldots,\phi_M'\in K_{\phi},
    \]
    the function
    \[
        f(y) = \sum_{m= 1}^M \pi'_m k(y|\, \th_m', \phi_m'),
    \]
    where \(\pi'_1,\ldots, \pi'_M \geq 0\) with \(\pi'_1 + \ldots + \pi'_M = 1\) defines an element of \(\mDY\), for which we have that
    \begin{align*}
        \int_{K_y} |f_M^\phi(y) - f(y)|\, dy &\leq \sum_{m=1}^M |\pi_m - \pi'_m| + \sum_{m=1}^M \pi_i \int_{K_y} |k(y|\, \th_m, \phi) - k(y|\, \th_m', \phi')|\, dy.
    \end{align*}
    Since \((y,\th,\phi)\mapsto k(y|\,\th,\phi)\) is continuous on \(K_{y}\times K_{\th}\times K_\phi\) there exists \(\delta > 0\) such that for \(y',y\in K_y\), \(\th',\th\in K_{\th}\) and \(\phi',\phi\in K_{\phi}\), we have that
    \[
        \max\set{|y' - y|,\,|\th'-\th|,\, |\phi'-\phi|} < \delta\,\,\Rightarrow\,\, |k(y'|\, \th',\phi') - k(y|\,\th,\phi)| < \frac{1}{32 |K_y|}\eps^4.
    \]
    Therefore, if \(|\th'_m - \th_m| < \delta\) for \(m\in \set{1,\ldots, M}\) and \(|\phi' - \phi| <\delta\), then
    \[
        \int_{K_y}|k(y|\,\th_m,\phi) - k(y|\, \th'_m,\phi')|\, dy < \frac{1}{32}\eps^4.
    \]
    Consider now the marginal \(G_{M,1}^{\phi}\in \mPTh\), given by
    \[
        G_{M,1}^{\phi}(\cdot) = \sum_{m=1}^M \pi_m \delta_{\th_m}(\cdot).
    \]
    Since \(M = 2^{n_0 - 1}\) and \(\pi_1,\ldots, \pi_M > 0\), this corresponds to a regular SBA of level \(n_0\). 
    
    We now proceed to prove the result for the DSBASp. Since \(M = 2^{n_0-1}\), the process DSBASp assigns positive probability to the event
    \begin{align*}
        \Omega_{\th} &:= \set{\omega\in\Omega:\,\, n(\omega) = n_0}\cap\\
        &\quad \bigcap_{m=1}^M\Lset{\omega\in\Omega:\,\, |\mu_{n_0,m}(\omega) - \th_m| < \delta,\, |w_{n_0, m}(\omega) - \pi_m| < \frac{1}{32M}\eps^4},
    \end{align*}
    and to the event
    \[
        \Omega_{\phi} :=\set{\omega\in\Omega:\,\, n(\omega) = n_0}\cap\bigcap_{m=1}^M\Lset{\omega\in\Omega:\,\, |\phi_{m}(\omega) - \phi| < \delta}.
    \]
    Since they are independent, we may define \(\Omega^{\star} = \Omega_{\th}\cap\Omega_{\phi}\). Then \(\Prob(\Omega^{\star}) > 0\) and for \(\omega \in \Omega^{\star}\) we have that for \(f(\omega) = T_k G(\omega)\), it holds that
    \begin{align*}
        \int_{K_y} |f_M^\phi(y) - f(\omega)(y)|\, dy &\leq \sum_{m=1}^M |\pi_m - w_{n_0,m}| \\
        &\quad +\: \sum_{m=1}^M \pi_i \int_{K_y} |k(y|\, \th_m, \phi) - k(y|\, \mu_{n_0,m}(\omega), \phi_m(\omega))|\, dy, \\
        &< \frac{1}{32M}\eps^4 M + \frac{1}{32}\eps^4,\\
        &= \frac{1}{16}\eps^4,
    \end{align*}
    whence
    \[
        \Hd(f_M^{\phi}, f(\omega))^2 < \Hd(f_M^{\phi}, \bar{f}_M^{\phi}) + \Hd(\bar{f}_M^{\phi}, f(\omega)) < \frac{1}{4}\eps^2 + \frac{1}{4}\eps^2 + \frac{1}{4}\eps^2 < \eps^2,
    \]
    proving the claim. 
    
    To prove the statement for the DSBASg remark that the same arguments show that the DSBASg assigns positive probability to the event
    \begin{align*}
        \Omega_{\th} &:= \set{\omega\in\Omega:\,\, m_1(\omega) = n_0}\cap\\
        &\quad \bigcap_{m=1}^M\Lset{\omega\in\Omega:\,\, |\mu_{n_0,m}(\omega) - \th_m| < \delta,\, |w^{\th}_{n_0, m}(\omega) - \pi_m| < \frac{1}{32M}\eps^4}
    \end{align*}
    and to the event
    \[
        \Omega_{\phi} :=\set{\omega\in\Omega:\,\, m_2(\omega) = 1}\cap\bigcap_{m=1}^M\Lset{\omega\in\Omega:\,\, |\phi_{1}(\omega) - \phi| < \delta}.
    \]
    Remark that for \(\omega\in \Omega_{\phi}\) we must have that \(w_{n_0,1}^{\phi}(\omega) = 1\) almost surely. Therefore, as before, we may define \(\Omega^{\star} = \Omega_{\th}\cap\Omega_{\phi}\). Then \(\Prob(\Omega^{\star}) > 0\) and for \(\omega \in \Omega^{\star}\) we have that for \(f(\omega) = T_k G(\omega)\) it holds that
    \begin{align*}
        \int_{K_y} |f_M^\phi(y) - f(\omega)(y)|\, dy &\leq \sum_{m=1}^M |\pi_m - w^{\th}_{n_0,m}| \\
        &\quad +\: \sum_{m=1}^M \pi_i \int_{K_y} |k(y|\, \th_m, \phi) - k(y|\, \mu_{n_0,m}(\omega), \phi_1(\omega))|\, dy, \\
        &< \frac{1}{8M}\eps^4 M + \frac{1}{8}\eps^4,\\
        &= \frac{1}{4}\eps^4,
    \end{align*}
    whence the same inequality as before shows that
    \[
        \Hd(f_M^{\phi}, f(\omega))^2 < \eps^2,
    \]
    proving the lemma.
\hfill$\square$

We can now prove the main theorem.

\noindent{\sc Proof of Theorem~\ref{thm:hellinger:support}:}
    As discussed earlier, it suffices to show that for any \(f_0\in \mDY\) and \(\eps > 0\), we have that
    \[
        \Prob(\set{\omega\in\Omega^*:\,\, T_k G(\omega)\in B_{\Hd}(f_0,\eps)}) > 0.
    \]
    By Proposition~\ref{prop:hellinger:densityOfCoc}, there exists \(\bar{f}_0 \in \mDsY\), such that
    \[
        \Hd(f_0, \bar{f}_0) < \frac{1}{4}\eps.
    \]
    In turn, Lemma~\ref{lem:hellinger:mixtureRangeDense} implies that for all the kernels under consideration there exists \(\phi > 0\), such that
    \[
        \Hd(\bar{f}_0, \bar{f}^{\phi}) < \frac{1}{4}\eps.
    \]
    Finally, from Lemma~\ref{lem:hellinger:discreteApproximation} we can choose \(n_0 \in \N\) such that \(M = 2^{n_0 - 1}\) is sufficiently large so that
    \[
        \Hd(\bar{f}^{\phi}, \bar{f}_M^{\phi}) < \frac{1}{4}\eps.
    \]
    Therefore,
    \[
        \Hd(f_0, \bar{f}_M^{\phi}) < \frac{3}{4}\eps.
    \]
    Now, it suffices to use Lemma~\ref{lem:hellinger:positiveProbability} to prove that there exists an event \(\Omega^{\star}\) such that DSBASp and DSBASg assigns positive probability to it, for which
    \[
        \omega\in \Omega^{\star}:\,\, \Hd(\bar{f}_M^{\phi}, T_kG(\omega)) < \frac{1}{4}\eps.
    \]
    From this statement the theorem follows.
\hfill$\square$

\bibliographystyle{jss}
\bibliography{ref}

@Article{arellano;castro;genton;gomez;2008,
  author =   {Arellano-Valle, R B and Castro, L M and Genton, M G and G\'omez, H},
  title =    {{Bayesian inference for shape mixtures of skewed distributions, with application to regression analysis}},
  journal =  {Bayesian Analysis},
  year =     {2008},
  volume =   {3},
  pages =    {513--540},
}

@Article{cifarelli;regazzini;90,
  author =       {Cifarelli, D M and Regazzini, E},
  title =        {Distribution functions of means of a {D}irichlet process},
  journal =      {The Annals of Statistics},
  year =         {1990},
  volume =       {18},
  pages =        {429--442},
}

@Book{cook;weisberg;94,
  author =   {Cook, D R and Weisberg, S},
  title =    {{An Introduction to Regression Graphics}},
  publisher = {John Wiley and Sons},
  year =     {1994},
  address =  {New York, USA},
}

@Article{ferguson;73,
  author =       {Ferguson, T S},
  title =        {A {B}ayesian  analysis of some nonparametric problems},
  journal =      {Annals of Statistics},
  year =         {1973},
  volume =       {1},
  pages =        {209--230}
}

@Article{geisser;eddy;79,
  author =       {Geisser, S and Eddy, W},
  title =        {A predictive approach to model selection},
  journal =      {Journal of the American Statistical Association},
  year =         {1979},
  volume =       {74},
  pages =        {153--160},
}

@book{GhosalVanDerVaart2017,
  author    = {Ghosal, Subhashis and van der Vaart, Aad},
  title     = {Fundamentals of Nonparametric {B}ayesian Inference},
  series    = {Cambridge Series in Statistical and Probabilistic Mathematics},
  volume    = {44},
  publisher = {Cambridge University Press},
  address   = {Cambridge, UK},
  year      = {2017},
  doi       = {10.1017/9781139029834},
  isbn      = {9781107034424}
}

@Article{green;richardson;97,
  author =       {Green, P J and Richardson, S},
  title =        {{Modelling heterogeneity with and without the Dirichlet process}},
  journal =      {Scandinavian Journal of Statistics},
  year =         {2001},
  volume =       {28},
  pages =        {355--375},
}

@Article{hill;monticino;98,
  author =   {Hill, T and Monticino, M},
  title =    {{ Constructions of random distributions via sequential barycenters}},
  journal =  {The Annals of Statistics},
  year =     {1998},
  volume =   {26},
  number = {4},
  pages =    {1242--1253},
 }

@Article{james;2005,
  author =       {James, L F},
  title =        {{Functionals of Dirichlet processes, the Cifarelli-Regazzini identity
and beta-gamma processes}},
  journal =      {The Annals of Statistics},
  year =         {2005},
  volume =       {33},
  pages =        {647--660},
}

@Article{james;lijoi;pruesnter;2008,
  author =       {James, L F and Lijoi, A and Pr\"{u}nster, I},
  title =        {{Distributions of linear functionals of two parameter Poisson-Dirichlet random measures}},
  journal =      {The Annals of Applied Probability},
  year =         {2008},
  volume =       {18},
  pages =        {521--551},
}

@Article{jara;etal;2011,
    title = {{DPpackage}: Bayesian Semi- and Nonparametric Modeling in
      {R}},
    author = {Alejandro Jara and Timothy Hanson and Fernando Quintana
      and Peter M\"uller and Gary Rosner},
    journal = {Journal of Statistical Software},
    year = {2011},
    volume = {40},
    number = {5},
    pages = {1--30},
  }

@Article{jara;2007,
    title = {Applied {B}ayesian Non- and Semi-parametric Inference
      Using {DPpackage}},
    author = {Alejandro Jara},
    journal = {R News},
    year = {2007},
    volume = {7},
    number = {3},
    pages = {17--26},
  }

@Article{kessler;hoff;dunson;2015,
  author =       {Kessler, D C and Hoff, P and Dunson, D B},
  title =        {{Marginally specified priors for non-parametric Bayesian estimation}},
  journal =      {Journal of the Royal Statistical Society, Series B},
  year =         {2015},
  volume =       {77},
  pages =        {35--58},
}

@Article{lijoi;pruenster;2009,
  author =       {Lijoi, A and Pr\"{u}nster, I},
  title =        {{Distributional properties of means of random probability measures}},
  journal =      {Statistics Surveys},
  year =         {2009},
  volume =       {3},
  pages =        {47--95},
}

@Article{lindsay;83,
  author =       {Lindsay, B G},
  title =        {{The geometry of mixture likelihoods, Part I: A general theory}},
  journal =      {The Annals of Statistics},
  year =         {1983},
  volume =       {11},
  pages =        {86--94},
}

@Book{mueller;quintana;jara;hanson;2015,
  author =       {M\"{u}ller, P and Quintana, F A and Jara, A and Hanson, T E},
  title =        {Bayesian Nonparametric Data Analysis},
  publisher = {Springer},
  address = {New York, USA},
  year = {2015}
}

@Article{neal;2003,
  author =       {Neal, R},
  title =        {{Slice sampling}},
  journal =      {The Annals of Statistics},
  year =         {2003},
  volume =       {31},
  pages =        {705--767},
}

@Article{nieto;pruenster;walker;2004,
  author =       {Nieto-Barajas, L E and Pr\"{u}nster, I and Walker, S},
  title =        {{Normalized random measures driven by increasing additive processes}},
  journal =      {The Annals of Statistics},
  year =         {2004},
  volume =       {32},
  pages =        {2343--2360},
}

@Article{pitman;yor;97,
  author =       {Pitman, J and Yor, M},
  title =        {{The two-parameter Poisson-Dirichlet distribution derived from a stable subordinator}},
  journal =      {The Annals of Probability},
  year =         {1997},
  volume =       {25},
  pages =        {855--900},
   }

@Article{roeder;1990,
  author =   {Roeder, K},
  title =    {Density estimation with confidence sets exemplified by superclusters and voids in the galaxies},
  journal =  {Journal of the American Statistical Association},
  year =     {1990},
  volume =   {85},
  pages =    {617--624},
}

@Article{silverman;81,
  author =       {Silverman, B W},
  title =        {{Using kernel density estimates to investigate multimodality}},
  journal =      {Journal of the Royal Society, Series B},
  year =         {1981},
  volume =       {43},
  pages =        {97--99},
}

@book{Villani2003,
  author    = {Villani, C{\'e}dric},
  title     = {Topics in Optimal Transportation},
  series    = {Graduate Studies in Mathematics},
  volume    = {58},
  publisher = {American Mathematical Society},
  address   = {Providence, RI},
  year      = {2003},
  doi       = {10.1090/gsm/058},
  isbn      = {978-0-8218-3312-4}
}

@book{villani2008optimal,
  title={Optimal Transport: Old and New},
  author={Villani, Cédric},
  volume={338},
  year={2008},
  publisher={Springer Science \& Business Media}
}

@article{GilLeyva2023,
author = {María F. Gil–Leyva and Ramsés H. Mena},
title = {Stick-Breaking Processes With Exchangeable Length Variables},
journal = {Journal of the American Statistical Association},
volume = {118},
number = {541},
pages = {537--550},
year = {2023},
publisher = {ASA Website}}

@MISC{Plummer_JAGS_2003,
    author = "Plummer, M",
    title = "{JAGS}: A Program for Analysis of Bayesian Graphical Models Using Gibbs Sampling",
    year = "2003",
    month = "March 20-22",
    howpublished = "Proceedings of the 3rd International Workshop on Distributed Statistical Computing (DSC 2003), Vienna",
    pages = "1--10",
}

@article{waic, author = {Watanabe, Sumio}, title = {A widely applicable {B}ayesian information criterion}, year = {2013}, issue_date = {January 2013}, publisher = {JMLR.org}, volume = {14}, number = {1}, issn = {1532-4435}, journal = {Journal of Machine Learning Research}, month = {mar}, pages = {867–897}, numpages = {31}}

@article{zellner1986gprior,
  title        = {On assessing prior distributions and Bayesian regression analysis with $g$-prior distributions},
  author       = {Zellner, Arnold},
  journal      = {Bayesian Inference and Decision Techniques: Essays in Honor of Bruno de Finetti},
  editor       = {Goel, Prem K. and Zellner, Arnold},
  pages        = {233--243},
  year         = {1986},
  publisher    = {North-Holland}
}

@book{Peebles1980,
  author    = {P. J. E. Peebles},
  title     = {The Large-Scale Structure of the Universe},
  publisher = {Princeton University Press},
  year      = {1980},
  address   = {Princeton, NJ}
}

@article{York2000,
  author    = {Donald G. York and John Adelman and John Anderson, Jr. and Scott F. Anderson and James Annis and Neta A. Bahcall and others},
  title     = {The Sloan Digital Sky Survey: Technical Summary},
  journal   = {The Astronomical Journal},
  year      = {2000},
  volume    = {120},
  number    = {3},
  pages     = {1579--1587},
  doi       = {10.1086/301513}
}

@article{Strauss2002,
  author    = {Michael A. Strauss and David H. Weinberg and Robert H. Lupton and Vladimir K. Narayanan and others},
  title     = {Spectroscopic Target Selection in the Sloan Digital Sky Survey: The Main Galaxy Sample},
  journal   = {The Astronomical Journal},
  year      = {2002},
  volume    = {124},
  number    = {3},
  pages     = {1810--1824},
  doi       = {10.1086/342343}
}

@article{Colless2001,
  author    = {Matthew Colless and Gavin Dalton and Steve Maddox and Will Sutherland and others},
  title     = {The 2dF Galaxy Redshift Survey: Spectra and Redshifts},
  journal   = {Monthly Notices of the Royal Astronomical Society},
  year      = {2001},
  volume    = {328},
  number    = {4},
  pages     = {1039--1063},
  doi       = {10.1046/j.1365-8711.2001.04902.x}
}

@article{GaffiLijoiPrunster2025,
  author       = {Gaffi, Francesco and Lijoi, Antonio and Pr{\"u}nster, Igor},
  title        = {Random probability measures with fixed mean distributions},
  journal      = {The Annals of Applied Probability},
  year         = {2025},
  volume       = {35},
  number       = {4},
  pages        = {2239--2261},
  doi          = {10.1214/24-AAP2130},
}

@article{LijoiRegazzini2004,
  author       = {Lijoi, Antonio and Regazzini, Eugenio},
  title        = {Means of a Dirichlet process and multiple hypergeometric functions},
  journal      = {The Annals of Probability},
  year         = {2004},
  volume       = {32},
  number       = {2},
  pages        = {1469--1495},
  doi          = {10.1214/009117904000000424}
}

@incollection{LijoiPrunster2010,
  author       = {Lijoi, Antonio and Pr{\"u}nster, Igor},
  title        = {Models beyond the Dirichlet process},
  booktitle    = {Bayesian Nonparametrics},
  editor       = {Hjort, Nils Lid and Holmes, Chris and M{\"u}ller, Peter and Walker, Stephen G.},
  pages        = {80--136},
  publisher    = {Cambridge University Press},
  address      = {Cambridge},
  year         = {2010},
  doi          = {10.1017/CBO9780511802478.004}
}

@Article{sanmartin;jara;rolin;mouchrat;2011,
     author = {San Mart\'{\i}n, E and Jara, A and Rolin, J-M and Mouchart, M},
     title = {{On the analysis of Bayesian semiparametric IRT-type models}},
     journal = {Psychometrika},
     year = {2011},
     volume = {76},
     pages = {385 -- 409}
}

@article{kuschinskiJara2025griduniform,
  title   = {Grid-uniform copulas and rectangle exchanges: Bayesian model and inference for a rich class of copula functions},
  author  = {Kuschinski, Nicol{\'a}s and Jara, Alejandro},
  journal = {Bayesian Analysis},
  year    = {2025},
  volume  = {20},
  number  = {1},
  pages   = {1357--1384},
  doi     = {10.1214/23-BA1396}
}

@misc{kuschinskiWarrJara2024bernsteinYettUniform,
  title         = {Bayesian copula density estimation using Bernstein Yett-Uniform priors},
  author        = {Kuschinski, Nicol{\'a}s and Warr, Richard and Jara, Alejandro},
  year          = {2024},
  eprint        = {2405.04475},
  archivePrefix = {arXiv},
  primaryClass  = {stat.ME},
  doi           = {10.48550/arXiv.2405.04475},
  url           = {https://arxiv.org/abs/2405.04475},
  note          = {Version v1}
}

@book{Schneider2013,
  author    = {Rolf Schneider},
  title     = {Convex Bodies: The Brunn–Minkowski Theory},
  edition   = {2nd},
  series    = {Encyclopedia of Mathematics and its Applications},
  volume    = {151},
  publisher = {Cambridge University Press},
  address   = {Cambridge},
  year      = {2013},
  doi       = {10.1017/CBO9781139003858},
  isbn      = {9781139003858}
}

@book{Santambrogio2015,
  author    = {Filippo Santambrogio},
  title     = {Optimal Transport for Applied Mathematicians: Calculus of Variations, PDEs, and Modeling},
  series    = {Progress in Nonlinear Differential Equations and Their Applications},
  volume    = {87},
  publisher = {Birkhäuser Cham},
  year      = {2015},
  doi       = {10.1007/978-3-319-20828-2},
  isbn      = {978-3-319-20827-5, 978-3-319-20828-2}
}

\end{document}